\documentclass[english,prl,aps,floatfix,twocolumn,longbibliography,superscriptaddress]{revtex4-1}
\usepackage[T1]{fontenc}
\usepackage[latin9]{inputenc}
\usepackage{color}
\usepackage{float}
\usepackage{bm}
\usepackage{amsmath}
\usepackage{amssymb}
\usepackage{graphicx}

\makeatletter

\usepackage[colorlinks=true,linkcolor=blue,anchorcolor=red,citecolor=blue,urlcolor=blue]{hyperref}

\makeatother

\usepackage{babel}
\begin{document}
\renewcommand{\figurename}{Fig.}
\title{Detection of second-order topological superconductors by Josephson junctions}
\author{Song-Bo Zhang}
\affiliation{Institut f\"ur Theoretische Physik und Astrophysik, Universit\"at
W\"urzburg, D-97074 W\"urzburg, Germany}
\author{Bj\"orn Trauzettel}
\affiliation{Institut f\"ur Theoretische Physik und Astrophysik, Universit\"at
W\"urzburg, D-97074 W\"urzburg, Germany}
\affiliation{W\"urzburg-Dresden Cluster of Excellence ct.qmat, Germany}
\date{\today}
\begin{abstract}
We study Josephson junctions based on second-order topological superconductors
(SOTSs) which can be realized in quantum spin Hall insulators with large
inverted gap in proximity to unconventional superconductors. We
find that tuning the chemical potential in the superconductor strongly
modifies the induced pairing of the helical edge states, resulting in topological phase transitions. In a corresponding Josephson junction, a $0$-$\pi$ transition is realized
by tuning the chemical potentials in the superconducting leads. This
striking feature is stable in junctions with respect to different sizes, doping the normal region, and the presence of disorder. Our transport results can serve
as novel experimental signatures of SOTSs. Moreover, the $0$-$\pi$
transition constitutes a fully electric way to create or annihilate
Majorana bound states in the junction without any magnetic manipulation.
\end{abstract}
\maketitle
\textit{Introduction.}\textendash The second-order topological superconductor
(SOTS) is a novel topological phase of matter and features Majorana
zero-dimensional (0D) corner or 1D hinge states which are two spatial dimensions
lower than the gapped bulk \citep{Langbehn17PRL,Khalaf18PRB,QYWang18PRL,ZBYan18PRL,TLiu18PRB, Geier18PRB,XYZhu18PRB,CHHsu18PRL,YXWang18PRB,RXZhang19PRL, SSQin19arxiv,Bultinck19PRB,Peng19PRB}.
They may form stable qubits for topological quantum computation
\citep{Kitaev01PU,kitaev03AP,Nayak08RMP,Alicea12PPP,Leijnse12SST,Beenakker13ARCMP,Elliott15RMP,sarma15npj,Sato16JPSJ}.
Recently, the SOTS has been discovered in a variety of realistic materials
and triggered tremendous interest \citep{QYWang18PRL,ZBYan18PRL,TLiu18PRB,Geier18PRB,SSQin19arxiv,Konig2019arXiv,XYZhu18PRB, CHHsu18PRL,Volpez19PRL,YXWang18PRB,RXZhang19PRL,Shapourian18PRB,XHPan18arXiv, Kheirkhah19arXiv,Ghorashi19arXiv}.
One way to mimic SOTSs in 2D is given by quantum spin Hall insulators
(QSHIs) in proximity to unconventional superconductors with $d_{x^{2}-y^{2}}$-wave
or $s_{\pm}$-wave pairing order \citep{QYWang18PRL,ZBYan18PRL,TLiu18PRB}.
The proximity effect of unconventional superconductivity in 2D systems has been intensively explored in theory \citep{Linder08PRB,Linder10PRL,Schaffer13PRB,FZhang13PRL,ZXLi15PRB,Zareapour16SCT,WJLi16PRB,XXWu16PRB,ZJWang15PRB,TZhou19PRB} and experiment \citep{Zareapour12nacom,EWang13npyhs,HZhao18PRB,Perconte18Nphys,SYXu14PRB,Yilmaz14PRL}.
To date, however, the only way proposed to detect 2D SOTSs is a
tunneling experiment without a concrete calculation of the observable signature. An alternative approach to probe SOTSs and manipulate the Majorana corner modes is thus needed.
In QSHIs, a finite doping is typically present, and the chemical potential can be far away from the Dirac points. Therefore, it is certainly interesting and experimentally relevant to explore the influence of the chemical potential in SOTSs.

In this Letter, we investigate superconductor-normal metal-superconductor (SNS) junctions formed by a 2D SOTS.
The SOTS can be realized in a QSHI with a large inverted gap in proximity to an unconventional superconductor.
We introduce a minimal model which is able to capture the essential physics of the SOTS.
We find that due to the nontrivial momentum-dependence of the pairing potential and mass, the chemical potential in the SOTS alters the pairing gap opened within the edge states significantly.
It can even switch the sign of the pairing gap, leading to a topological phase transition.
While the supercurrent across the SNS junction is insensitive to the chemical potential in the N region, it depends strongly on the filling in the superconductors.
Strikingly, tuning the chemical potentials in the superconductors gives rise to a $0$-$\pi$ transition, which is absent in junctions based on conventional $s$-wave pairing.
These features are robust against disorder in junctions with different sizes.
They offer novel signatures to detect the SOTS with Majorana corner states.
Furthermore, while Majorana bound states (MBSs) emerge in the  $0$-junction when the phase difference across the junction is $\phi=\pm\pi$, they appear at vanishing $\phi$ in the $\pi$-junction.
Thus, Josephson junctions with such a doping-induced $0$-$\pi$ transition provide an innovative platform to create or annihilate MBSs by electric gating in the absence of $\phi$.
These predictions are applicable to a number of candidate systems including high-temperature QSHIs \citep{XFQian14science,SJTang17nphys,ZYFei17nphys,SFWu18science,PChen18ncomm,HMWeng15PRB,SChen16NL,Reis17science,Hsu15NJP,Wrasse14NL,JWLiu15NL,WHWan17AM}
in proximity to high-temperature cuprate or iron-based superconductors.

\textit{Minimal model for SOTSs.\textendash }We consider the minimal model
for SOTSs realized in QSHIs in proximity to superconductors,
\begin{eqnarray}
H({\bf k})& = & H_{0}({\bf k})+\Delta({\bf k})\tau_{y}s_{y},\nonumber \\
H_{0}({\bf k})& = & m({\bf k})\tau_{z}\sigma_{z}+v_{x}k_{x}s_{z}\sigma_{x}+v_{y}k_{y}\tau_{z}\sigma_{y}-\mu\tau_{z}\label{eq:General-model}
\end{eqnarray}
written in the Nambu basis $(c_{a,\uparrow,{\bf k}},c_{b,\uparrow,{\bf k}},c_{a,\downarrow,{\bf k}},c_{b,\downarrow,{\bf k}},$
$c_{a,\uparrow,-{\bf k}}^{\dagger},c_{b,\uparrow,-{\bf k}}^{\dagger},c_{a,\downarrow,-{\bf k}}^{\dagger},c_{b,\downarrow,-{\bf k}}^{\dagger})$, where $c_{\sigma,s,{\bf k}}^{\dagger}$($c_{\sigma,s,{\bf k}}$)
creates(annihilates) an electron with spin $s\in\{\uparrow,\downarrow\}$,
orbital $\sigma\in\{a,b\}$ and
the momentum ${\bf k}=(k_{x},k_{y})$ measured from the band inversion point of the QSHI. $\bm{\tau}$,
$\bm{\sigma}$ and $\bm{s}$ are Pauli matrices acting on Nambu, orbital
and spin spaces, respectively. $m({\bf k})=m_{0}-m_{x}k_{x}^{2}-m_{y}k_{y}^{2}$
is the mass term of the QSHI and $\mu$ is the chemical potential.
The band inversion implies the conditions $m_{0}m_{x}>0$ and $m_{0}m_{y}>0$ \citep{Bernevig06science}.
The pairing potential is written in general as $\Delta({\bf k})=\Delta_{0}+\Delta_{2}\left(k_{x}^{2}-k_{y}^{2}\right)$.
When $\Delta_{0}\neq0$ and $\Delta_{2}=0$, it refers to conventional
$s$-wave pairing. When $\Delta_{0}=0$ and $\Delta_{2}\neq0$, the
pairing is formally $d_{x^{2}-y^{2}}$-wave. It can be induced in
a QSHI with band inversion at the $\Gamma$ point via the proximity
to a cuprate superconductor \citep{ZBYan18PRL}. When $0\leqslant|\Delta_{0}|<m_{0}|\Delta_{2}|/2m_{x(y)}$,
the system possesses a mixture of $s$-wave
and $d_{x^{2}-y^{2}}$-wave pairing. It can also describe effectively
a QSHI with band inversion at the $X$ point \citep{WHWan17AM,Wrasse14NL,JWLiu15NL}
and $s_{\pm}$-wave pairing induced from an iron-based superconductor
\citep{Stewart11RMP,Hirschfeld11RPP,PZhang18science,DFWang18science,PZhang19nphys}.

In the absence of $\Delta({\bf k})$, the system hosts gapless helical edge states across
the bulk gap, which are protected by time-reversal symmetry. The pairing term with $\Delta_{2}\neq0$ induces a pairing gap of the edge states. The gap may switch sign at the corners, resulting in Majorana corner modes \citep{ZBYan18PRL,QYWang18PRL,TLiu18PRB}.
We note that although the model \eqref{eq:General-model} is a low-energy effective model, it captures the essential physics of the SOTS. Based on this model, we can understand the second-order topology more intuitively from the picture of edge states and show
that it can be strongly altered by changing $\mu$.

\begin{figure}[htp]
\centering

\includegraphics[width=8.5cm]{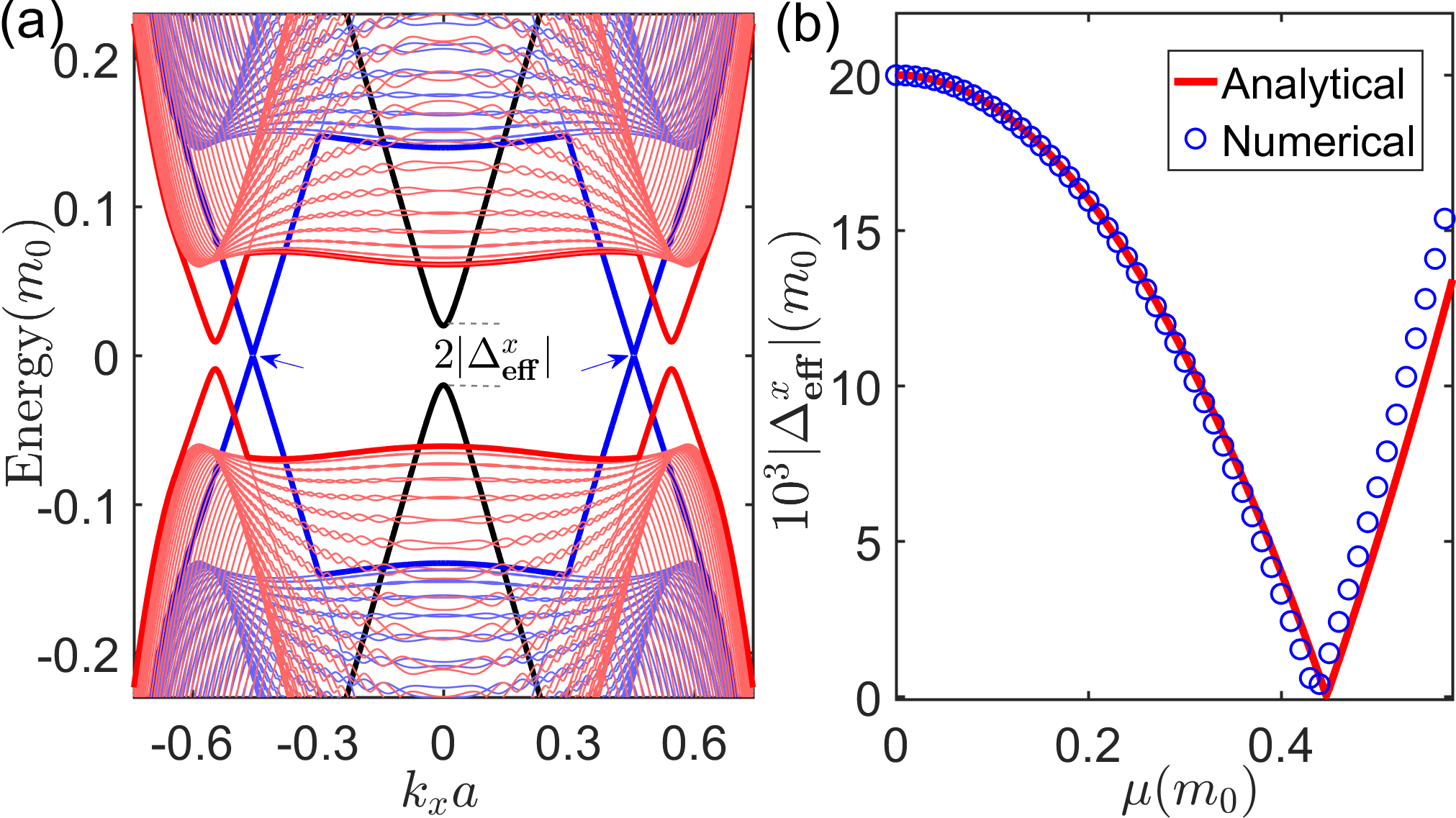}

\caption{(a) Energy spectra of the model (\ref{eq:General-model}) in a ribbon along $x$ direction for chemical potential $\mu=0$ (black), $0.44m_0$ (blue), and $0.52m_0$ (red), respectively. The thick curves close to zero energy are edge bands. The pairing gap $\Delta_{\text{eff}}^{x}$ vanishes at around $\mu=0.44m_0$, as pointed by the blue arrows. (b) $|\Delta_{\text{eff}}^{x}|$ as a function of $\mu$. The blue circles are numerical results from tight-binding calculation while the red curve is the plot of Eq.\ (\ref{eq:General-formula}). Other parameters are $m_{x(y)}=2.5,$
$v_{x(y)}=1,$ $\Delta_{0}=0,$ $\Delta_{2}=0.05$ and $200$
lattice layers in $y$ direction. The units for energy and wavenumber
are $m_{0}$ and $a^{-1}$, respectively.}

\label{fig:CaseII-gap-closing}
\end{figure}

\textit{Pairing gaps of edge states and topological phase transitions.}\textendash To analyze the Majorana corner states and the influence of $\mu$ on the SOTS, we analytically derive the effective model for
edge states. For illustration, we consider the edge along $x$ direction of the SOTS in the half-plane $y\leqslant0$ and assume hard-wall boundary conditions \citep{Note5}. As in realistic systems, we assume weak pairing. We
first calculate the edge states of $H_{0}$, following the approach of Ref.\ \citep{ZhangSB16NJP}. In this model, $k_{x}$ is a good quantum number.
The helical electron and hole edge bands are found explicitly as
\begin{eqnarray}
E_{e(h),\uparrow/\downarrow}(k_{x}) =\pm\text{sgn}(m_{y}v_{y})v_{x}k_{x}-(+)\mu.
\end{eqnarray}
The wavefunctions in the orbital basis $\{a,b\}$ read
\begin{eqnarray}
\Psi_{e,\uparrow,k_{x}}({\bf r}) =\mathcal{N}e^{ik_{x}x}(e^{\lambda_{1}y}-e^{\lambda_{2}y})\left(\text{sgn}(m_{y}v_{y}),1\right)^{T}.
\end{eqnarray}
They fulfill $\Psi_{e,\downarrow,k_{x}}({\bf r})=\Psi_{e,\uparrow,-k_{x}}^{*}({\bf r})$
and $\Psi_{h,\downarrow/\uparrow,k_{x}}({\bf r})=\Psi_{e,\downarrow/\uparrow,-k_{x}}^{*}({\bf r})$,
due to time-reversal and particle-hole symmetries;  $\lambda_{1(2)}=|v_{y}/2m_{y}|-(+)(v_{y}^{2}/4m_{y}^{2}-m_{0}/m_{y}+m_{x}k_{x}^{2}/m_{y})^{1/2}$
and $\mathcal{N}$ is the normalization factor. The decaying length
of edge states is given by $\xi_{\text{edge}}=\text{max}[1/\text{Re}(\lambda_{1}),1/\text{Re}(\lambda_{2})]$. At zero energy, the electron and hole bands touch at $k_{x}=\pm k_{c}$ with $k_{c}=\text{sgn}(m_{y}v_{y})$ $\mu/v_{x}$. For $\mu=0$, $k_{c}=0$. However, for $\mu\neq0$, the touching points  shift to finite $\pm k_{c}$. Projecting the pairing term onto the edge states, the resulting Bogoliubov de-Gennes (BdG) Hamiltonian for edge
states is obtained as
\begin{eqnarray}
h_{\text{BdG}}^{x}=\text{sgn}(m_{y}v_{y})v_{x}k_{x}\tau_{z}s_{z}-\mu\tau_{z}+\Delta_{\text{eff}}^{x}\tau_{x}s_{z}\label{eq:edgeModel}
\end{eqnarray}
in the basis $(\Psi_{e,\uparrow},\Psi_{e,\downarrow},\Psi_{h,\downarrow},\Psi_{h,\uparrow})$, and
the pairing gap is given by
\begin{eqnarray}
\Delta_{\text{eff}}^{x}=-\Delta_{0}+\Delta_{2}\left[m_{0}/m_{y}-(1+m_x/m_y)\mu^{2}/v_{x}^{2}\right].\label{eq:General-formula}
\end{eqnarray}
Without loss of generality, a real $\Delta({\bf k})$ has been assumed
\citep{Note1}.  We provide the derivation in detail in the
Supplemental Material\ \citep{SupplementalMaterial}. Similarly, for an edge along $y$ direction, we find the BdG Hamiltonian
of the same form but with a different pairing gap
\begin{eqnarray}
\Delta_{\text{eff}}^{y}=-\Delta_{0}-\Delta_{2}\left[m_{0}/m_{x}-(1+m_y/m_x)\mu^{2}/v_{y}^{2}\right].
\end{eqnarray}
The combination of $\Delta_{\text{eff}}^{x}$ and $\Delta_{\text{eff}}^{y}$
(with opposite signs) in Eq.\ (\ref{eq:edgeModel}) mimics the Jackiw-Rebbi
model \citep{Jackiw76PRD} at corners of $x$ and $y$ axes. Thus,
Majorana corner states at zero energy appear if $\Delta_{\text{eff}}^{x}\Delta_{\text{eff}}^{y}<0$.

For $s$-wave pairing, $\Delta_{\text{eff}}^{x}$ and $\Delta_{\text{eff}}^{y}$
are identical and constant. Thus, no corner state emerge.
In contrast, for unconventional pairings with $|\Delta_{0}|<m_{0}|\Delta_{2}|/2m_{x(y)}$, we obtain corner states at small $\mu$.
When $\mu=0$, the SOTS has two reflection symmetries. When $m_x=m_y$, $v_x=v_y$ and $\Delta_0=0$, it possesses a  fourfold rotation symmetry. In these particular cases, the system can be characterized by a topological invariant calculated from the bulk Hamiltonian \citep{Benalcazar17science,Benalcazar17PRB,ZDSong17PRL,SupplementalMaterial}. However, the corner states in our model are not restricted to any crystalline symmetries.
Interestingly, $\Delta_{\text{eff}}^{x(y)}$ depends strongly on $\mu$. The $\mu$ dependence stems from the quadratic terms in the model (\ref{eq:General-model}), which are crucial for the topological properties of the SOTS.
Moreover, $\Delta_{\text{eff}}^{x(y)}$ vanishes at $\mu=\pm\mu_{x(y)}^c$, where
\begin{eqnarray}
\mu_{x(y)}^{c} =|v_{x(y)}|\sqrt{[m_{0}/m_{y(x)}\mp \Delta_{0}/\Delta_{2}]/(1+m_{x(y)}/m_{y(x)})}. \label{eq:transitionPoint}
\end{eqnarray}
This behavior indicates that we can switch the
sign of $\Delta_{\text{eff}}^{x(y)}$ by varying $\mu$.
Without loss of generality, we suppose $\mu_{x}^{c}\leqslant \mu_{y}^{c}$. The system is in a SOTS phase in
the parameter regions $0\leqslant|\mu|<\mu_{x}^{c}$ and $\mu_{y}^{c}<|\mu|<m_{\text{g}}$ with $m_{\text{g}}$ being the bulk gap \cite{Note2},
whereas if $\mu_{x}^{c}<|\mu|<\mu_{y}^{c}$, it becomes a trivial superconductor with no corner
state. For the particular case with $v_{x(y)}=v$, $m_{x(y)}= m$,
and $\Delta_{0}=0$, $\Delta_{\text{eff}}^{x}$ and $\Delta_{\text{eff}}^{y}$ are always opposite. They both close at $\mu = \pm v\sqrt{m_{0}/2m}$. Thus, there is no parameter space for the trivial phase. Nevertheless, the sign
of $\Delta_{\text{eff}}^{x(y)}$ can still be changed by a finite
$\mu$ inside the bulk gap \citep{Note3} if
\begin{eqnarray}
2m_{0}m>v^{2}.\label{eq:condition-gap-closing}
\end{eqnarray}
This condition indeed corresponds to a QSHI phase with a large inverted
gap or equivalently an indirect bulk gap. It is likely realized in
the inverted InAs/GaSb bilayer \citep{CXLiu08PRLb,Knez11prl,Krishtopenkoeaap18science},
WTe$_{2}$ monolayer \citep{XFQian14science,SJTang17nphys,ZYFei17nphys,SFWu18science,PChen18ncomm},
functionalized MXene \citep{HMWeng15PRB,SChen16NL}, Bismuthene on
SiC \citep{Reis17science,Hsu15NJP}, and PbS monolayer \citep{WHWan17AM,Wrasse14NL,JWLiu15NL}.

To test our analytical results, we discretize the model (\ref{eq:General-model}), put it on a square lattice, choose a proper set of parameters (satisfying
the inequality (\ref{eq:condition-gap-closing})) and calculate the energy
spectrum in a ribbon geometry{\ \citep{Note6}}. For concreteness, we consider $\Delta_{0}=0$ and set the lattice constant $a$ to unity. As shown in Fig.\ \ref{fig:CaseII-gap-closing}(a), the edge states for $\mu=0$ open a gap at $k_{x}=0$. As $\mu$ is increased, the gap splits to two points away from $k_{x}=0$. The magnitude of the gap first decreases, vanishes at a critical $\mu$ and then reopens, which explicitly demonstrates a topological phase transition. This
behavior is in perfect agreement with Eq.\ (\ref{eq:General-formula}),
cf. Fig.\ \ref{fig:CaseII-gap-closing}(b).

\begin{figure}[htp]
\centering
\includegraphics[width=8.5cm]{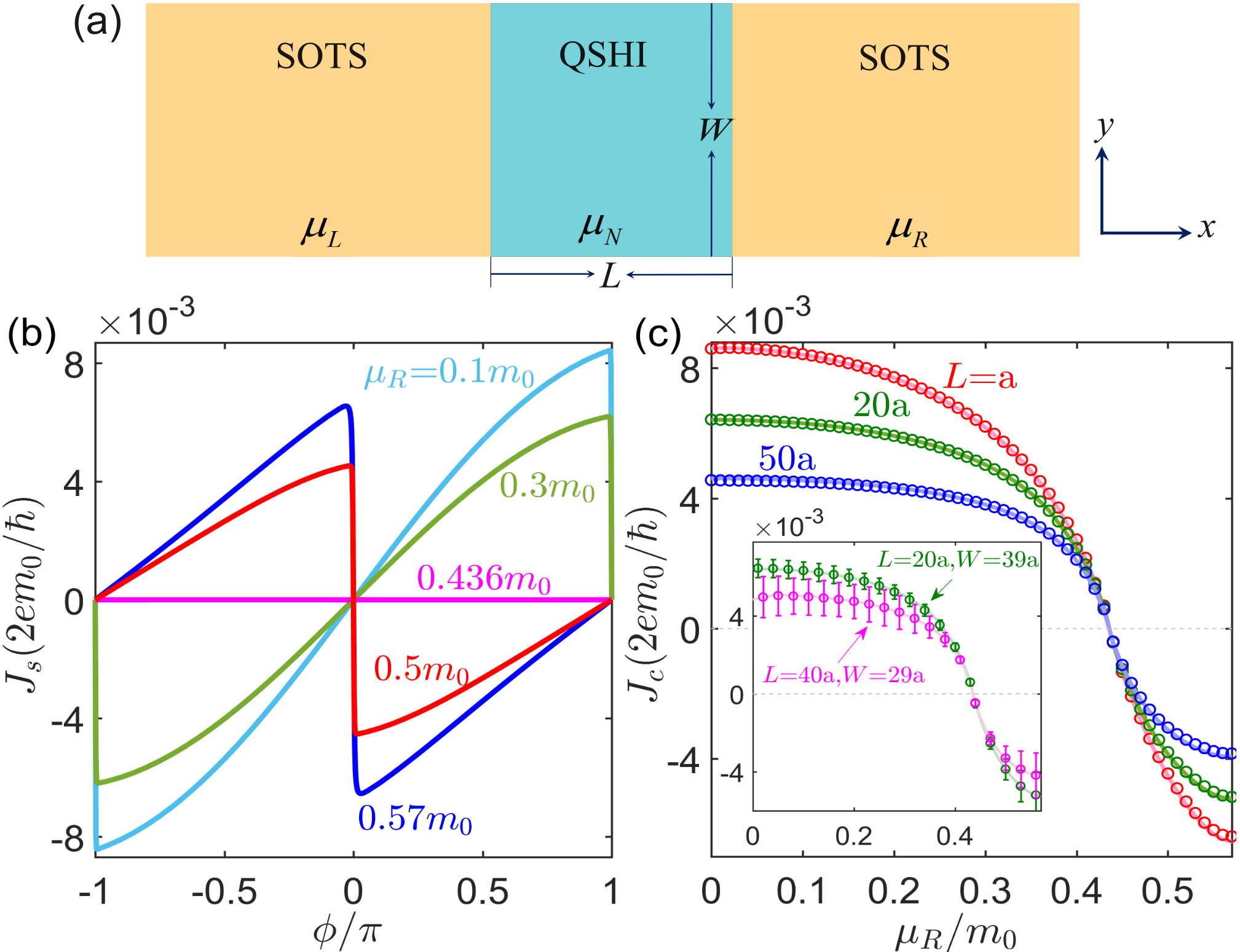}
\caption{(a) Schematic for the SNS setup; (b) Current-phase relations for
$L=a$ and $\mu_{R}=0.1m_{0}$, $0.3m_{0},$ $0.436m_{0},$ $0.5m_{0}$ and $0.57m_{0}$, respectively. (b) Critical
current $J_{c}$ as a function of $\mu_{R}$ for $L=a$ (red), $20a$ (green) and
$50a$ (blue), respectively. The inset displays the results in the
presence of disorder of strength $V_{\text{dis}}=m_0$ for $W=39a$ and
$L=20a$ (green), $W=29a$ and $L=40a$ (purple), respectively. The error bars are $500$
times enlarged for visibility. For all solid curves, $\mu_{L}=\mu_{N}=0.1m_0$,
$W=39a$, $k_BT=10^{-3}\Delta_{L}$ and other parameters are the same
as those in Fig.\ \ref{fig:CaseII-gap-closing}. The circled dots
in (c) are the same as the solid curves but for $\mu_{N}=-0.3m_0$.}
\label{fig:CPR-caseII-0pitransition}
\end{figure}

\textit{0-$\pi$ transition and its robustness.\textendash }We
now consider an SNS junction in which two SOTSs (also called S leads below)
are connected by a QSHI with length $L$ in $x$ direction, as sketched in Fig.\ \ref{fig:CPR-caseII-0pitransition}(a). The width
of the junction ribbon is $W$. For simplicity, we assume the chemical
and pairing potentials in step-like forms. $\mu_{L(R)}$ and $\mu_{N}$ denote the chemical potentials in the left(right) S lead and N (QSHI) region, respectively. $\phi$ is the phase difference across the junction. We calculate the supercurrent $J_{s}$ by the lattice Green's function technique \citep{Asano01PRB,Rodero94PRL,Furusaki94PB}
and provide the details in the Supplemental Material\ \citep{SupplementalMaterial}.

At low temperatures, the transport in the junction is conducted dominantly
by helical edge channels. Perfect Andreev reflection occurs
at the NS interfaces. Thus, the current-phase relation (CPR) takes
a sawtooth shape with a sudden jump, see Fig.\ \ref{fig:CPR-caseII-0pitransition}(b). The sawtooth-like CPR is insensitive to $\mu_{N}$ and stays stable in junctions of different sizes ($L$ and $W$), provided that the two edges at $y=\pm W/2$ are well separated, $W\gg\xi_{\text{edge}}$. The sudden jump can be related to the fermion parity anomaly at each edge \citep{LFu09PRB,Crepin14PRL}. It indicates the formation of degenerate MBSs in the junction discussed below. $J_{s}$ decreases monotonically with increasing $L$, see Fig.\ \ref{fig:CPR-caseII-0pitransition}(b). The critical current $J_{c}$ (maximal value of $J_{s}$) decays as $\text{\ensuremath{\sim}}1/L$ in long junctions, similar to junctions based on conventional s-wave pairing. In short junctions, $J_{c}$ is of the same order of magnitude but always
smaller than $e\sqrt{|\Delta_{L}\Delta_{R}|}/\hbar$, in contrast to the case of $s$-wave pairing. In this estimate,
$\Delta_{L(R)}$ is the induced pairing gap of edge states in the
left(right) S lead and determined by Eq.\ (\ref{eq:General-formula}).
We attribute this difference to the inhomogeneity of the superconducting pairing at the boundaries of our setup.

The CPRs for a fixed $\mu_{L}$ and various values of $\mu_{R}$ are
displayed in Fig.\ \ref{fig:CPR-caseII-0pitransition}(a). Since $J_{s}$ is even in $\mu_{L(R)}$, we present only the results for $\mu_{L(R)}>0$. While $J_{s}$ is insensitive to $\mu_{N}$, it decreases significantly when we increase $\mu_{R}$. This behavior can be understood as a result of the reduction of
$|\Delta_{R}|$ by $\mu_{R}$, see Eq.\ (\ref{eq:General-formula}).
Strikingly, increasing $\mu_{R}$ further, we observe a clear $0$-$\pi$
transition for the parameters satisfying the inequality (\ref{eq:condition-gap-closing}). While $J_{s}(\phi)$ in the region $0<\phi<\pi$ is positive for
$\mu_{R}<\mu^{c}$, it becomes negative for $\mu_{R}>\mu^{c}$.
We coin the former case a $0$-junction and the latter one a $\pi$-junction.
Meanwhile, the sudden jump of the CPR is switched to $\phi=0$ in
the $\pi$-junction, which is in strong contrast to the $0$-junction
where the jump is at $\phi=\pm\pi$. In Fig.\ \ref{fig:CPR-caseII-0pitransition}(c), we plot $J_{c}$ as a function of $\mu_{R}$. The critical value $\mu^{c}$ for the transition is approximately given by $v\sqrt{m_{0}/2m}$, in accord with our
analytical result. Close to $\mu^{c}$, $J_{c}$ drops quickly and
switches sign. These features are generic and apply to junctions of
different lengths and widths. They are also robust with respect to
nonmagnetic disorder in the N region. To illustrate this, we model
the disorder as random on-site potentials in the range $[-V_{\text{dis}}/2,V_{\text{dis}}/2]$
\citep{CALi18PRB,SupplementalMaterial} and calculate 200 random
disorder configurations in the inset of Fig.\ \ref{fig:CPR-caseII-0pitransition}(c). There is no qualitative difference in the features compared to those
in clean junctions. This can be expected since the helical edge channels
which mediate the transport are less sensitive to backscattering.
Similar effects can be observed by tuning $\mu_{L}$ and fixing $\mu_{R}$.
Finally, it is important to note that the variation of $J_{s}$ and the $0$-$\pi$ transition by tuning $\mu_{L(R)}$ are directly related to the strong $\mu_{L(R)}$-dependence in $\Delta_{L(R)}$ in the SOTS, and absent in conventional junctions based on $s$-wave pairing.

\begin{figure}[htp]
\centering

\includegraphics[width=8.6cm]{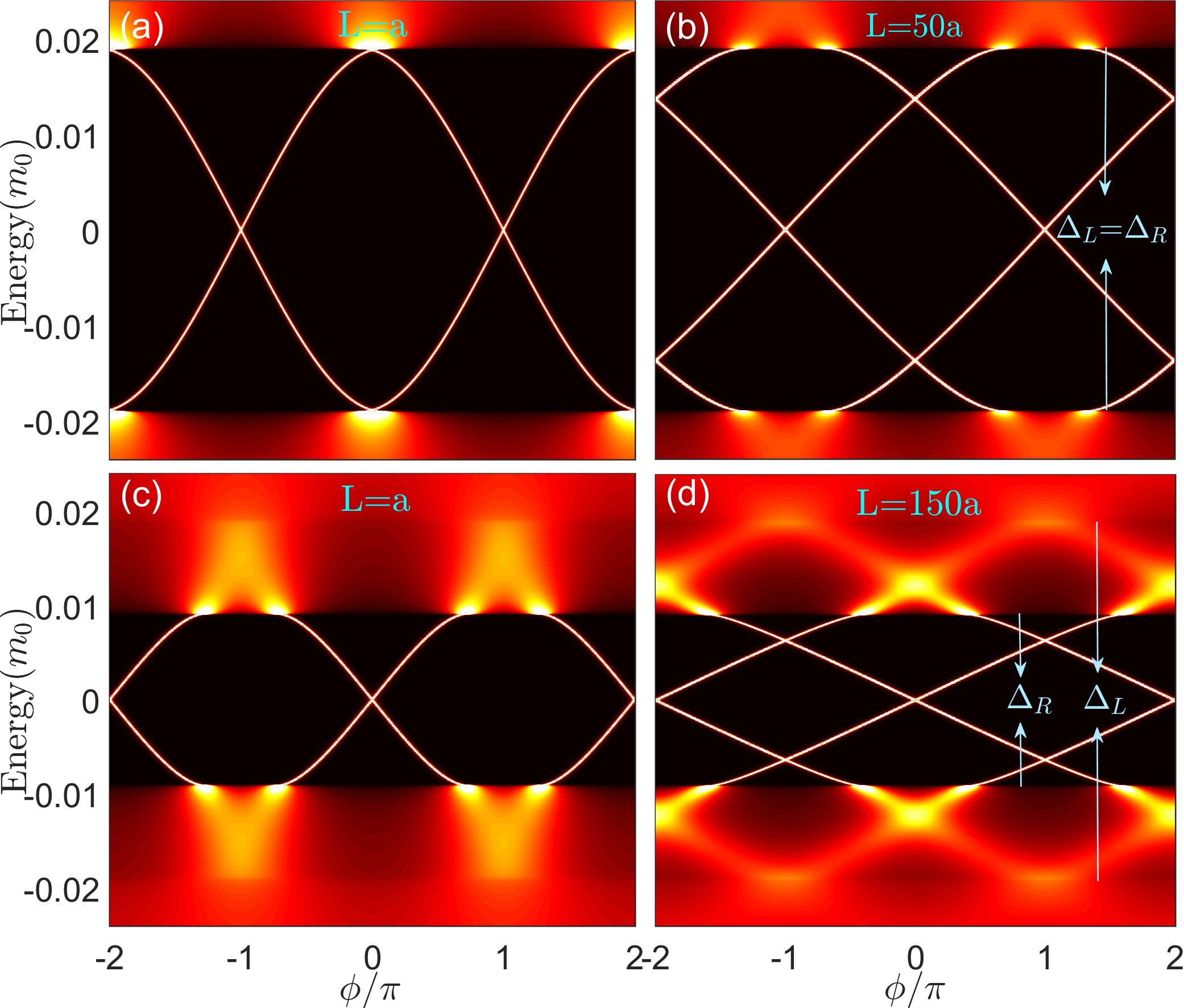}

\caption{Andreev bound states for $\mu_{R}=0.1m_0$ (a,b) and $\mu_{R}=0.52m_0$
(c,d), respectively. (a,c) are for short junctions with $L=a$, while
(b,d) for long junctions with $L=50a$ and $150a$, respectively. For all plots, $W=39a$
and other parameters are the same as those in Fig.\ \ref{fig:CaseII-gap-closing}.}

\label{fig:ABS-0junction}
\end{figure}

\textit{Majorana bound states.\textendash }Next, we discuss the Andreev
bound states (ABSs) formed in the junction, which can be obtained from the lattice Green's function.
In short junctions, there are two bands of ABSs
with opposite energies, see Fig.\ \ref{fig:ABS-0junction}(a,c).
When the sudden jump of the CPR occurs, the positive and negative
bands touch at zero energy. This degeneracy is robust and protected by time-reversal and particle-hole symmetries. It resembles Kramers pairs of MBSs. This can be best understood
from the effective Hamiltonian (\ref{eq:edgeModel}) for edge states. In the short junction limit, two ABS bands at a given edge can be described by
\begin{eqnarray}
E_{\pm}(\phi)=\pm\dfrac{\Delta_{L}\Delta_{R}\sin\phi}{\sqrt{\Delta_{L}^{2}+\Delta_{R}^{2}-2\Delta_{L}\Delta_{R}\cos\phi}}.\label{eq:ABS-bands}
\end{eqnarray}
Notably, the ABSs are confined in the pairing gaps for $\phi$ satisfying
$(\cos\phi$$-$$\Delta_{L}/\Delta_{R})(\cos\phi$$-$$\Delta_{R}/\Delta_{L})>0$,
as verified in Fig.\ \ref{fig:ABS-0junction}(a,c). Noticing $\Delta_{L}\Delta_{R}>0$ in the $0$-junction for $\mu_{R}<\mu^{c}$, whereas $\Delta_{L}\Delta_{R}<0$ in the $\pi$-junction for $\mu_{R}\ensuremath{>}\mu^{c}$, we can see that $E_{\pm}(\phi)$ touch at $\phi=\pm\pi$ and $0$, respectively. Using the valid formula at zero temperature, $J_{s}(\phi)=\partial|E_{+}(\phi)|/\partial\phi$ \citep{Beenakker91PRL}, we also reproduce the sudden jump in the CPR. The wavefunctions of the zero modes can be written as
\begin{eqnarray}
\gamma_{+}(x)& = & \Psi_+\eta(x) +\Psi_-\eta^*(x) \nonumber \\
\gamma_{-}(x)& = & i \Psi_+\eta(x) -i \Psi_-\eta^*(x),
\end{eqnarray}
where $\Psi_{\pm} = \text{sgn}(\Delta_{L})\Psi_{h/e,\downarrow}\mp i\Psi_{e/h,\uparrow}$ and the spatial dependence is $\eta(x)=\exp\{[\theta(x)(i\mu_{R}$$-$$|\Delta_{R}|)+\theta(-x)(i\mu_{L}+|\Delta_{L}|)]x/v\}$
\citep{SupplementalMaterial}. Since $\Psi_{h,\downarrow/\uparrow}=\Psi_{e,\downarrow/\uparrow}^{*}$,
the zero modes have self-adjoint wavefunctions, $\ensuremath{\gamma}_{\pm}(x)=\gamma_{\pm}^{*}(x)$. They are Majorana fermions. Under the time-reversal operation $\mathcal{T}$,
$\mathcal{T}\Psi_{e,\downarrow/\uparrow}=\pm\Psi_{e\uparrow/\downarrow}^{*}$
and $\mathcal{T}\Psi_{h,\downarrow/\uparrow}=\pm\Psi_{h,\uparrow/\downarrow}^{*}$.
Therefore, $\ensuremath{\gamma}_{\pm}$ are related by time-reversal symmetry,
$\mathcal{T}\gamma_{+}=\gamma_{-}$. A similar analysis can be applied
to the other edge where another Kramers pair of MBSs are located. In long
junctions, all the features persist but with more pairs of discrete ABS
bands emerging from the continuum spectrum, see Fig.\ \ref{fig:ABS-0junction}(b,d).

At $\phi=0$, the MBSs emerge for $\mu>\mu^{c}$, whereas they disappear for $\mu<\mu^{c}$. In this sense, we are able to switch between the presence and absence of MBSs by gating the S leads in the absence of $\phi$. Our setup indeed realizes fully electrically controllable MBSs without
fine tuning of magnetic field or threaded flux. This is an important advantage compared to previous proposals based on conventional
$s$-wave superconductivity \citep{Fu08PRL,LFu09PRB,Lutchyn10PRL,Sau10PRL,Oreg10PRL,Volpez19PRL}. Moreover,
since the localization lengths of the MBSs in the S leads are determined
by $\xi_{L(R)}=|v/\Delta_{L(R)}|$, we are also able to control the
spatial profiles of the MBSs by $\mu_{L(R)}$.

\textit{Experimental relevance and summary.\textendash }Now we briefly
discuss the experimental relevance of our proposal. QSHIs with large
inverted gaps \citep{XFQian14science,SJTang17nphys,ZYFei17nphys,SFWu18science,PChen18ncomm,HMWeng15PRB,SChen16NL,Reis17science,Hsu15NJP,Wrasse14NL,JWLiu15NL,WHWan17AM,CXLiu08PRLb,Knez11prl,Krishtopenkoeaap18science}
in proximity to cuprate or iron-based superconductors \citep{Stewart11RMP,Hirschfeld11RPP,PZhang18science,DFWang18science,HZhao18PRB,Zareapour12nacom,EWang13npyhs,PZhang19nphys}
could provide promising platforms to verify our predictions.
For concreteness, we take the inverted InAs/GaSb bilayer and WTe$_{2}$ monolayer to estimate $\mu_{x(y)}^{c}$.
For simplicity, we consider $\Delta_{0}=0$ such that $\mu^{c}_x(y)$ is independent of the magnitude of the pairing potential.
For the inverted InAs/GaSb bilayer, $m_{0} = 0.0055$ eV, $m_{x(y)} = 81.9$ eV$\cdot\mathring{\text{A}}$$^{2}$, $v_{x(y)} = 0.72$
eV$\cdot\mathring{\text{A}}$ \citep{CXLiu13models}.
To realize the $0$-$\pi$ transition, one can fabricate the Josephson junction in
any direction and find that $\mu_{x(y)}^{c} = 0.0042$ eV which is smaller than the bulk gap $m_{\text{gap}} = 0.005$ eV.
For the WTe$_{2}$ monolayer with $m_{0} = 0.33$ eV, $m_{x} = 4.6$ eV$\cdot\mathring{\text{A}}{}^{2}$, $m_{y} = 16.9$ eV$\cdot\mathring{\text{A}}$$^{2}$, $v_{x} = 2.55$ eV$\cdot\mathring{\text{A}}$ and $v_{y} = 0.3$ eV$\cdot\mathring{\text{A}}$ \citep{XFQian14science}, we have $\mu_{x}^{c} = 0.252$ eV, $\mu_{y}^{c} = 0.057$ eV and $m_{\text{gap}} = 0.08$ eV.
Thus, it is better to design the junction
in $y$ direction in our model \citep{Note7}. According to Eq.\ (\ref{eq:transitionPoint}),
the inclusion of a small $\Delta_{0}$ would suppress $\mu_{x}^{c}$
or $\mu_{y}^{c}$ and hence make it more feasible to observe the $0$-$\pi$
transition.
A particle-hole symmetry breaking term, which is neglected
here, breaks the symmetry with respect to $\mu$ but does not qualitatively
change our main results.

We note in passing that there have been experimental efforts trying to incorporate unconventional superconductivity in topological systems \citep{Zareapour12nacom,EWang13npyhs,HZhao18PRB,PZhang18science,DFWang18science,PZhang19nphys}. Moreover, large proximity-induced pairing gaps in 2D systems from unconventional superconductors have been probed \citep{Zareapour12nacom,EWang13npyhs,HZhao18PRB,Perconte18Nphys}.

In summary, we have found that the chemical potentials in superconductors can be used to modulate the supercurrent and realize a $0$-$\pi$ transition in Josephson junctions based on SOTSs. These features are attributed to the dependence of the pairing gap of edge states on the chemical potential. They could serve as novel experimental signatures of the SOTS.
We have predicted the $0$-$\pi$ transition as a fully electric way to create or annihilate MBSs at elevated temperatures.

\begin{acknowledgements}
We thank Fernando Dominguez, Feng Liu, Frank Schindler, Gaomin Tang, Xianxin Wu and Wenbin Rui for valuable
discussion. This work was supported by the DFG (SPP1666, SFB1170 ``ToCoTronics''),
the W\"urzburg-Dresden Cluster of Excellence ct.qmat, EXC2147, project-id
39085490, and the Elitenetzwerk Bayern Graduate School on ``Topological
insulators''.
\end{acknowledgements}


%

\appendix

\section{Derivation of the effective BdG model for edge states\label{sec:Derivation-of-edge}}

\subsection{Edges in $x$ or $y$ direction}

In the absence of the pairing potential, the Bogoliubov-de Gennes
(BdG) Hamiltonian {[}Eq.\ (1) in the main text{]} decouples into
four blocks. Each block can be analyzed separately. Let us take the
block for spin-up electrons for illustration, following the approach
of Ref.\ \citep{ZhangSB16NJP}. The block for spin-up electrons reads
\begin{eqnarray}
h_{e,\uparrow}({\bf k}) & = & \begin{pmatrix}m({\bf k})-\mu & v_{x}k_{x}-iv_{y}k_{y}\\
v_{x}k_{x}+iv_{y}k_{y} & -m({\bf k})-\mu
\end{pmatrix}
\end{eqnarray}
in the basis $(c_{a,\uparrow},c_{b,\uparrow})$, where $m({\bf k})=m_{0}-m_{x}k_{x}^{2}-m_{y}k_{y}^{2}$.

Consider the edge in $x$ direction of a semi-infinite SOTS in the
half-plane $y\leqslant0$ and impose hard-wall boundary conditions.
Then, $k_{x}$ is a good quantum number. We assume the trial wavefunction
of the form,
\begin{eqnarray}
\Psi_{\lambda}({\bf r}) & = & e^{ik_{x}x}e^{\lambda y}\psi_{\lambda},\label{eq:trial-WF}
\end{eqnarray}
where $\psi_{\lambda}$ is a two-component spinor. Plugging Eq.\ (\ref{eq:trial-WF})
in $h_{e,\uparrow}(-i\nabla_{{\bf r}})\Psi_{\lambda}=E\Psi_{\lambda}$,
we obtain the secular equation
\begin{eqnarray}
\det|h_{e,\uparrow}(k_{x},-i\lambda)-E-\mu| & = & 0\label{eq:secular-eq}
\end{eqnarray}
for a nontrivial solution of $\psi_{\lambda}$. Solving Eq.\ (\ref{eq:secular-eq})
gives four $\lambda$, denoted as $\beta\lambda_{\alpha}$ with $\beta=\pm$
and $\alpha=1,2$,
\begin{eqnarray}
\lambda_{\alpha}^{2} & = & (-1)^{\alpha}\sqrt{v_{y}^{4}-4m_{y}^{2}[v_{y}^{2}m_{k}+v_{x}^{2}k_{x}^{2}+(E+\mu)^{2}]}/2m_{y}^{2}\nonumber \\
 &  & +v_{y}^{2}/2m_{y}^{2}-m_{k},\label{eq:Lambdas}
\end{eqnarray}
where $m_{k}=(m_{0}-m_{x}k_{x}^{2})/m_{y}.$ Each $\beta\lambda_{\alpha}$
corresponds to a spinor state written as
\begin{eqnarray}
\psi_{\alpha\beta} & = & \begin{pmatrix}m_{y}(m_{k}+\lambda_{\alpha}^{2})+E+\mu\\
v_{x}k_{x}+v_{y}\beta\lambda_{\alpha}
\end{pmatrix},\label{State1}
\end{eqnarray}
or alternatively,
\begin{eqnarray}
\psi_{\alpha\beta} & = & \begin{pmatrix}-v_{x}k_{x}+v_{y}\beta\lambda_{\alpha}\\
m_{y}(m_{k}+\lambda_{\alpha}^{2})-E-\mu
\end{pmatrix}.\label{State2}
\end{eqnarray}
Then, a general wavefunction is given by
\begin{eqnarray}
\Psi_{e,\uparrow,k_{x}}(E,{\bf r}) & = & e^{ik_{x}x}\sum_{\alpha=1,2}\sum_{\beta=\pm}C_{\alpha\beta}\psi_{\alpha\beta}e^{\beta\lambda_{\alpha}y},
\end{eqnarray}
where the energy $E$ and coefficients $C_{\alpha\beta}$ are found
from the boundary conditions. The hard-wall boundary conditions read
\begin{eqnarray}
\Psi_{e,\uparrow,k_{x}}(y & = & -\infty)=\Psi_{e,\uparrow,k_{x}}(y=0)=0.
\end{eqnarray}
The condition $\Psi_{e,\uparrow,k_{x}}(y=-\infty)=0$ requires that
$\Psi_{e,\uparrow,k_{x}}$ contains only the terms with $\beta=+$
and $\mathrm{Re}(\lambda_{\alpha})>0$, i.e, $C_{1-}=C_{2,-}=0$.
The other condition $\Psi_{e,\uparrow,k_{x}}(y=0)=0$ then leads to
\begin{eqnarray}
\begin{vmatrix}(\psi_{1+} & \psi_{2+})\end{vmatrix} & = & 0.\label{secular}
\end{eqnarray}
Plugging Eqs.~(\ref{State1}) and (\ref{State2}) into Eq.~(\ref{secular}),
respectively, and considering $\lambda_{1}\neq\lambda_{2}$, we obtain
\begin{eqnarray}
E+\mu= & v_{x} & m_{y}k_{x}\left(\lambda_{1}+\lambda_{2}\right)/v_{y}-m_{y}\left(m_{k}-\lambda_{1}\lambda_{2}\right),\nonumber \\
E+\mu= & v_{x} & m_{y}k_{x}\left(\lambda_{1}+\lambda_{2}\right)/v_{y}+m_{y}\left(m_{k}-\lambda_{1}\lambda_{2}\right).\label{SeEq2}
\end{eqnarray}
By comparing these two equations, we identify
\begin{eqnarray}
\lambda_{1}\lambda_{2} & = & m_{k}.\label{Lambdasquare0}
\end{eqnarray}
According to Eq.\ (\ref{eq:Lambdas}), there are two cases of $\lambda_{1,2}$,
one is $\lambda_{1(2)}>0$ and the other $\lambda_{1}=\lambda_{2}^{*}$.
In both cases, $\lambda_{1}\lambda_{2}>0.$ This determines the region
for well-localized edge states:
\begin{eqnarray}
k_{x}^{2} & < & m_{0}/m_{x}.
\end{eqnarray}
From Eq.~(\ref{eq:Lambdas}), we derive
\begin{eqnarray}
\lambda_{1}^{2}+\lambda_{2}^{2} & = & v_{y}^{2}/m_{y}^{2}-2m_{k}.\label{Lambdasquare-1}
\end{eqnarray}
By exploiting Eqs.~(\ref{Lambdasquare0}) and (\ref{Lambdasquare-1})
and $\lambda_{1}+\lambda_{2}>0$, we obtain $\lambda_{1}+\lambda_{2}=|v_{y}/m_{y}|$.
With this result in Eq.~(\ref{SeEq2}), we find the dispersion as
\begin{eqnarray}
E_{e,\uparrow}(k_{x}) & \equiv & E=\mathrm{sgn}(m_{y}v_{y})v_{x}k_{x}-\mu,\label{SurfaceEnergy}
\end{eqnarray}
and consequently the wavefunction as
\begin{eqnarray}
\Psi_{e,\uparrow,k_{x}}({\bf r}) & = & \mathcal{N}e^{ik_{x}x}\begin{pmatrix}\text{sgn}(m_{y}v_{y})\\
1
\end{pmatrix}(e^{\lambda_{1}y}-e^{\lambda_{2}y}),\label{eq:wavefunction}
\end{eqnarray}
where the two penetration depths $\lambda_{1,2}$ and the normalization
factor are given, respectively, by
\begin{eqnarray}
\lambda_{1(2)} & = & |v_{y}/2m_{y}|\mp\sqrt{v_{y}^{2}/4m_{y}^{2}-(m_{0}-m_{x}k_{x}^{2})/m_{y}},\nonumber \\
\dfrac{1}{2\mathcal{N}^{2}} & = & \dfrac{1}{\lambda_{1}+\lambda_{1}^{*}}+\dfrac{1}{\lambda_{2}+\lambda_{2}^{*}}-\dfrac{1}{\lambda_{1}+\lambda_{2}^{*}}-\dfrac{1}{\lambda_{2}+\lambda_{1}^{*}}.
\end{eqnarray}
The decaying length of the edge states is determined by
\begin{eqnarray}
\xi_{\text{edge}} & = & \text{max}\left(1/\text{Re}(\lambda_{1}),1/\text{Re}(\lambda_{2})\right).
\end{eqnarray}
Note that in contrast to previous studies \citep{ZBYan18PRL,TLiu18PRB},
we neither neglect nor treat the quadratic terms as perturbations.

Similarly, the edge states for the other three blocks are found as
\begin{eqnarray}
E_{e,\downarrow}(k_{x}) & = & -\text{sgn}(m_{y}v_{y})v_{x}k_{x}-\mu,\nonumber \\
E_{h,\uparrow/\downarrow}(k_{x}) & = & \pm\text{sgn}(m_{y}v_{y})v_{x}k_{x}+\mu.
\end{eqnarray}
Their wavefunctions in the orbital basis $\{a,b\}$ can be related
to Eq.\ (\ref{eq:wavefunction}) by exploiting time-reversal and
particle-hole symmetries, i.e.,
\begin{eqnarray}
\Psi_{e,\downarrow,k_{x}}({\bf r}) & = & \Psi_{e,\uparrow,-k_{x}}^{*}({\bf r}),\nonumber \\
\Psi_{h,\uparrow/\downarrow,k_{x}}({\bf r}) & = & \Psi_{e,\uparrow/\downarrow,-k_{x}}^{*}({\bf r}),\label{eq:PH-TRS}
\end{eqnarray}

Next, we calculate the pairing gap $\Delta_{\text{eff}}^{x}$ of edge
states. At the Fermi energy ($E=0$), $E_{e,\uparrow}$ crosses $E_{h,\downarrow}$
at $k_{c}$, while $E_{e,\uparrow}$ crosses $E_{h,\downarrow}$ at
$-k_{c}$, where
\begin{eqnarray}
k_{c} & = & \text{sgn}(m_{y}v_{y})\mu/v_{x}.
\end{eqnarray}
At the crossing point $k_{c}$, $\Delta_{\text{eff}}^{x}$ is given
by
\begin{eqnarray}
\Delta_{\text{eff}}^{x} & = & -\int_{-\infty}^{0}dy\Psi_{e,\uparrow,k_{c}}^{\dagger}({\bf r})\Delta(-i\partial_{{\bf r}})\Psi_{h,\downarrow,k_{c}}({\bf r}).
\end{eqnarray}
Using Eqs.\ (\ref{eq:wavefunction}) and (\ref{eq:PH-TRS}), it is
found explicitly as
\begin{eqnarray}
\Delta_{\text{eff}}^{x}= & - & \Delta_{0}-\Delta_{2}\Big(1+\dfrac{m_{x}}{m_{y}}\Big)\dfrac{\mu^{2}}{v_{x}^{2}}+\dfrac{\Delta_{2}m_{0}}{m_{y}}.
\end{eqnarray}
Similarly, the pairing gap between $E_{e,-}$ and $E_{h,-}$ at $-k_{c}$
is found as $-\Delta_{\text{eff}}^{x}$. Therefore, the full BdG Hamiltonian
for the edge states in $x$ direction can be written as
\begin{eqnarray}
h_{\text{edge}}^{x} & = & \text{\text{sgn}(\ensuremath{m_{y}v_{y}})\ensuremath{v_{x}k_{x}}}\tau_{z}s_{z}-\mu\tau_{z}+\Delta_{\text{eff}}^{x}\tau_{x}s_{z}\label{eq:BdG-edge}
\end{eqnarray}
in the basis $(\Psi_{e,\uparrow},\Psi_{e,\downarrow},\Psi_{h,\downarrow},\Psi_{h,\uparrow})$,
where $\bm{\tau}$ and ${\bf s}$ are Pauli matrices acting on Nambu
and spin spaces, respectively. Notably, this BdG Hamiltonian is only
effective for the excitation near the crossing points $k_{x}=\pm k_{c}$.

\subsection{Edges in an arbitrary direction}

In this subsection, we will show that the corner states are more generic
and not restricted to a specific choice of directions (i.e., $x$
or $y$ direction) of the edges. To this end, we consider the edge
in an arbitrary direction $x_{1}$ which has the angle $\theta$ relative
to  $x$ direction. For simplicity, we consider the isotropic QSHI
case, $m_{x}=m_{y}=m$ and $v_{x}=v_{y}=v$. We note that the main
conclusion persists in the general anisotropic case. To derive the
edge states, it is convenient to use the $x_{1}$ and $x_{2}$ (normal
to $x_{1}$) coordinates. The $x_{1}\text{-}x_{2}$ and $x\text{-}y$
coordinates are related by the rotations
\begin{eqnarray}
\begin{pmatrix}k_{x}\\
k_{y}
\end{pmatrix} & = & \begin{pmatrix}\cos\theta & -\sin\theta\\
\sin\theta & \cos\theta
\end{pmatrix}\begin{pmatrix}q_{1}\\
q_{2}
\end{pmatrix},\nonumber \\
\begin{pmatrix}s_{x}\\
s_{y}\\
s_{z}
\end{pmatrix} & = & \begin{pmatrix}\cos\theta & -\sin\theta & 0\\
\sin\theta & \cos\theta & 0\\
0 & 0 & 1
\end{pmatrix}\begin{pmatrix}\rho_{1}\\
\rho_{2}\\
\rho_{3}
\end{pmatrix}.
\end{eqnarray}
In the $x_{1}\text{-}x_{2}$ coordinates, the BdG Hamiltonian becomes
\begin{eqnarray}
H_{\text{BdG}}({\bf q}) & = & m({\bf q})\tau_{z}\sigma_{z}+v\left(q_{1}\cos\theta-q_{2}\sin\theta\right)\rho_{3}\sigma_{x}\nonumber \\
 &  & +v\left(q_{1}\sin\theta+q_{2}\cos\theta\right)\tau_{z}\sigma_{y}-\mu\tau_{z}\nonumber \\
 &  & +\Delta({\bf q})\tau_{y}(\rho_{1}\sin\theta+\rho_{2}\cos\theta)\label{eq:Full-BdG}
\end{eqnarray}
in the new basis $(c_{a,\uparrow'},c_{b,\uparrow'},c_{a,\downarrow'},c_{b,\downarrow'},c_{a,\uparrow'}^{\dagger},c_{b,\uparrow'}^{\dagger},c_{a,\downarrow'}^{\dagger},c_{b,\downarrow'}^{\dagger})$,
where the subscript $\prime$ implies that the direction of spin polarization
is also rotated, and
\begin{eqnarray}
m({\bf q}) & = & m_{0}-m(q_{1}^{2}+q_{2}^{2})\nonumber \\
\Delta({\bf q}) & = & \Delta_{0}+\Delta_{2}\cos2\theta(q_{1}^{2}-q_{2}^{2})\nonumber \\
 &  & -2\Delta_{2}\sin2\theta q_{1}q_{2}.
\end{eqnarray}
The full BdG Hamiltonian (\ref{eq:Full-BdG}) always decouples into
two blocks, $H_{0}$ and its time-reversal counterpart, similar to
that before the rotation.

In the absence of the pairing potential, each BdG block further decouples
into two sub-blocks, one for electrons and one for holes. The sub-blocks
take exactly the same form as those without rotation. Following the
same approach, the dispersion of spin-up electrons and spin-down holes
are given, respectively, by
\begin{eqnarray}
E_{e,\uparrow}(q_{1}) & = & -E_{h,\downarrow}(q_{1})=\text{sgn}(mv)q_{1}-\mu,\label{eq:electron-band}
\end{eqnarray}
Accordingly, the wavefunctions are
\begin{eqnarray}
\Psi_{e,\uparrow,q_{1}}({\bf x}) & = & \mathcal{N}e^{iq_{1}x_{1}}(e^{\lambda_{1}x_{2}}-e^{\lambda_{2}x_{2}})(\text{sgn}(mv),e^{i\theta})^{T},\nonumber \\
\Psi_{h,\downarrow,q_{1}}({\bf x}) & = & \Psi_{e,\uparrow,q_{1}}({\bf x})\label{eq:wave-function}
\end{eqnarray}
in the orbital basis $(a,b)$, where
\begin{eqnarray}
\lambda_{1(2)} & = & |v/2m|\mp\sqrt{v^{2}/4m^{2}-m_{0}/m+q_{1}^{2}},\label{eq:lambda}\\
\dfrac{1}{2\mathcal{N}^{2}} & = & \dfrac{1}{\lambda_{1}+\lambda_{1}^{*}}+\dfrac{1}{\lambda_{2}+\lambda_{2}^{*}}-\dfrac{1}{\lambda_{1}+\lambda_{2}^{*}}-\dfrac{1}{\lambda_{2}+\lambda_{1}^{*}}.\label{eq:C-normalization}
\end{eqnarray}

With the wavefunctions in Eqs.\ (\ref{eq:wave-function}), the pairing
interaction between the edge electrons and holes is calculated as
\begin{eqnarray}
\Delta_{\text{edge}}^{X_{1}} & = & -e^{i\theta}\int_{-\infty}^{0}dx_{2}\Psi_{e,\uparrow,q_{1}}^{\dagger}({\bf x})\Delta(-i\partial_{{\bf x}})\Psi_{h,\downarrow,q_{1}}({\bf x})\nonumber \\
 & = & -e^{i\theta}[\Delta_{0}+\Delta_{2}\cos2\theta(q_{1}^{2}+F_{1})\nonumber \\
 &  & -2iq_{1}\sin2\theta\Delta_{2}F_{2}],
\end{eqnarray}
where
\begin{eqnarray}
F_{1} & = & 2\mathcal{N}^{2}\Big(\dfrac{\lambda_{1}^{2}}{\lambda_{1}+\lambda_{1}^{*}}+\dfrac{\lambda_{2}^{2}}{\lambda_{2}+\lambda_{2}^{*}}-\dfrac{\lambda_{1}^{2}}{\lambda_{1}+\lambda_{2}^{*}}-\dfrac{\lambda_{2}^{2}}{\lambda_{2}+\lambda_{1}^{*}}\Big),\nonumber \\
F_{2} & = & 2\mathcal{N}^{2}\Big(\dfrac{\lambda_{1}}{\lambda_{1}+\lambda_{1}^{*}}+\dfrac{\lambda_{2}}{\lambda_{2}+\lambda_{2}^{*}}-\dfrac{\lambda_{1}}{\lambda_{1}+\lambda_{2}^{*}}-\dfrac{\lambda_{2}}{\lambda_{2}+\lambda_{1}^{*}}\Big).
\end{eqnarray}
Using the expressions (\ref{eq:lambda}) of $\lambda_{1}$ and $\lambda_{2}$,
we derive
\begin{eqnarray}
F_{1} & = & -\lambda_{1}\lambda_{2}=q_{1}^{2}-m_{0}/m,\ \ F_{2}=0.
\end{eqnarray}
Therefore, the pairing interaction is given by
\begin{eqnarray}
\Delta_{\text{edge}}^{X_{1}}= & - & e^{i\theta}[\Delta_{0}+\Delta_{2}\cos2\theta(2q_{1}^{2}-m_{0}/m)].
\end{eqnarray}
According to Eqs.\ (\ref{eq:electron-band}), the crossing point
between the electron and hole bands is
\begin{eqnarray}
q_{c} & = & \text{sgn}(mv)\mu/v.
\end{eqnarray}
Thus, the pairing gap at $q_{1}=q_{c}$ is
\begin{eqnarray}
\Delta_{\text{gap}}^{X_{1}} & = & -e^{i\theta}[\Delta_{0}+\Delta_{2}\cos2\theta(2\mu^{2}/v^{2}-m_{0}/m)].
\end{eqnarray}
Here, the phase factor $e^{i\theta}$ stems from the rotation of spin,
while the $\theta$ dependence in the brackets comes from the rotation
of coordinates. Note that in this derivation, the SOTS is on the right
hand side while the vacuum is on the left hand side. In the spin basis
for $\theta=0$, the phase factor $e^{i\theta}$ is discarded. Thus,
in this basis, the pairing gap reads
\begin{eqnarray}
\widetilde{\Delta}_{\text{gap}}^{X_{1}} & = & -[\Delta_{0}+\Delta_{2}\cos2\theta(2\mu^{2}/v^{2}-m_{0}/m)].
\end{eqnarray}
When $\theta=0$ and $\pi/2$, we recover the results for the edge
in  $x$ and $y$ directions, respectively:
\begin{eqnarray}
\widetilde{\Delta}_{\text{gap}}^{x} & = & -(\Delta_{0}-\Delta_{2}m_{0}/m+2\Delta_{2}\mu^{2}/v^{2}),\nonumber \\
\widetilde{\Delta}_{\text{gap}}^{y} & = & -(\Delta_{0}+\Delta_{2}m_{0}/m-2\Delta_{2}\mu^{2}/v^{2}).\label{eq:y-direction}
\end{eqnarray}

\begin{figure}[H]
\centering

\includegraphics[width=8.5cm]{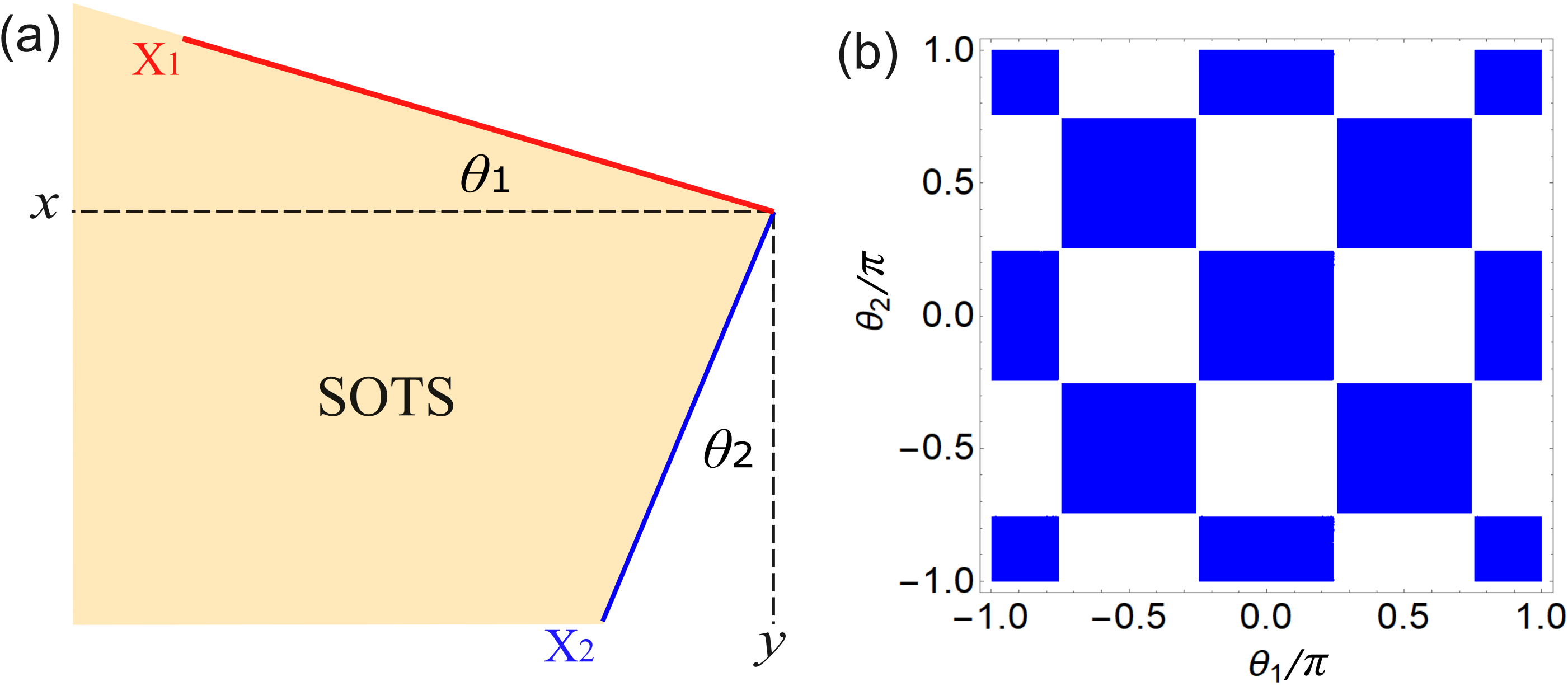}

\caption{(a) schematic of two edges with a common corner. (b) The phase diagram
for the presence (blue region) and absence (white region) as functions
of the two edge directions, $\theta_{1}$ and $\theta_{2}$ (with
respect to the $x$ and $y$ axies, respectively).}

\label{fig:diagram}
\end{figure}

To form corner states, we need another edge. Let us consider the other
edge in  $x_{2}$ direction and the SOTS in the $x_{1}<0$ half plane.
Along the lines of that we did for the $x_{1}$ edge, we can find
analytically the electron and hole edge bands as
\begin{eqnarray}
E_{e,\uparrow}(q_{2}) & = & -E_{h,\downarrow}(q_{2})=\text{sgn}(mv)q_{2}-\mu,
\end{eqnarray}
and their wavefunctions as
\begin{eqnarray}
\Psi_{e,\uparrow,q_{2}}({\bf x}) & = & \mathcal{N}e^{iq_{2}x_{2}}(e^{\lambda_{1}x_{1}}-e^{\lambda_{1}x_{1}})(\text{sgn}(mv),e^{i\theta})^{T},\nonumber \\
\Psi_{h,\downarrow,q_{2}}({\bf x}) & = & \Psi_{e,\uparrow,q_{2}}({\bf x}),
\end{eqnarray}
where $\lambda_{1(2)}$ and $\mathcal{N}$ are given by Eqs.\ (\ref{eq:lambda})
and (\ref{eq:C-normalization}), respectively. The pairing interaction
between the electron and hole bands is found as
\begin{eqnarray}
\Delta_{\text{edge}}^{X_{2}} & = & -e^{-i\theta}[\Delta_{0}+\Delta_{2}\cos2\theta(m_{0}/m-2q_{2}^{2})].
\end{eqnarray}
At the crossing point $q_{2}=q_{c}=\text{sgn}(mv)\mu/v$ and in the
spin basis for $\theta=0$, the pairing gap is given by
\begin{eqnarray}
\widetilde{\Delta}_{\text{gap}}^{X_{2}} & = & -[\Delta_{0}+\Delta_{2}\cos2\theta(m_{0}/m-2\mu^{2}/v^{2})].\label{eq:Delta-X2}
\end{eqnarray}
When $\theta=-\pi/2$ and $0$, we recover again the results for the
edges in $y$ and $x$ direction, respectively.

Denote the angle between  $x_{1}$ and  $x$ direction by $\theta_{1}$,
and the angle between $x_{2}$ and  $y$ direction by $\theta_{2}$.
The pairing gap of the edge states in $x_{1}$ and $x_{2}$ directions
are
\begin{eqnarray}
\widetilde{\Delta}_{\text{gap}}^{X_{1}}(\theta_{1})= & - & [\Delta_{0}-\Delta_{2}\cos2\theta_{1}(m_{0}/m-2\mu^{2}/v^{2})],\nonumber \\
\widetilde{\Delta}_{\text{gap}}^{X_{2}}(\theta_{2})= & - & [\Delta_{0}+\Delta_{2}\cos2\theta_{2}(m_{0}/m-2\mu^{2}/v^{2})].
\end{eqnarray}
Note that the spin basis is the same for all edge states (i.e., the
spin basis in the particular $x$-$y$ coordinates). The existence
of corner states yields that $\widetilde{\Delta}_{\text{gap}}^{X_{1}}(\theta_{1})$
and $\widetilde{\Delta}_{\text{gap}}^{X_{2}}(\theta_{2})$ have different
signs,
\begin{align}
[\Delta_{0}-\Delta_{2}\cos2\theta_{1}(m_{0}/m-2\mu^{2}/v^{2})]\nonumber \\
\times[\Delta_{0}+\Delta_{2}\cos2\theta_{2}(m_{0}/m-2\mu^{2}/v^{2})] & <0.\label{eq:angle-condition}
\end{align}
For $\Delta_{0}=0$ and considering, in general, $m_{0}/m-2\mu^{2}/v^{2}\neq0$,
Eq.\ (\ref{eq:angle-condition}) simplifies to
\begin{eqnarray}
\cos2\theta_{1}\cos2\theta_{2} & > & 0.\label{eq:angle-condition-1}
\end{eqnarray}
The phase diagram for corner states is displayed in Fig.\ \ref{fig:diagram}.
One can see that the corner states exist in a wide range of the angles
$\theta_{1}$ and $\theta_{2}$ (see the blue areas). This indicates
that the corner states, in general, do not require a crystalline symmetry.

\section{Calculations of Josephson current\label{sec:Calculations-of-Josephson}}

There are different tight-binding lattice models having the low-energy
minimal Hamiltonian we consider. In the customary regularization,
we can obtain a tight-binding model by replacing $k_{x(y)}\rightarrow\sin k_{x(y)}$
and $k_{x(y)}^{2}\rightarrow2[1-\cos k_{x(y)}]$. For convenience,
the lattice constant is set to unity. Then, Fourier transforming into
lattice space, the BdG Hamiltonian is given by
\begin{eqnarray}
\mathcal{H}= & \dfrac{1}{2} & \sum_{l,l'}\sum_{j,j'}C_{l,j}^{\dagger}\{(M_{0}\tau_{z}\sigma_{z}-\mu)\delta_{l,l'}\delta_{j,j'}+\tau_{z}\sigma_{z}\nonumber \\
 & \times & [m_{x}\delta_{j,j'}(\delta_{l,l'-1}+\delta_{l,l'+1})+m_{y}\delta_{l,l'}\left(\delta_{j,j'-1}+\delta_{j,j'+1}\right)]\nonumber \\
 & + & i\delta_{j,j'}(\delta_{l,l'-1}-\delta_{l,l'+1})v_{x}\sigma_{x}s_{z}/2+i\delta_{l,l'}(\delta_{j,j'-1}\nonumber \\
 & - & \delta_{j,j'+1})v_{y}\tau_{z}\sigma_{y}/2+\hat{d}_{0}\delta_{l,l'}\delta_{j,j'}+\hat{d}_{2}\delta_{j,j'}(\delta_{l,l'-1}\nonumber \\
 & + & \delta_{l,l'+1})-\hat{d}_{2}\delta_{l,l'}(\delta_{j,j'-1}+\delta_{j,j'+1})\}C_{l',j'}
\end{eqnarray}
with
\begin{eqnarray}
\hat{d}_{t} & = & \begin{pmatrix}0 & -i\Delta_{t}s_{y}\\
i\Delta_{t}^{*}s_{y} & 0
\end{pmatrix},\ t\in\{0,2\},
\end{eqnarray}
where the spinor operators are $C_{l,j}^{\dagger}=(c_{a,\uparrow;l,j}^{\dagger},c_{b,\uparrow;l,j}^{\dagger},$
$c_{a,\downarrow;l,j}^{\dagger},c_{b,\downarrow;l,j}^{\dagger},c_{a,\uparrow;l,j},c_{b,\uparrow;l,j},c_{a,\downarrow;l,j},c_{b,\downarrow;l,j})$;
$M_{0}=m_{0}-2m_{x}-2m_{y}$; $\{l,l'\}$ and $\{j,j'\}$ denote the
lattice sites in $x$ and $y$ directions, respectively. The identity
matrices for spin, Nambu and orbital spaces are omitted for ease of
notation.

For an SNS junction, the BdG Hamiltonian can be written as
\begin{eqnarray}
\mathcal{H}= & \dfrac{1}{2} & \sum_{l,l'=-\infty}^{\infty}\sum_{j,j'=1}^{W}C_{l,j}^{\dagger}\{(M_{0}\tau_{z}\sigma_{z}-\mu_{l})\delta_{l,l'}\delta_{j,j'}+\tau_{z}\sigma_{z}\nonumber \\
 & \times & [m_{x}\delta_{j,j'}(\delta_{l,l'-1}+\delta_{l,l'+1})+m_{y}\delta_{l,l'}\left(\delta_{j,j'-1}+\delta_{j,j'+1}\right)]\nonumber \\
 & + & i\delta_{j,j'}(\delta_{l,l'-1}-\delta_{l,l'+1})v_{x}\sigma_{x}s_{z}/2+i\delta_{l,l'}(\delta_{j,j'-1}\nonumber \\
 & - & \delta_{j,j'+1})v_{y}\tau_{z}\sigma_{y}/2+[\hat{d}_{0,l}\delta_{l,l'}\delta_{j,j'}+\hat{d}_{2,l}\delta_{j,j'}(\delta_{l,l'-1}\nonumber \\
 & + & \delta_{l,l'+1})-\hat{d}_{2,l}\delta_{l,l'}(\delta_{j,j'-1}+\delta_{j,j'+1})]\nonumber \\
 & \times & \left[\Theta(1/2-l')+\Theta(l'+1/2-L)\right]\}C_{l',j'},
\end{eqnarray}
where
\begin{eqnarray}
\hat{d}_{t,l} & = & \begin{pmatrix}0 & -i\Delta_{t,l}s_{y}\\
i\Delta_{t,l}^{*}s_{y} & 0
\end{pmatrix},\ t\in\{0,2\},
\end{eqnarray}
and $\Theta(x)$ is the Heaviside step function; $L$ and $W$ (in
units of the lattice constant $a$) are the length and width of the
junction, respectively. The chemical and pairing potentials are modeled
as
\begin{align}
\mu_{l} & =\begin{cases}
\mu_{L}, & l\leqslant0\\
\mu_{N}, & 1\leqslant l\leqslant L,\\
\mu_{R}, & l\geqslant L+1
\end{cases}\ \Delta_{t,l}=\begin{cases}
\Delta_{t}, & l\leqslant0\\
0, & 1\leqslant l\leqslant L.\\
\Delta_{t}e^{i\phi}, & l\geqslant L+1
\end{cases}
\end{align}

The dc Josephson current can be calculated as \citep{Furusaki94PB,Rodero94PRL,Asano01PRB}
\begin{align}
J_{s} & =\dfrac{ie}{\hbar}\Big\langle\Big[\sum_{j}c_{\sigma,s,l,j}^{\dagger}c_{\sigma,s,l,j}^{\dagger},\mathcal{H}\Big]\Big\rangle\\
 & =\dfrac{ieT}{2\hbar}\sum_{\omega_{\nu}}\text{Tr}\{\check{\tau}_{3}[\check{H}_{t}\check{\mathcal{G}}(l,l+1,i\omega_{\nu})-\check{H}_{t}^{\dagger}\check{\mathcal{G}}(l+1,i\omega_{\nu})]\},\nonumber
\end{align}
where the over-script $\check{...}$ indicates $8W\text{\text{\ensuremath{\times}}}8W$
matrices expanded in Nambu, spin, orbital, and lattice ($j=1,...,W$)
spaces,
\begin{eqnarray}
\check{H}_{t} & = & (m_{x}\tau_{z}\sigma_{z}-iv_{x}\sigma_{x}s_{z}/2)\text{\ensuremath{\mathbb{I}_{W\times W},}}\nonumber \\
\check{\tau}_{3} & = & \tau_{3}\text{\ensuremath{\mathbb{I}_{W\times W}.}}
\end{eqnarray}
$\check{\mathcal{G}}(l,l',i\omega_{\nu})$ is the Matsubara Green's
function, $\omega_{\nu}=(2\nu+1)\pi k_{B}T$ with $\nu=0,\pm1,...$,
is the Matsubara frequency and $T$ is the temperature. $\mathbb{I}_{W\times W}$
denotes the $W\text{\text{\ensuremath{\times}}}W$ identity matrix.
We find $\check{\mathcal{G}}(l,l',i\omega_{\nu})$ numerically by
the recursive Green's function technique\,\citep{PALee81PRL}. The
trace is taken over Nambu, spin, orbital, and lattice degrees of freedom.
The supercurrent $J_{s}$ is independent of $l$ \citep{Asano01PRB}.
Thus, it is convenient to calculate $J_{s}$ at $l=L$.

By performing the analytical continuation $i\omega_{\nu}\rightarrow E+i\Gamma$
with a positive infinitesimal $\Gamma$, we obtain the retarded Green's
function $\check{G}^{R}(l,l,E+i\Gamma)$. The density of states is
then calculated as
\begin{eqnarray}
\rho(E) & = & -\dfrac{1}{\pi}\text{Im}[\text{Tr}\check{G}^{R}(l,l,E+i\Gamma)],
\end{eqnarray}
where $1\leqslant l\leqslant L$. The energy of Andreev bound states
(ABSs) can be found as the peaks of $\rho(E)$. Note that $\rho(E)$
is the same for $l\in\{1,...,L\}$. A small but finite bandwidth $\Gamma$
is employed for the calculation of $\rho(E)$. In this work, we use
$\Gamma=10^{-5}m_{0}$ throughout.

\begin{figure}[h]
\centering

\includegraphics[width=8.5cm]{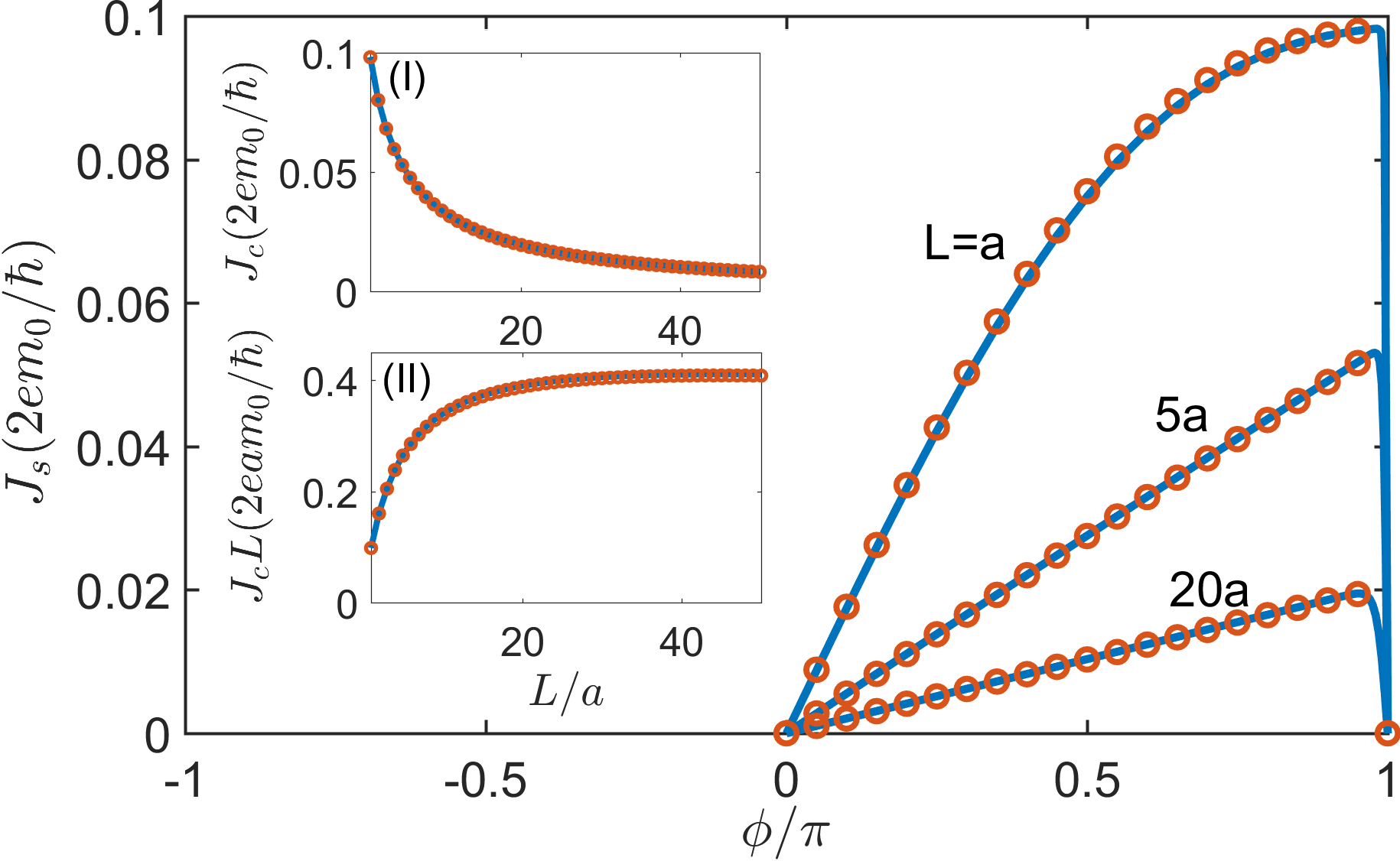}

\caption{Current-phase relations $J_{s}(\phi)$ for different lengths $L$
and widths $W$. Here, $L=a,$ 5a, and $20a$ as indicated in the
figure. The blue solid curves are for $W=19a$, and the orange circled
dots are for $W=9a$. The insets (I) and (II) show the critical current
$J_{c}$ and $J_{c}L$ as functions of $L$, respectively. They imply
that $J_{c}\sim1/L$ for long junctions. To quickly converge to the
long junction regime, we choose the parameters: $v_{x}=v_{y}=2$,
$m_{x}=m_{y}=1$, $\mu_{N}=\mu_{L}=\mu_{R}=0$, $\Delta_{0}=0$ and
$\Delta_{2}=0.2$. The units for energy and length(wavenumber) are
$m_{0}$ and $a$($a^{-1}$), respectively.}

\label{fig:CPR-L}
\end{figure}

To show the $L$-dependence in the supercurrent $J_{s}(\phi)$, we
plot in Fig.\ \ref{fig:CPR-L} the current-phase relation for different
values of $L$, and $J_{c}$ and $J_{c}L$ as functions of $L$ in
the insets (I) and (II), respectively. We can observe that $J_{s}(\phi)$
decays monotonically as we increase $L$. For long junctions $L\gg\text{max}\{v_{x}/|\Delta_{L}|,v_{x}/|\Delta_{R}|\}$,
$J_{c}L$ saturates to a constant. This indicates that $J_{c}$ scales
as $\sim1/L$. Moreover, $J_{s}(\phi)$ is insensitive to the width
$W$ as long as $W\gg\xi_{\text{edge}}$.

\section{Supercurrent from the edge BdG model\label{sec:Supercurrent-from-the}}

In this section, we look at the edge BdG Hamiltonian\ (\ref{eq:BdG-edge})
and derive analytically the ABSs and Majorana bound states (MBSs).
Without loss of generality, we assume $m_{y}v_{y}>0$. Then, one edge
of the Josephson junction, say the upper one, is described by
\begin{eqnarray}
h_{\text{edge}}^{x} & = & v_{x}k_{x}\tau_{z}s_{z}-\mu(x)\tau_{z}\nonumber \\
 &  & +\begin{pmatrix}0 & 0 & \Delta(x) & 0\\
0 & 0 & 0 & -\Delta(x)\\
\Delta^{*}(x) & 0 & 0 & 0\\
0 & -\Delta^{*}(x) & 0 & 0
\end{pmatrix}.
\end{eqnarray}
where the spatially dependent chemical and pairing potentials are
\begin{eqnarray}
\mu(x) & = & \mu_{N}\Theta(L/2-|x|)+\mu_{L}\Theta(-x-L/2)\nonumber \\
 &  & +\mu_{R}\Theta(x-L/2),\nonumber \\
\Delta(x) & = & \Delta_{L}\Theta(-x-L/2)+\Delta_{R}e^{i\phi}\Theta(x-L/2).
\end{eqnarray}

In the N region, the basis functions can be written as
\begin{align}
\varphi_{e,+}(x) & =(1,0,0,0)^{T}e^{ik_{e}x},\nonumber \\
\varphi_{e,-}(x) & =(0,1,0,0)^{T}e^{-ik_{e}x},\nonumber \\
\varphi_{e,+}(x) & =(0,0,1,0)^{T}e^{ik_{h}x},\nonumber \\
\varphi_{e,-}(x) & =(0,0,0,1)^{T}e^{-ik_{h}x},
\end{align}
where $k_{e(h)}=(\mu_{N}\pm E)/v_{x}.$ Thus, the wavefunction in
the N region is expanded as
\begin{eqnarray}
\Psi_{\text{N}}(x) & = & \sum_{\eta=\pm}\left[A_{e,\eta}\varphi_{e,\eta}(x)+A_{h,\eta}\varphi_{h,\eta}(x)\right],\ |x|<L/2.
\end{eqnarray}
In the S leads, the basis functions are
\begin{eqnarray}
\varphi_{qe,+}^{s}(x) & = & (E+\Omega_{s},0,\Delta_{s}e^{-i\phi_{s}},0)^{T}e^{ik_{eq}^{s}x},\nonumber \\
\varphi_{qh,-}^{s}(x) & = & (\Delta_{s}e^{i\phi_{s}},0,E+\Omega_{s},0)^{T}e^{ik_{hq}^{s}x},\nonumber \\
\varphi_{qe,-}^{s}(x) & = & (0,-E-\Omega_{s},0,\Delta_{s}e^{-i\phi_{s}})^{T}e^{-ik_{eq}^{s}x},\nonumber \\
\varphi_{qh,+}^{s}(x) & = & (0,\Delta_{s}e^{i\phi_{s}},0,-E-\Omega_{s})^{T}e^{-ik_{hq}^{s}x},
\end{eqnarray}
where $k_{qe(qh)}^{s}=\mu_{s}\pm\Omega_{s}$ and
\begin{eqnarray*}
\Omega_{s} & = & \begin{cases}
\text{sgn}(E)\sqrt{E^{2}-\Delta_{s}^{2}}, & E\geqslant\Delta_{s}\\
i\sqrt{\Delta_{s}^{2}-E^{2}}, & E<\Delta_{s}
\end{cases}
\end{eqnarray*}
with $s\in\{L,R\}$ distinguishing the left and right S leads. $\phi_{L}=0$
and $\phi_{R}=\phi$. We are most interested in the ABSs whose energies
satisfy $|E|<\Delta_{s}$. Thus, the wavefunction in the S leads is
given by
\begin{eqnarray}
\Psi_{\text{S}}(x) & = & \begin{cases}
B_{eL}\varphi_{e,-}^{L}(x)+B_{hL}\varphi_{qh,-}^{L}(x), & x<-L/2\\
B_{eR}\varphi_{qe,+}^{R}(x)+B_{hR}\varphi_{qh,+}^{R}(x), & x>L/2
\end{cases}
\end{eqnarray}
The energies $E$ of ABSs and the coefficients $A_{e(h),\pm}$, $B_{e(h)L}$
and $B_{e(h)R}$ are found from the continuity of the wavefunction,
i.e.,
\begin{eqnarray}
\Psi_{\text{N}}(-L/2) & = & \Psi_{\text{S}}(-L/2),\ \Psi_{\text{N}}(L/2)=\Psi_{\text{S}}(L/2).
\end{eqnarray}
A nontrivial solution of these equations yields
\begin{eqnarray}
0= & [ & e^{i(\phi-2EL/v_{x})}\Delta_{L}\Delta_{R}-(E+\Omega_{L})(E+\Omega_{R})]\nonumber \\
 & \times & [e^{i(\phi+2EL/v_{x})}\Delta_{L}\Delta_{R}-(E+\Omega_{L})(E+\Omega_{R})].
\end{eqnarray}
This can be recast to the transcendental equations
\begin{eqnarray}
E & = & \pm\dfrac{\Delta_{L}\Delta_{R}\sin(\phi\pm2EL/v_{x})}{\sqrt{\Delta_{L}^{2}+\Delta_{R}^{2}-2\Delta_{L}\Delta_{R}\cos(\phi\pm2EL/v_{x}})}\label{eq:Self-eq}
\end{eqnarray}
with $\phi$ satisfying
\begin{eqnarray}
[\cos\left(\phi\pm2EL/v_{x}\right)-\Delta_{L}/\Delta_{R}]\nonumber \\
\times[\cos\left(\phi\pm2EL/v_{x}\right)-\Delta_{R}/\Delta_{L}] & = & 0.\label{eq:ABS-condition}
\end{eqnarray}
The solutions of $E$ can be found self-consistently from Eq.\,(\ref{eq:Self-eq}).
With the obtained $E$, the coefficients are also obtained. Several
salient features of ABSs are obvious: (i) ABSs appear in pairs with
opposite energies; (ii) the ABS spectrum is independent $\mu_{N}$;
(iii) more ABS branches appear for a longer $L$.

In the short junction limit $L=0$, the ABS spectrum can be found
analytically as
\begin{eqnarray}
E_{\pm}(\phi) & = & \pm\dfrac{\Delta_{L}\Delta_{R}\sin\phi}{\sqrt{\Delta_{L}^{2}+\Delta_{R}^{2}-2\Delta_{L}\Delta_{R}\cos\phi}}.\label{eq:short-limit-spectrum}
\end{eqnarray}
Correspondingly, equation\ (\ref{eq:ABS-condition}) defines the
parameter range for the existence of ABSs
\begin{eqnarray}
\left(\cos\phi-\Delta_{L}/\Delta_{R}\right)\left(\cos\phi-\Delta_{R}/\Delta_{L}\right) & = & 0.\label{eq:ABS-condition-1}
\end{eqnarray}
Note that for the 0-junction, $\Delta_{L}\Delta_{R}>0$, while for
the $\pi$-junction, $\Delta_{L}\Delta_{R}<0$. From Eqs.\ (\ref{eq:short-limit-spectrum})
and (\ref{eq:ABS-condition-1}), it is easy to see that the zero-energy
modes are at $\phi=\pm\pi$ in the 0-junction, while they switched
to be at $\phi=0$ in the $\pi$-junction. Note that this result holds
also in longer junctions. In both junctions, the wavefunctions of
two zero-energy modes can be written in a compact form
\begin{eqnarray}
\Psi_{+}(x) & = & \eta(x)(-i,0,\text{sgn}(\Delta_{L}),0)^{T},\nonumber \\
\Psi_{-}(x) & = & \eta^{*}(x)(0,\text{sgn}(\Delta_{L}),0,i)^{T},
\end{eqnarray}
where
\begin{eqnarray}
\eta(x) & = & e^{(i\mu_{L}+|\Delta_{L}|)x\Theta(-x)+(i\mu_{R}-|\Delta_{R}|)x\Theta(x)}.
\end{eqnarray}
Restoring the basis $(\Psi_{e,\uparrow},\Psi_{e,\downarrow},\Psi_{h,\downarrow},\Psi_{h,\uparrow}),$
we can write
\begin{eqnarray}
\Psi_{+}(x) & = & -i\eta(x)\Psi_{e,\uparrow}+\text{sgn}(\Delta_{L})\eta(x)\Psi_{h,\downarrow},\nonumber \\
\Psi_{-}(x) & = & \text{sgn}(\Delta_{L})\eta^{*}(x)\Psi_{e,\downarrow}+i\eta^{*}(x)\Psi_{h,\uparrow},
\end{eqnarray}
Recall Eqs.\ (\ref{eq:PH-TRS}), $\Psi_{h,\uparrow/\downarrow}=\Psi_{e,\uparrow/\downarrow}^{*}.$
Hence, the two zero-energy modes obey
\begin{eqnarray}
\Psi_{+}(x) & = & \Psi_{-}^{*}(x).
\end{eqnarray}
This indicates that they are related by particle-hole symmetry. We
can recombine them and obtain
\begin{eqnarray}
\gamma_{+}(x) & = & \Psi_{+}(x)+\Psi_{-}(x),\nonumber \\
\gamma_{-}(x) & = & i[\Psi_{+}(x)-\Psi_{-}(x)].
\end{eqnarray}
The new zero-energy modes have self-adjoint wavefunctions
\begin{eqnarray}
\ensuremath{\gamma}_{\pm}(x) & = & \gamma_{\pm}^{*}(x),
\end{eqnarray}
and behave like MBSs. Under time-reversal operation $\mathcal{T}$,
\begin{eqnarray}
\mathcal{T}\Psi_{e,\downarrow} & = & \Psi_{e,\uparrow}^{*},\ \mathcal{T}\Psi_{e,\uparrow}=-\Psi_{e,\downarrow}^{*},\nonumber \\
\mathcal{T}\Psi_{h,\downarrow} & = & \Psi_{h\uparrow}^{*},\ \mathcal{T}\Psi_{h,\uparrow}=-\Psi_{h,\downarrow}^{*}.
\end{eqnarray}
This shows that the two MBSs are connected by time-reversal symmetry,
\begin{eqnarray}
\mathcal{T}\ensuremath{\gamma}_{+}(x) & = & \gamma_{-}(x).
\end{eqnarray}
Hence, they are Kramers partners.

\section{Symmetries and quadrupole moment\label{sec:Topological-invariant-from}}

In this section, we analyze the symmetries and calculate the quadrupole
moment of the SOTS.

\subsection{Symmetries}

The BdG Hamiltonian Eq.\,(1) in the main text
\begin{eqnarray}
H_{\text{BdG}}({\bf k}) & = & m({\bf k})\tau_{z}\sigma_{z}+v_{x}k_{x}s_{z}\sigma_{x}+v_{y}k_{y}\tau_{z}\sigma_{y}\nonumber \\
 &  & +\Delta({\bf k})\tau_{y}s_{y}-\mu\tau_{z}\label{eq:kp-Hamiltonain}
\end{eqnarray}
satisfies the following symmetries:\\
$\bullet$ time-reversal symmetry, $\mathcal{T}=is_{y}\mathcal{K}$
with $\mathcal{K}$ the complex conjugation:
\begin{equation}
\mathcal{T}H_{\text{BdG}}({\bf k})\mathcal{T}^{-1}=H_{\text{BdG}}({\bf -k});
\end{equation}
$\bullet$ particle-hole symmetry, $\Xi=\tau_{x}\mathcal{K}$:
\begin{equation}
\Xi H_{\text{BdG}}({\bf k})\Xi^{-1}=-H_{\text{BdG}}({\bf -k});
\end{equation}
$\bullet$ inversion symmetry, $\mathcal{P}=\sigma_{z}$:
\begin{align}
\mathcal{P}H_{\text{BdG}}({\bf k})\mathcal{P}^{-1} & =H_{\text{BdG}}({\bf -k});
\end{align}
$\bullet$ if $\mu=0$, combined reflection symmetries, $\mathcal{M}_{x}=\tau_{x}s_{x}\sigma_{x}$
and $\mathcal{\mathcal{M}}_{y}=\tau_{x}s_{x}\sigma_{y}$:
\begin{align}
\mathcal{M}_{x}H_{\text{BdG}}({\bf k})\mathcal{\mathcal{M}}_{x}^{-1} & =H_{\text{BdG}}(-k_{x},k_{y}),\nonumber \\
\mathcal{\mathcal{M}}_{y}H_{\text{BdG}}({\bf k})\mathcal{\mathcal{M}}_{y}^{-1} & =H_{\text{BdG}}(k_{x},-k_{y});
\end{align}
$\bullet$ if $m_{x}=m_{y}$, $v_{x}=v_{y}=v$ and $\Delta_{0}=0$,
combined four-fold rotation symmetry, $\mathcal{C}_{4}=\tau_{z}e^{i\pi\sigma_{z}/4}$:
\begin{equation}
\mathcal{C}_{4}H_{\text{BdG}}({\bf k})\mathcal{C}_{4}^{-1}=H_{\text{BdG}}(k_{y},-k_{x}).
\end{equation}

\subsection{Quadrupole moment}

To find the quadrupole moment by the Wilson-loop approach of \citet{Benalcazar17science,Benalcazar17PRB},
we need to consider a periodic lattice model. As the previous numerical
calculations, we consider the lattice model by replacing $k_{x(y)}\rightarrow\sin k_{x(y)}$
and $k_{x(y)}^{2}\rightarrow2[1-\cos k_{x(y)}]$ in Eq.\ (\ref{eq:kp-Hamiltonain})

The projected position operator $P^{\text{occ}}\hat{x}P^{\text{occ}}$
into the occupied bands can define a Wilson line. In $x$ direction,
the Wilson line operator is given by
\begin{equation}
\mathcal{W}_{k_{f}\leftarrow k_{i}}=F_{k_{f}-\delta_{k}}F_{k_{f}-2\delta_{k}}\cdots F_{k_{i}+\delta_{k}}F_{k_{i}},
\end{equation}
where $[F_{k}]^{nn'}=\langle u_{k+\delta_{k_{x}}}^{n}|u_{k}^{n'}\rangle$,
$\delta_{k_{x}}=2\pi/N_{x}$ and $|u_{k}^{n}\rangle$ are the eigenstates
of the lattice Hamiltonian. $N_{x}$ is the number of lattice sites
in $x$ direction. For the limit $N_{x}\rightarrow\infty$, $[F_{k}]^{nn'}\approx e^{-iA_{nn'}\delta_{k}}$
with $A_{nn'}=i\langle u_{k}^{n}|\partial_{k_{x}}u_{k}^{n'}\rangle$
the non-Abelian gauge field.

\begin{figure}[h]
\centering

\includegraphics[width=8.5cm]{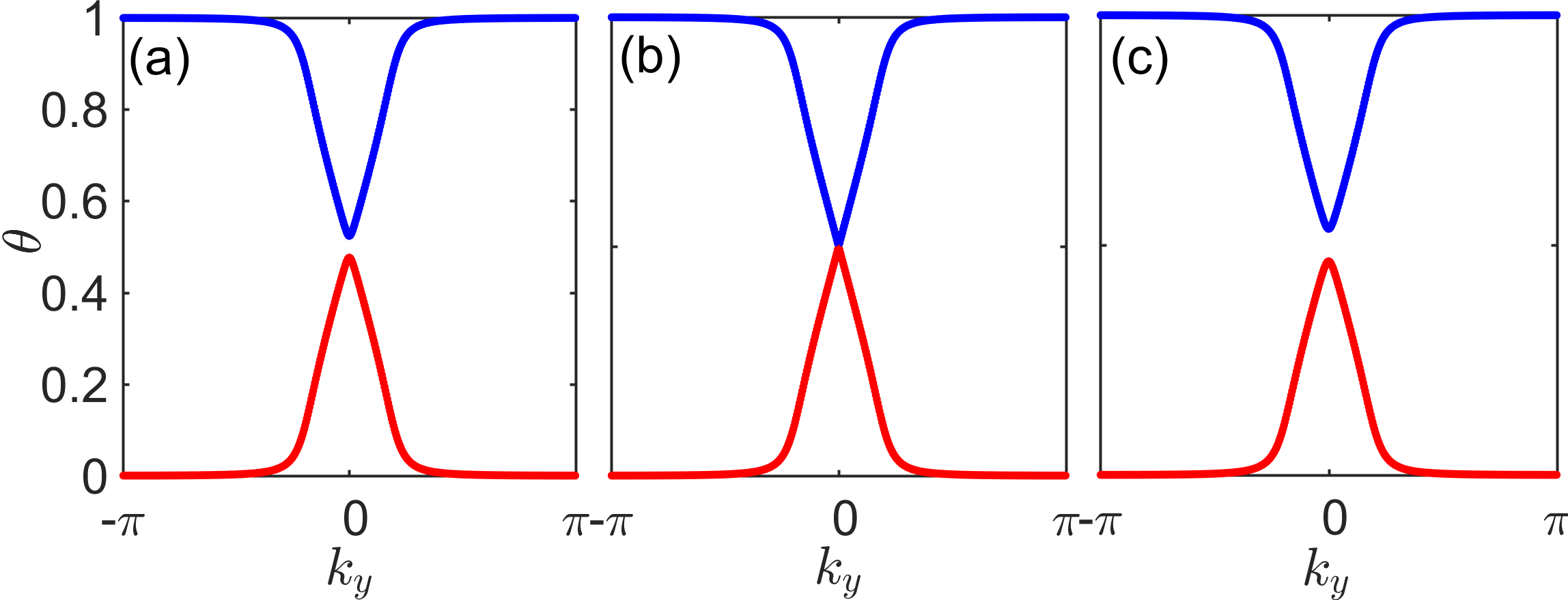}

\caption{Wannier bands $\theta_{x}^{+}$ (blue) and $\theta_{x}^{-}$ (red)
of occupied states in $x$ direction for $\Delta_{0}=0$ (a), $0.04m_{0}$
(b) and $0.1m_{0}$ (c). The Wannier bands do not touch at any point
over $k_{y}\in(-\pi,\pi]$ except for the special case with $\Delta_{0}=2\Delta_{2}m_{0}/m$.
Other parameters are $\mu=0$, $m=2.5$, $m_{0}=1$, $A=1$ and $\Delta_{2}=0.2$
(we choose a larger $\Delta_{2}$ in order to make the Wannier gap
more visible).}

\label{fig:diagram-1}
\end{figure}

A Wilson loop $\mathcal{W}_{k_{x}+2\pi\leftarrow k_{x}}$ is defined
as a Wilson line that goes across the entire Brillouin zone. It is
unitary and its eigenvalues take the form
\begin{equation}
E_{j,R}=e^{i\delta_{k}(\theta_{x}^{j}+R)},
\end{equation}
where $R\in\{0,1,...,N-1\}$. The phases $\theta_{x}^{j}$ of the
eigenvalues are called the Wannier centers. They correspond to the
position of the electrons relative to the center of the unit cell
\citep{RYu11PRB}. The Wilson loop can be connected to the Wannier
Hamiltonian $H_{\mathcal{W}}(k)$ of the edge, $\mathcal{W}_{x,k}\equiv e^{iH_{\mathcal{W}}(k)}.$
It can be adiabatically related to the physical Hamiltonian of the
$x$ edge \citep{Fidkowski11PRL}. Correspondingly, $\theta_{x}^{j}(k_{y})$
with $j\in\{1,...,N_{\text{occ}}\}$ are refereed to the Wannier spectrum
(or bands). It depends on the $k_{y}$ coordinate. Given the normalized
Wilson-loop eigenstate, the eigenstates of $\mathcal{W}_{k_{x}+2\pi\leftarrow k_{x}}$
are written as
\begin{align}
|w_{k}^{j}\rangle & =\sum_{n=1}^{N_{\text{occ}}}[v_{k}^{j}]^{n}\gamma_{n,k}^{\dagger}|0\rangle,\label{eq:Wannier-functions}
\end{align}
where $[v_{k}^{j}]^{n}$ is the $n$th component of the $j$th Wilson-loop
eigenstate $|v_{k}^{j}\rangle$.  Note that while the Wilson-loop
eigenvalues $\theta_{x}^{j}$ do not depend on the base point $k_{x}$,
their eigenstates $|v^{j}\rangle$ do. The electronic contribution
to the dipole moment, called polarization, is proportional to
\begin{equation}
p_{x}(k_{y})=\sum_{j}\theta_{x}^{j}(k_{y})\ \text{mod\ }1.
\end{equation}
Consider the SOTS with $\mu=0$ and vary $\Delta_{0}$. There are
$N_{\text{occ}}=2$ Wannier bands $\theta_{x}^{\pm}(k_{y})$ corresponding
to the two occupied bands, as shown in Fig.\ \ref{fig:diagram-1}.
These two Wannier bands, in general, do not touch at any point over
$k_{y}$ except for $\Delta_{0}=2\Delta_{2}m_{0}/m$. They obey $\theta_{x}^{+}(k_{y})+\theta_{x}^{-}(k_{y})=0$
mod 1. Thus, the total polarization is always zero. We can define
the two Wannier bands as $\theta_{x}^{+}\in[0,1/2)$ and $\theta_{x}^{-}\in(1/2,1]$.

\begin{figure}[ht]
\centering

\includegraphics[width=8.5cm]{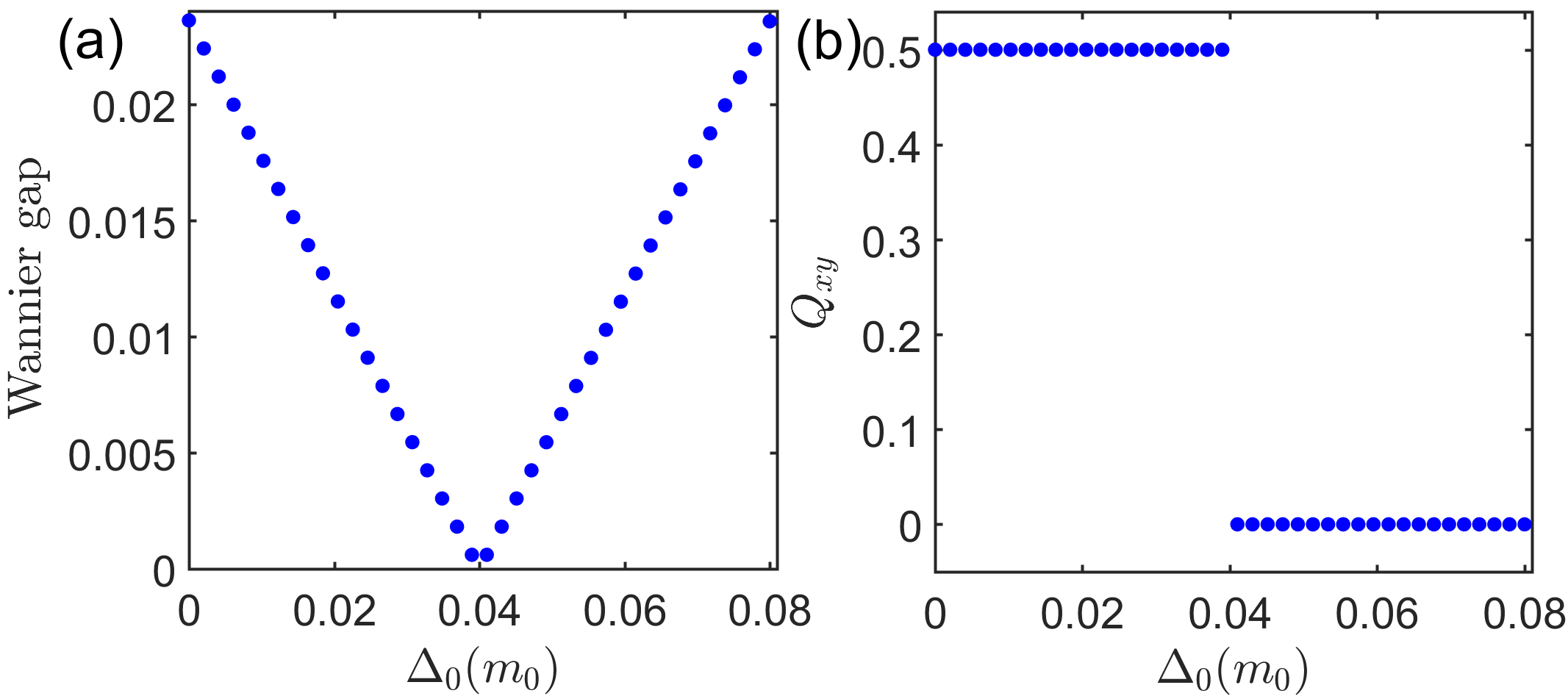}

\caption{(a) Wannier gap, $\text{min}(|\theta_{x}^{+}-\theta_{x}^{-}|)$, as
a function of $\Delta_{0}$; (b) quadrupole moment as a function of
$\Delta_{0}$. Other parameters are the same as those in Fig.\ \ref{fig:diagram-1}.}

\label{fig:diagram-1-1}
\end{figure}

Following a similar approach as that for the lattice model, we can
define a nested Wilson loop $\mathcal{\widetilde{W}}_{k_{y}+2\pi\leftarrow k_{y}}$
for the Wannier bands with the Wannier functions given by Eq.\ (\ref{eq:Wannier-functions})
and calculate the associated polarization $p_{y}^{\theta_{x}^{\pm}}$.
Under reflections $\mathcal{M}_{x}$, $\mathcal{M}_{y}$, and inversion
$\mathcal{I}$, the Wannier sector polarization obey
\begin{eqnarray}
p_{y}^{\theta_{x}^{+}} & \overset{\mathcal{I}}{=} & -p_{y}^{\theta_{x}^{-}},\ p_{y}^{\theta_{x}^{+}}\overset{\mathcal{M}_{x}}{=}p_{y}^{\theta_{x}^{-}},\ p_{y}^{\theta_{x}^{\pm}}\overset{\mathcal{M}_{y}}{=}-p_{y}^{\theta_{x}^{\pm}}
\end{eqnarray}
mod 1. Thus, $p_{y}^{\theta_{x}^{\pm}}$ must quantize (at 0 or $1/2$)
in the presence of the symmetries $\mathcal{M}_{x}$, $\mathcal{M}_{y}$
and $\mathcal{I}$. The relations and result for the other Wannier
polarization $p_{x}^{\theta_{y}^{\pm}}$ are the same as above but
with exchanging $x\leftrightarrow y$. In reflection symmetric insulators,
the Wannier polarization can be alternatively computed from the eigenvalues
of symmetry operators at the reflection-invariant momenta \citep{Benalcazar17PRB}.
The existence of corner states can be associated with the quantized
polarization $p_{y}^{\theta_{x}^{\pm}},p_{x}^{\theta_{y}^{\pm}}=1/2$.

The quadrupole moment $Q_{xy}$ can be written as
\begin{equation}
Q_{xy}=2p_{y}^{\theta_{x}^{\pm}}p_{x}^{\theta_{y}^{\pm}}.
\end{equation}
The SOTS at $\mu=0$ preserves $\mathcal{M}_{x}$ and $\mathcal{M}_{y}$
symmetries. Thus, $Q_{xy}$ is always quantized and can be identified
as the topological invariant for the SOTS. When increasing $\Delta_{0}$,
the quadrupole moment changes from 1/2 to 0 at $\Delta_{0}=2\Delta_{2}m_{0}/m$
where the Wannier gap closes, as shown in Fig.\ \ref{fig:diagram-1-1}.
However, when $\mu\neq0$, both $\mathcal{M}_{x}$ and $\mathcal{M}_{y}$
symmetries are broken. Then, $Q_{xy}$ is no longer quantized. Nevertheless,
since we can smoothly vary $\mu$ to the particular limit (with $\mathcal{M}_{x}$
and $\mathcal{M}_{y}$ symmetries) without closing either the bulk
or edge gap, the general SOTS phase is topologically equivalent to
the SOTSs that preserve these reflection symmetries.

\begin{figure}[hb]
\centering

\includegraphics[width=8.6cm]{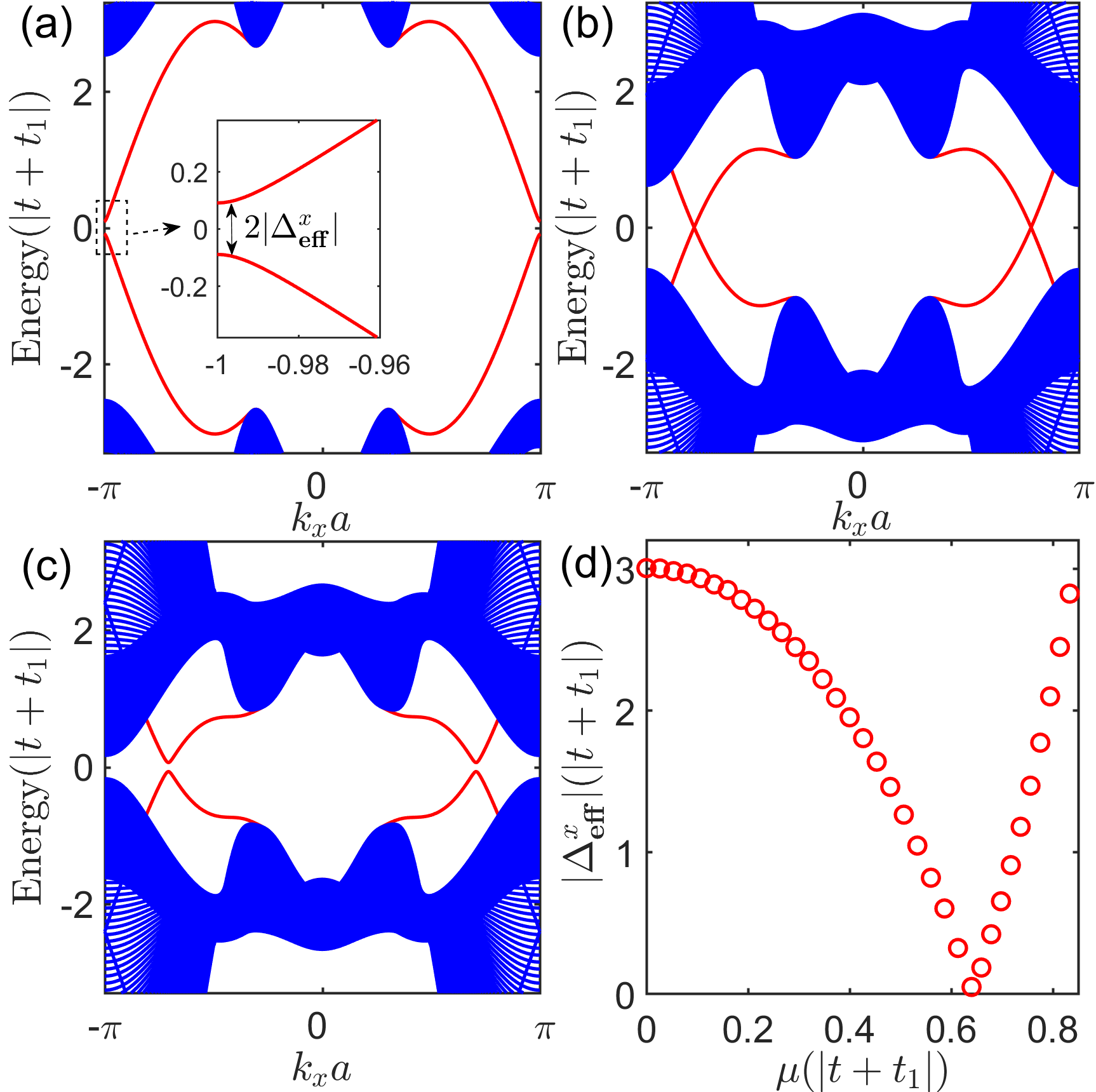}

\caption{(a-c) Energy spectra of the lattice model (\ref{eq:LM-different})
in a ribbon geometry along $x$ direction for the chemical potential
$\mu=0$, $0.64|t+t_{1}|$ and $0.8|t+t_{1}|$, respectively. The
red curves corresponds to the edge bands while the blue ones are the
bulk bands. (d) Edge pairing gap $|\Delta_{\text{eff}}^{x}|$ as a
function of $\mu$. One can observe a gap closing of $\Delta_{\text{eff}}^{x}$
as increasing $\mu$. Other parameters are $t=2$, $t_{1}=1$, $\lambda=1.5$,
$a=1$, $\Delta_{0}=0.12|t+t_{1}|$ and $\Delta_{2}=0.05|t+t_{1}|$.
The number of lattice sites in $y$ direction is $L_{y}=200$.}

\label{fig:LM-spectrum}
\end{figure}
\section{Results from another typical lattice model\label{sec:Results-from-another}}

In this section, we will show that the minimal model can be derived
from different lattice models for SOTSs at low energies. For instance,
we consider another typical lattice model for SOTSs given by \citep{QYWang18PRL}
\begin{eqnarray}
H_{\text{BdG}} & = & M({\bf k})\tau_{z}\sigma_{z}+2\lambda(\sin k_{x}s_{y}-\sin k_{y}s_{x})\tau_{z}\sigma_{x}\nonumber \\
 &  & -\mu\tau_{z}+\Delta({\bf k})\tau_{x}\label{eq:LM-different}
\end{eqnarray}
in the basis $(c_{a,\uparrow},c_{a,\downarrow},c_{b,\uparrow},c_{b,\downarrow},c_{a,\uparrow}^{\dagger},c_{a,\downarrow}^{\dagger},c_{b,\uparrow}^{\dagger},c_{b,\downarrow}^{\dagger})$,
where $M({\bf k})=2t(\cos k_{x}-\cos k_{y})+4t_{1}\cos k_{x}\cos k_{y}$
and $\Delta({\bf k})=\Delta_{0}+2\Delta_{2}(\cos k_{x}+\cos k_{y})$
with $|\Delta_{0}|<2|\Delta_{2}|$. The Pauli matrices $\bm{\tau}$,
$\bm{\sigma}$ and ${\bf s}$ act on Nambu, orbital and spin spaces,
respectively. This lattice model describes a QSHI with $s_{\pm}$-wave
pairing potential but with band inversion at the $X=(\pi,0)$ point.

 Around the $X$ point, we expand all terms up to quadratic order
in momentum ${\bf k}$:
\begin{eqnarray}
M({\bf k}) & \rightarrow & m({\bf k})=-4(t+t_{1})+(t+2t_{1})|{\bf k}|^{2},\nonumber \\
\sin k_{x} & \rightarrow & -k_{x},\ \ \sin k_{y}\rightarrow k_{y},\nonumber \\
\Delta({\bf k}) & \rightarrow & \Delta_{0}+\Delta_{2}(k_{x}^{2}-k_{y}^{2}).
\end{eqnarray}
Here, ${\bf k}$ is measured from the $X$ point. Rearranging the
basis to $(c_{a,\uparrow},-c_{b,\downarrow},c_{a,\downarrow},c_{b,\uparrow},c_{a,\downarrow}^{\dagger},c_{b,\uparrow}^{\dagger},-c_{a,\uparrow}^{\dagger},c_{b,\downarrow}^{\dagger})$,
we then obtain the low-energy effective model as \begin{widetext}
\begin{eqnarray}
H_{\text{BdG}} & \rightarrow & \begin{pmatrix}m({\bf k})-\mu & 2\lambda k_{-} & 0 & 0 & 0 & 0 & -\Delta({\bf k}) & 0\\
2\lambda k_{+} & -m({\bf k})-\mu & 0 & 0 & 0 & 0 & 0 & -\Delta({\bf k})\\
0 & 0 & m({\bf k})-\mu & -2\lambda k_{+} & \Delta({\bf k}) & 0 & 0 & 0\\
0 & 0 & -2\lambda k_{-} & -m({\bf k})-\mu & 0 & \Delta({\bf k}) & 0 & 0\\
0 & 0 & \Delta({\bf k}) & 0 & -m({\bf k})+\mu & 2\lambda k_{+} & 0 & 0\\
0 & 0 & 0 & \Delta({\bf k}) & 2\lambda k_{-} & m({\bf k})+\mu & 0 & 0\\
-\Delta({\bf k}) & 0 & 0 & 0 & 0 & 0 & -m({\bf k})+\mu & -2\lambda k_{-}\\
0 & -\Delta({\bf k}) & 0 & 0 & 0 & 0 & -2\lambda k_{+} & m({\bf k})+\mu
\end{pmatrix},
\end{eqnarray}
\end{widetext}where $k_{\pm}=k_{y}\pm ik_{x}$. It takes the some
form of the minimal model (1) in the main text.

Taking a set of parameters that satisfy the large inverted gap condition,
the energy spectra of the lattice model (\ref{eq:LM-different}) of
a ribbon geometry in $x$ direction are presented in Fig.\ \ref{fig:LM-spectrum}.
At $k_{x}=\pm\pi$, band inversion happens and edge states appear
nearby. More importantly, with increasing the chemical potential from
zero, we can again observe a gap closing and reopening at the edge
states without closing the bulk gap.


\begin{thebibliography}{84}%
\makeatletter
\providecommand \@ifxundefined [1]{%
 \@ifx{#1\undefined}
}%
\providecommand \@ifnum [1]{%
 \ifnum #1\expandafter \@firstoftwo
 \else \expandafter \@secondoftwo
 \fi
}%
\providecommand \@ifx [1]{%
 \ifx #1\expandafter \@firstoftwo
 \else \expandafter \@secondoftwo
 \fi
}%
\providecommand \natexlab [1]{#1}%
\providecommand \enquote  [1]{``#1''}%
\providecommand \bibnamefont  [1]{#1}%
\providecommand \bibfnamefont [1]{#1}%
\providecommand \citenamefont [1]{#1}%
\providecommand \href@noop [0]{\@secondoftwo}%
\providecommand \href [0]{\begingroup \@sanitize@url \@href}%
\providecommand \@href[1]{\@@startlink{#1}\@@href}%
\providecommand \@@href[1]{\endgroup#1\@@endlink}%
\providecommand \@sanitize@url [0]{\catcode `\\12\catcode `\$12\catcode
  `\&12\catcode `\#12\catcode `\^12\catcode `\_12\catcode `\%12\relax}%
\providecommand \@@startlink[1]{}%
\providecommand \@@endlink[0]{}%
\providecommand \url  [0]{\begingroup\@sanitize@url \@url }%
\providecommand \@url [1]{\endgroup\@href {#1}{\urlprefix }}%
\providecommand \urlprefix  [0]{URL }%
\providecommand \Eprint [0]{\href }%
\providecommand \doibase [0]{http://dx.doi.org/}%
\providecommand \selectlanguage [0]{\@gobble}%
\providecommand \bibinfo  [0]{\@secondoftwo}%
\providecommand \bibfield  [0]{\@secondoftwo}%
\providecommand \translation [1]{[#1]}%
\providecommand \BibitemOpen [0]{}%
\providecommand \bibitemStop [0]{}%
\providecommand \bibitemNoStop [0]{.\EOS\space}%
\providecommand \EOS [0]{\spacefactor3000\relax}%
\providecommand \BibitemShut  [1]{\csname bibitem#1\endcsname}%
\let\auto@bib@innerbib\@empty
\bibitem [{\citenamefont {Langbehn}\ \emph {et~al.}(2017)\citenamefont
  {Langbehn}, \citenamefont {Peng}, \citenamefont {Trifunovic}, \citenamefont
  {von Oppen},\ and\ \citenamefont {Brouwer}}]{Langbehn17PRL}%
  \BibitemOpen
  \bibfield  {author} {\bibinfo {author} {\bibfnamefont {J.}~\bibnamefont
  {Langbehn}}, \bibinfo {author} {\bibfnamefont {Y.}~\bibnamefont {Peng}},
  \bibinfo {author} {\bibfnamefont {L.}~\bibnamefont {Trifunovic}}, \bibinfo
  {author} {\bibfnamefont {F.}~\bibnamefont {von Oppen}}, \ and\ \bibinfo
  {author} {\bibfnamefont {P.~W.}\ \bibnamefont {Brouwer}},\ }\bibfield
  {title} {\enquote {\bibinfo {title} {{Reflection-Symmetric Second-Order
  Topological Insulators and Superconductors}},}\ }\href {\doibase
  10.1103/PhysRevLett.119.246401} {\bibfield  {journal} {\bibinfo  {journal}
  {Phys. Rev. Lett.}\ }\textbf {\bibinfo {volume} {119}},\ \bibinfo {pages}
  {246401} (\bibinfo {year} {2017})}\BibitemShut {NoStop}%
\bibitem [{\citenamefont {Khalaf}(2018)}]{Khalaf18PRB}%
  \BibitemOpen
  \bibfield  {author} {\bibinfo {author} {\bibfnamefont {E.}~\bibnamefont
  {Khalaf}},\ }\bibfield  {title} {\enquote {\bibinfo {title} {Higher-order
  topological insulators and superconductors protected by inversion
  symmetry},}\ }\href {\doibase 10.1103/PhysRevB.97.205136} {\bibfield
  {journal} {\bibinfo  {journal} {Phys. Rev. B}\ }\textbf {\bibinfo {volume}
  {97}},\ \bibinfo {pages} {205136} (\bibinfo {year} {2018})}\BibitemShut
  {NoStop}%
\bibitem [{\citenamefont {Wang}\ \emph
  {et~al.}(2018{\natexlab{a}})\citenamefont {Wang}, \citenamefont {Liu},
  \citenamefont {Lu},\ and\ \citenamefont {Zhang}}]{QYWang18PRL}%
  \BibitemOpen
  \bibfield  {author} {\bibinfo {author} {\bibfnamefont {Q.}~\bibnamefont
  {Wang}}, \bibinfo {author} {\bibfnamefont {C.-C.}\ \bibnamefont {Liu}},
  \bibinfo {author} {\bibfnamefont {Y.-M.}\ \bibnamefont {Lu}}, \ and\ \bibinfo
  {author} {\bibfnamefont {F.}~\bibnamefont {Zhang}},\ }\bibfield  {title}
  {\enquote {\bibinfo {title} {{High-Temperature Majorana Corner States}},}\
  }\href {\doibase 10.1103/PhysRevLett.121.186801} {\bibfield  {journal}
  {\bibinfo  {journal} {Phys. Rev. Lett.}\ }\textbf {\bibinfo {volume} {121}},\
  \bibinfo {pages} {186801} (\bibinfo {year} {2018}{\natexlab{a}})}\BibitemShut
  {NoStop}%
\bibitem [{\citenamefont {Yan}\ \emph {et~al.}(2018)\citenamefont {Yan},
  \citenamefont {Song},\ and\ \citenamefont {Wang}}]{ZBYan18PRL}%
  \BibitemOpen
  \bibfield  {author} {\bibinfo {author} {\bibfnamefont {Z.}~\bibnamefont
  {Yan}}, \bibinfo {author} {\bibfnamefont {F.}~\bibnamefont {Song}}, \ and\
  \bibinfo {author} {\bibfnamefont {Z.}~\bibnamefont {Wang}},\ }\bibfield
  {title} {\enquote {\bibinfo {title} {{Majorana Corner Modes in a
  High-Temperature Platform}},}\ }\href {\doibase
  10.1103/PhysRevLett.121.096803} {\bibfield  {journal} {\bibinfo  {journal}
  {Phys. Rev. Lett.}\ }\textbf {\bibinfo {volume} {121}},\ \bibinfo {pages}
  {096803} (\bibinfo {year} {2018})}\BibitemShut {NoStop}%
\bibitem [{\citenamefont {Liu}\ \emph {et~al.}(2018)\citenamefont {Liu},
  \citenamefont {He},\ and\ \citenamefont {Nori}}]{TLiu18PRB}%
  \BibitemOpen
  \bibfield  {author} {\bibinfo {author} {\bibfnamefont {T.}~\bibnamefont
  {Liu}}, \bibinfo {author} {\bibfnamefont {J.~J.}\ \bibnamefont {He}}, \ and\
  \bibinfo {author} {\bibfnamefont {F.}~\bibnamefont {Nori}},\ }\bibfield
  {title} {\enquote {\bibinfo {title} {Majorana corner states in a
  two-dimensional magnetic topological insulator on a high-temperature
  superconductor},}\ }\href {\doibase 10.1103/PhysRevB.98.245413} {\bibfield
  {journal} {\bibinfo  {journal} {Phys. Rev. B}\ }\textbf {\bibinfo {volume}
  {98}},\ \bibinfo {pages} {245413} (\bibinfo {year} {2018})}\BibitemShut
  {NoStop}%
\bibitem [{\citenamefont {Geier}\ \emph {et~al.}(2018)\citenamefont {Geier},
  \citenamefont {Trifunovic}, \citenamefont {Hoskam},\ and\ \citenamefont
  {Brouwer}}]{Geier18PRB}%
  \BibitemOpen
  \bibfield  {author} {\bibinfo {author} {\bibfnamefont {M.}~\bibnamefont
  {Geier}}, \bibinfo {author} {\bibfnamefont {L.}~\bibnamefont {Trifunovic}},
  \bibinfo {author} {\bibfnamefont {M.}~\bibnamefont {Hoskam}}, \ and\ \bibinfo
  {author} {\bibfnamefont {P.~W.}\ \bibnamefont {Brouwer}},\ }\bibfield
  {title} {\enquote {\bibinfo {title} {Second-order topological insulators and
  superconductors with an order-two crystalline symmetry},}\ }\href {\doibase
  10.1103/PhysRevB.97.205135} {\bibfield  {journal} {\bibinfo  {journal} {Phys.
  Rev. B}\ }\textbf {\bibinfo {volume} {97}},\ \bibinfo {pages} {205135}
  (\bibinfo {year} {2018})}\BibitemShut {NoStop}%
\bibitem [{\citenamefont {Zhu}(2018)}]{XYZhu18PRB}%
  \BibitemOpen
  \bibfield  {author} {\bibinfo {author} {\bibfnamefont {X.}~\bibnamefont
  {Zhu}},\ }\bibfield  {title} {\enquote {\bibinfo {title} {{Tunable Majorana
  corner states in a two-dimensional second-order topological superconductor
  induced by magnetic fields}},}\ }\href {\doibase 10.1103/PhysRevB.97.205134}
  {\bibfield  {journal} {\bibinfo  {journal} {Phys. Rev. B}\ }\textbf {\bibinfo
  {volume} {97}},\ \bibinfo {pages} {205134} (\bibinfo {year}
  {2018})}\BibitemShut {NoStop}%
\bibitem [{\citenamefont {Wang}\ \emph {et~al.}(2018)\citenamefont {Wang},
   \citenamefont {Lin},\ and\ \citenamefont
  {Hughes}}]{YXWang18PRB}%
  \BibitemOpen
  \bibfield  {author} {\bibinfo {author} {\bibfnamefont {Y.}\ \bibnamefont
  {Wang}}, \bibinfo {author} {\bibfnamefont {M.}~\bibnamefont {Lin}}, \ and\ \bibinfo
  {author} {\bibfnamefont {T. L.}~\bibnamefont {Hughes}},\ }\bibfield  {title}
  {\enquote {\bibinfo {title} {{Weak-pairing higher order topological superconductors}},}\ }\href {\doibase 10.1103/PhysRevB.98.165144}
  {\bibfield  {journal} {\bibinfo  {journal} {Phys. Rev. B}\ }\textbf
  {\bibinfo {volume} {98}},\ \bibinfo {pages} {165144} (\bibinfo {year}
  {2018})}\BibitemShut {NoStop}%
\bibitem [{\citenamefont {Hsu}\ \emph {et~al.}(2018)\citenamefont {Hsu},
  \citenamefont {Stano}, \citenamefont {Klinovaja},\ and\ \citenamefont
  {Loss}}]{CHHsu18PRL}%
  \BibitemOpen
  \bibfield  {author} {\bibinfo {author} {\bibfnamefont {C.-H.}\ \bibnamefont
  {Hsu}}, \bibinfo {author} {\bibfnamefont {P.}~\bibnamefont {Stano}}, \bibinfo
  {author} {\bibfnamefont {J.}~\bibnamefont {Klinovaja}}, \ and\ \bibinfo
  {author} {\bibfnamefont {D.}~\bibnamefont {Loss}},\ }\bibfield  {title}
  {\enquote {\bibinfo {title} {{Majorana Kramers Pairs in Higher-Order
  Topological Insulators}},}\ }\href {\doibase 10.1103/PhysRevLett.121.196801}
  {\bibfield  {journal} {\bibinfo  {journal} {Phys. Rev. Lett.}\ }\textbf
  {\bibinfo {volume} {121}},\ \bibinfo {pages} {196801} (\bibinfo {year}
  {2018})}\BibitemShut {NoStop}%
\bibitem [{\citenamefont {Zhang}\ \emph
  {et~al.}(2019{\natexlab{a}})\citenamefont {Zhang}, \citenamefont {Cole},\
  and\ \citenamefont {Das~Sarma}}]{RXZhang19PRL}%
  \BibitemOpen
  \bibfield  {author} {\bibinfo {author} {\bibfnamefont {R.-X.}\ \bibnamefont
  {Zhang}}, \bibinfo {author} {\bibfnamefont {W.~S.}\ \bibnamefont {Cole}}, \
  and\ \bibinfo {author} {\bibfnamefont {S.}~\bibnamefont {Das~Sarma}},\
  }\bibfield  {title} {\enquote {\bibinfo {title} {{Helical Hinge Majorana
  Modes in Iron-Based Superconductors}},}\ }\href {\doibase
  10.1103/PhysRevLett.122.187001} {\bibfield  {journal} {\bibinfo  {journal}
  {Phys. Rev. Lett.}\ }\textbf {\bibinfo {volume} {122}},\ \bibinfo {pages}
  {187001} (\bibinfo {year} {2019}{\natexlab{a}})}\BibitemShut {NoStop}%
\bibitem [{\citenamefont {{Qin}}\ \emph {et~al.}()\citenamefont {{Qin}},
  \citenamefont {{Hu}}, \citenamefont {{Le}}, \citenamefont {{Zeng}},
  \citenamefont {{Zhang}}, \citenamefont {{Fang}},\ and\ \citenamefont
  {{Hu}}}]{SSQin19arxiv}%
  \BibitemOpen
  \bibfield  {author} {\bibinfo {author} {\bibfnamefont {S.}~\bibnamefont
  {{Qin}}}, \bibinfo {author} {\bibfnamefont {L.}~\bibnamefont {{Hu}}},
  \bibinfo {author} {\bibfnamefont {C.}~\bibnamefont {{Le}}}, \bibinfo {author}
  {\bibfnamefont {J.}~\bibnamefont {{Zeng}}}, \bibinfo {author} {\bibfnamefont
  {F.-C.}\ \bibnamefont {{Zhang}}}, \bibinfo {author} {\bibfnamefont
  {C.}~\bibnamefont {{Fang}}}, \ and\ \bibinfo {author} {\bibfnamefont
  {J.}~\bibnamefont {{Hu}}},\ }\bibfield  {title} {\enquote {\bibinfo {title}
  {{Quasi 1D topological nodal vortex line phase in doped superconducting 3D
  Dirac Semimetals}},}\ }\href@noop {} {\ }\Eprint
  {http://arxiv.org/abs/1901.04932} {arXiv:1901.04932} \BibitemShut {NoStop}%
\bibitem [{\citenamefont {Bultinck}\ \emph {et~al.}(2019)\citenamefont
  {Bultinck}, \citenamefont {Bernevig},\ and\ \citenamefont
  {Zaletel}}]{Bultinck19PRB}%
  \BibitemOpen
  \bibfield  {author} {\bibinfo {author} {\bibfnamefont {N.}~\bibnamefont
  {Bultinck}}, \bibinfo {author} {\bibfnamefont {B.~A.}\ \bibnamefont
  {Bernevig}}, \ and\ \bibinfo {author} {\bibfnamefont {M.~P.}\ \bibnamefont
  {Zaletel}},\ }\bibfield  {title} {\enquote {\bibinfo {title}
  {Three-dimensional superconductors with hybrid higher-order topology},}\
  }\href {\doibase 10.1103/PhysRevB.99.125149} {\bibfield  {journal} {\bibinfo
  {journal} {Phys. Rev. B}\ }\textbf {\bibinfo {volume} {99}},\ \bibinfo
  {pages} {125149} (\bibinfo {year} {2019})}\BibitemShut {NoStop}%
  \bibitem [{\citenamefont {Peng}\ \emph {et~al.}(2019)\citenamefont
  {Peng}, \ and\ \citenamefont
  {Xu}}]{Peng19PRB}%
  \BibitemOpen
  \bibfield  {author} {\bibinfo {author} {\bibfnamefont {Y.}~\bibnamefont
  {Peng}} \ and\ \bibinfo {author} {\bibfnamefont {Y.}\ \bibnamefont
  {Xu}},\ }\bibfield  {title} {\enquote {\bibinfo {title}
  {{Proximity-induced Majorana hinge modes in antiferromagnetic topological insulators}},}\
  }\href {\doibase 10.1103/PhysRevB.99.195431} {\bibfield  {journal} {\bibinfo
  {journal} {Phys. Rev. B}\ }\textbf {\bibinfo {volume} {99}},\ \bibinfo
  {pages} {195431} (\bibinfo {year} {2019})}\BibitemShut {NoStop}%
\bibitem [{\citenamefont {Kitaev}(2001)}]{Kitaev01PU}%
  \BibitemOpen
  \bibfield  {author} {\bibinfo {author} {\bibfnamefont {A.~Y.}\ \bibnamefont
  {Kitaev}},\ }\bibfield  {title} {\enquote {\bibinfo {title} {{Unpaired
  Majorana fermions in quantum wires}},}\ }\href
  {http://stacks.iop.org/1063-7869/44/i=10S/a=S29} {\bibfield  {journal}
  {\bibinfo  {journal} {Phys. Usp.}\ }\textbf {\bibinfo {volume} {44}},\
  \bibinfo {pages} {131} (\bibinfo {year} {2001})}\BibitemShut {NoStop}%
\bibitem [{\citenamefont {Kitaev}(2003)}]{kitaev03AP}%
  \BibitemOpen
  \bibfield  {author} {\bibinfo {author} {\bibfnamefont {A~Y.}\ \bibnamefont
  {Kitaev}},\ }\bibfield  {title} {\enquote {\bibinfo {title} {Fault-tolerant
  quantum computation by anyons},}\ }\href {\doibase
  10.1016/S0003-4916(02)00018-0} {\bibfield  {journal} {\bibinfo  {journal}
  {Ann. Phys.}\ }\textbf {\bibinfo {volume} {303}},\ \bibinfo {pages} {2}
  (\bibinfo {year} {2003})}\BibitemShut {NoStop}%
\bibitem [{\citenamefont {Nayak}\ \emph {et~al.}(2008)\citenamefont {Nayak},
  \citenamefont {Simon}, \citenamefont {Stern}, \citenamefont {Freedman},\ and\
  \citenamefont {Das~Sarma}}]{Nayak08RMP}%
  \BibitemOpen
  \bibfield  {author} {\bibinfo {author} {\bibfnamefont {C.}~\bibnamefont
  {Nayak}}, \bibinfo {author} {\bibfnamefont {S.~H.}\ \bibnamefont {Simon}},
  \bibinfo {author} {\bibfnamefont {A.}~\bibnamefont {Stern}}, \bibinfo
  {author} {\bibfnamefont {M.}~\bibnamefont {Freedman}}, \ and\ \bibinfo
  {author} {\bibfnamefont {S.}~\bibnamefont {Das~Sarma}},\ }\bibfield  {title}
  {\enquote {\bibinfo {title} {{Non-Abelian anyons and topological quantum
  computation}},}\ }\href {\doibase 10.1103/RevModPhys.80.1083} {\bibfield
  {journal} {\bibinfo  {journal} {Rev. Mod. Phys.}\ }\textbf {\bibinfo {volume}
  {80}},\ \bibinfo {pages} {1083} (\bibinfo {year} {2008})}\BibitemShut
  {NoStop}%
\bibitem [{\citenamefont {Alicea}(2012)}]{Alicea12PPP}%
  \BibitemOpen
  \bibfield  {author} {\bibinfo {author} {\bibfnamefont {J.}~\bibnamefont
  {Alicea}},\ }\bibfield  {title} {\enquote {\bibinfo {title} {{New directions
  in the pursuit of Majorana fermions in solid state systems}},}\ }\href
  {http://stacks.iop.org/0034-4885/75/i=7/a=076501} {\bibfield  {journal}
  {\bibinfo  {journal} {Rep. Prog. Phys.}\ }\textbf {\bibinfo {volume} {75}},\
  \bibinfo {pages} {076501} (\bibinfo {year} {2012})}\BibitemShut {NoStop}%
\bibitem [{\citenamefont {Leijnse}\ and\ \citenamefont
  {Flensberg}(2012)}]{Leijnse12SST}%
  \BibitemOpen
  \bibfield  {author} {\bibinfo {author} {\bibfnamefont {M.}~\bibnamefont
  {Leijnse}}\ and\ \bibinfo {author} {\bibfnamefont {K.}~\bibnamefont
  {Flensberg}},\ }\bibfield  {title} {\enquote {\bibinfo {title} {{Introduction
  to topological superconductivity and Majorana fermions}},}\ }\href {\doibase
  10.1088/0268-1242/27/12/124003} {\bibfield  {journal} {\bibinfo  {journal}
  {Semicond. Sci. Technol.}\ }\textbf {\bibinfo {volume} {27}},\ \bibinfo
  {pages} {124003} (\bibinfo {year} {2012})}\BibitemShut {NoStop}%
\bibitem [{\citenamefont {Beenakker}(2013)}]{Beenakker13ARCMP}%
  \BibitemOpen
  \bibfield  {author} {\bibinfo {author} {\bibfnamefont {C.~W.~J.}\
  \bibnamefont {Beenakker}},\ }\bibfield  {title} {\enquote {\bibinfo {title}
  {{Search for Majorana Fermions in Superconductors}},}\ }\href {\doibase
  10.1146/annurev-conmatphys-030212-184337} {\bibfield  {journal} {\bibinfo
  {journal} {Annu. Rev. Condens. Matter Phys.}\ }\textbf {\bibinfo {volume}
  {4}},\ \bibinfo {pages} {113} (\bibinfo {year} {2013})}\BibitemShut {NoStop}%
\bibitem [{\citenamefont {Elliott}\ and\ \citenamefont
  {Franz}(2015)}]{Elliott15RMP}%
  \BibitemOpen
  \bibfield  {author} {\bibinfo {author} {\bibfnamefont {S.~R.}\ \bibnamefont
  {Elliott}}\ and\ \bibinfo {author} {\bibfnamefont {M.}~\bibnamefont
  {Franz}},\ }\bibfield  {title} {\enquote {\bibinfo {title} {Colloquium:
  Majorana fermions in nuclear, particle, and solid-state physics},}\ }\href
  {\doibase 10.1103/RevModPhys.87.137} {\bibfield  {journal} {\bibinfo
  {journal} {Rev. Mod. Phys.}\ }\textbf {\bibinfo {volume} {87}},\ \bibinfo
  {pages} {137} (\bibinfo {year} {2015})}\BibitemShut {NoStop}%
\bibitem [{\citenamefont {Sarma}\ \emph {et~al.}(2015)\citenamefont {Sarma},
  \citenamefont {Freedman},\ and\ \citenamefont {Nayak}}]{sarma15npj}%
  \BibitemOpen
  \bibfield  {author} {\bibinfo {author} {\bibfnamefont {S.~Das}\ \bibnamefont
  {Sarma}}, \bibinfo {author} {\bibfnamefont {M.}~\bibnamefont {Freedman}}, \
  and\ \bibinfo {author} {\bibfnamefont {C.}~\bibnamefont {Nayak}},\ }\bibfield
   {title} {\enquote {\bibinfo {title} {Majorana zero modes and topological
  quantum computation},}\ }\href {\doibase 10.1038/npjqi.2015.1} {\bibfield
  {journal} {\bibinfo  {journal} {npj Quantum Inf.}\ }\textbf {\bibinfo
  {volume} {1}},\ \bibinfo {pages} {15001} (\bibinfo {year}
  {2015})}\BibitemShut {NoStop}%
\bibitem [{\citenamefont {Sato}\ and\ \citenamefont
  {Fujimoto}(2016)}]{Sato16JPSJ}%
  \BibitemOpen
  \bibfield  {author} {\bibinfo {author} {\bibfnamefont {M.}~\bibnamefont
  {Sato}}\ and\ \bibinfo {author} {\bibfnamefont {S.}~\bibnamefont
  {Fujimoto}},\ }\bibfield  {title} {\enquote {\bibinfo {title} {Majorana
  fermions and topology in superconductors},}\ }\href {\doibase
  10.7566/JPSJ.85.072001} {\bibfield  {journal} {\bibinfo  {journal} {J. Phys.
  Soc. Jpn.}\ }\textbf {\bibinfo {volume} {85}},\ \bibinfo {pages} {072001}
  (\bibinfo {year} {2016})}\BibitemShut {NoStop}%
\bibitem [{\citenamefont {{K{\"o}nig}}\ and\ \citenamefont
  {{Coleman}}()}]{Konig2019arXiv}%
  \BibitemOpen
  \bibfield  {author} {\bibinfo {author} {\bibfnamefont {E.~J.}\ \bibnamefont
  {{K{\"o}nig}}}\ and\ \bibinfo {author} {\bibfnamefont {P.}~\bibnamefont
  {{Coleman}}},\ }\bibfield  {title} {\enquote {\bibinfo {title} {{Helical
  Majorana modes in iron based Dirac superconductors}},}\ }\href@noop {} {\
  }\Eprint {http://arxiv.org/abs/1901.03692} {arXiv:1901.03692} \BibitemShut
  {NoStop}%
\bibitem [{\citenamefont {Volpez}\ \emph {et~al.}(2019)\citenamefont {Volpez},
  \citenamefont {Loss},\ and\ \citenamefont {Klinovaja}}]{Volpez19PRL}%
  \BibitemOpen
  \bibfield  {author} {\bibinfo {author} {\bibfnamefont {Y.}~\bibnamefont
  {Volpez}}, \bibinfo {author} {\bibfnamefont {D.}~\bibnamefont {Loss}}, \ and\
  \bibinfo {author} {\bibfnamefont {J.}~\bibnamefont {Klinovaja}},\ }\bibfield
  {title} {\enquote {\bibinfo {title} {{Second-Order Topological
  Superconductivity in $\ensuremath{\pi}$-Junction Rashba Layers}},}\ }\href
  {\doibase 10.1103/PhysRevLett.122.126402} {\bibfield  {journal} {\bibinfo
  {journal} {Phys. Rev. Lett.}\ }\textbf {\bibinfo {volume} {122}},\ \bibinfo
  {pages} {126402} (\bibinfo {year} {2019})}\BibitemShut {NoStop}%
\bibitem [{\citenamefont {Shapourian}\ \emph {et~al.}(2018)\citenamefont
  {Shapourian}, \citenamefont {Wang},\ and\ \citenamefont
  {Ryu}}]{Shapourian18PRB}%
  \BibitemOpen
  \bibfield  {author} {\bibinfo {author} {\bibfnamefont {H.}~\bibnamefont
  {Shapourian}}, \bibinfo {author} {\bibfnamefont {Y.}~\bibnamefont {Wang}}, \
  and\ \bibinfo {author} {\bibfnamefont {S.}~\bibnamefont {Ryu}},\ }\bibfield
  {title} {\enquote {\bibinfo {title} {Topological crystalline
  superconductivity and second-order topological superconductivity in
  nodal-loop materials},}\ }\href {\doibase 10.1103/PhysRevB.97.094508}
  {\bibfield  {journal} {\bibinfo  {journal} {Phys. Rev. B}\ }\textbf {\bibinfo
  {volume} {97}},\ \bibinfo {pages} {094508} (\bibinfo {year}
  {2018})}\BibitemShut {NoStop}%
\bibitem [{\citenamefont {{Pan}}\ \emph {et~al.}()\citenamefont {{Pan}},
  \citenamefont {{Yang}}, \citenamefont {{Chen}}, \citenamefont {{Xu}},
  \citenamefont {{Liu}},\ and\ \citenamefont {{Liu}}}]{XHPan18arXiv}%
  \BibitemOpen
  \bibfield  {author} {\bibinfo {author} {\bibfnamefont {X.-H.}\ \bibnamefont
  {{Pan}}}, \bibinfo {author} {\bibfnamefont {K.-J.}\ \bibnamefont {{Yang}}},
  \bibinfo {author} {\bibfnamefont {L.}~\bibnamefont {{Chen}}}, \bibinfo
  {author} {\bibfnamefont {G.}~\bibnamefont {{Xu}}}, \bibinfo {author}
  {\bibfnamefont {C.-X.}\ \bibnamefont {{Liu}}}, \ and\ \bibinfo {author}
  {\bibfnamefont {X.}~\bibnamefont {{Liu}}},\ }\bibfield  {title} {\enquote
  {\bibinfo {title} {{Lattice symmetry assisted second order topological
  superconductors and Majorana patterns}},}\ }\href@noop {} {\ }\Eprint
  {http://arxiv.org/abs/1812.10989} {arXiv:1812.10989} \BibitemShut {NoStop}%
\bibitem [{\citenamefont {{Kheirkhah}}\ \emph {et~al.}()\citenamefont
  {{Kheirkhah}}, \citenamefont {{Nagai}}, \citenamefont {{Chen}},\ and\
  \citenamefont {{Marsiglio}}}]{Kheirkhah19arXiv}%
  \BibitemOpen
  \bibfield  {author} {\bibinfo {author} {\bibfnamefont {M.}~\bibnamefont
  {{Kheirkhah}}}, \bibinfo {author} {\bibfnamefont {Y.}~\bibnamefont
  {{Nagai}}}, \bibinfo {author} {\bibfnamefont {C.}~\bibnamefont {{Chen}}}, \
  and\ \bibinfo {author} {\bibfnamefont {F.}~\bibnamefont {{Marsiglio}}},\
  }\bibfield  {title} {\enquote {\bibinfo {title} {{Majorana corner flat bands
  in two-dimensional second-order topological superconductors}},}\ }\href@noop
  {} {\ }\Eprint {http://arxiv.org/abs/1904.00990} {arXiv:1904.00990}
  \BibitemShut {NoStop}%
\bibitem [{\citenamefont {{Ghorashi}}\ \emph {et~al.}()\citenamefont
  {{Ghorashi}}, \citenamefont {{Hu}}, \citenamefont {{Hughes}},\ and\
  \citenamefont {{Rossi}}}]{Ghorashi19arXiv}%
  \BibitemOpen
  \bibfield  {author} {\bibinfo {author} {\bibfnamefont {S.~A.~A.}\
  \bibnamefont {{Ghorashi}}}, \bibinfo {author} {\bibfnamefont
  {X.}~\bibnamefont {{Hu}}}, \bibinfo {author} {\bibfnamefont {T.~L.}\
  \bibnamefont {{Hughes}}}, \ and\ \bibinfo {author} {\bibfnamefont
  {E.}~\bibnamefont {{Rossi}}},\ }\bibfield  {title} {\enquote {\bibinfo
  {title} {{Second-order Dirac superconductors and magnetic field induced
  Majorana hinge modes}},}\ }\href@noop {} {\ }\Eprint
  {http://arxiv.org/abs/1901.07579} {arXiv:1901.07579} \BibitemShut {NoStop}%
\bibitem [{\citenamefont {Linder}\ and\ \citenamefont
  {Sudb\o{}}(2008)}]{Linder08PRB}%
  \BibitemOpen
  \bibfield  {author} {\bibinfo {author} {\bibfnamefont {J.}~\bibnamefont
  {Linder}}\ and\ \bibinfo {author} {\bibfnamefont {A.}~\bibnamefont
  {Sudb\o{}}},\ }\bibfield  {title} {\enquote {\bibinfo {title} {{Tunneling
  conductance in $s$- and $d$-wave superconductor-graphene junctions: Extended
  Blonder-Tinkham-Klapwijk formalism}},}\ }\href {\doibase
  10.1103/PhysRevB.77.064507} {\bibfield  {journal} {\bibinfo  {journal} {Phys.
  Rev. B}\ }\textbf {\bibinfo {volume} {77}},\ \bibinfo {pages} {064507}
  (\bibinfo {year} {2008})}\BibitemShut {NoStop}%
\bibitem [{\citenamefont {Linder}\ \emph {et~al.}(2010)\citenamefont {Linder},
  \citenamefont {Tanaka}, \citenamefont {Yokoyama}, \citenamefont {Sudb\o{}},\
  and\ \citenamefont {Nagaosa}}]{Linder10PRL}%
  \BibitemOpen
  \bibfield  {author} {\bibinfo {author} {\bibfnamefont {J.}~\bibnamefont
  {Linder}}, \bibinfo {author} {\bibfnamefont {Y.}~\bibnamefont {Tanaka}},
  \bibinfo {author} {\bibfnamefont {T.}~\bibnamefont {Yokoyama}}, \bibinfo
  {author} {\bibfnamefont {A.}~\bibnamefont {Sudb\o{}}}, \ and\ \bibinfo
  {author} {\bibfnamefont {N.}~\bibnamefont {Nagaosa}},\ }\bibfield  {title}
  {\enquote {\bibinfo {title} {{Unconventional Superconductivity on a
  Topological Insulator}},}\ }\href {\doibase 10.1103/PhysRevLett.104.067001}
  {\bibfield  {journal} {\bibinfo  {journal} {Phys. Rev. Lett.}\ }\textbf
  {\bibinfo {volume} {104}},\ \bibinfo {pages} {067001} (\bibinfo {year}
  {2010})}\BibitemShut {NoStop}%
\bibitem [{\citenamefont {Black-Schaffer}\ and\ \citenamefont
  {Balatsky}(2013)}]{Schaffer13PRB}%
  \BibitemOpen
  \bibfield  {author} {\bibinfo {author} {\bibfnamefont {A.~M.}\ \bibnamefont
  {Black-Schaffer}}\ and\ \bibinfo {author} {\bibfnamefont {A.~V.}\
  \bibnamefont {Balatsky}},\ }\bibfield  {title} {\enquote {\bibinfo {title}
  {Proximity-induced unconventional superconductivity in topological
  insulators},}\ }\href {\doibase 10.1103/PhysRevB.87.220506} {\bibfield
  {journal} {\bibinfo  {journal} {Phys. Rev. B}\ }\textbf {\bibinfo {volume}
  {87}},\ \bibinfo {pages} {220506(R)} (\bibinfo {year} {2013})}\BibitemShut
  {NoStop}%
\bibitem [{\citenamefont {Zhang}\ \emph {et~al.}(2013)\citenamefont {Zhang},
  \citenamefont {Kane},\ and\ \citenamefont {Mele}}]{FZhang13PRL}%
  \BibitemOpen
  \bibfield  {author} {\bibinfo {author} {\bibfnamefont {F.}~\bibnamefont
  {Zhang}}, \bibinfo {author} {\bibfnamefont {C.~L.}\ \bibnamefont {Kane}}, \
  and\ \bibinfo {author} {\bibfnamefont {E.~J.}\ \bibnamefont {Mele}},\
  }\bibfield  {title} {\enquote {\bibinfo {title} {{Time-Reversal-Invariant
  Topological Superconductivity and Majorana Kramers Pairs}},}\ }\href
  {\doibase 10.1103/PhysRevLett.111.056402} {\bibfield  {journal} {\bibinfo
  {journal} {Phys. Rev. Lett.}\ }\textbf {\bibinfo {volume} {111}},\ \bibinfo
  {pages} {056402} (\bibinfo {year} {2013})}\BibitemShut {NoStop}%
\bibitem [{\citenamefont {Li}\ \emph {et~al.}(2015)\citenamefont {Li},
  \citenamefont {Chan},\ and\ \citenamefont {Yao}}]{ZXLi15PRB}%
  \BibitemOpen
  \bibfield  {author} {\bibinfo {author} {\bibfnamefont {Z.-X.}\ \bibnamefont
  {Li}}, \bibinfo {author} {\bibfnamefont {C.}~\bibnamefont {Chan}}, \ and\
  \bibinfo {author} {\bibfnamefont {H.}~\bibnamefont {Yao}},\ }\bibfield
  {title} {\enquote {\bibinfo {title} {Realizing majorana zero modes by
  proximity effect between topological insulators and $d$-wave high-temperature
  superconductors},}\ }\href {\doibase 10.1103/PhysRevB.91.235143} {\bibfield
  {journal} {\bibinfo  {journal} {Phys. Rev. B}\ }\textbf {\bibinfo {volume}
  {91}},\ \bibinfo {pages} {235143} (\bibinfo {year} {2015})}\BibitemShut
  {NoStop}%
\bibitem [{\citenamefont {Zareapour}\ \emph {et~al.}(2016)\citenamefont
  {Zareapour}, \citenamefont {Xu}, \citenamefont {Zhao}, \citenamefont {Jain},
  \citenamefont {Xu}, \citenamefont {Liu}, \citenamefont {Gu},\ and\
  \citenamefont {Burch}}]{Zareapour16SCT}%
  \BibitemOpen
  \bibfield  {author} {\bibinfo {author} {\bibfnamefont {P.}~\bibnamefont
  {Zareapour}}, \bibinfo {author} {\bibfnamefont {J.}~\bibnamefont {Xu}},
  \bibinfo {author} {\bibfnamefont {S.~Yang~F.}\ \bibnamefont {Zhao}}, \bibinfo
  {author} {\bibfnamefont {A.}~\bibnamefont {Jain}}, \bibinfo {author}
  {\bibfnamefont {Z.}~\bibnamefont {Xu}}, \bibinfo {author} {\bibfnamefont
  {T.~S.}\ \bibnamefont {Liu}}, \bibinfo {author} {\bibfnamefont {G.~D.}\
  \bibnamefont {Gu}}, \ and\ \bibinfo {author} {\bibfnamefont {K.~S.}\
  \bibnamefont {Burch}},\ }\bibfield  {title} {\enquote {\bibinfo {title}
  {Modeling tunneling for the unconventional superconducting proximity
  effect},}\ }\href {\doibase 10.1088/0953-2048/29/12/125006} {\bibfield
  {journal} {\bibinfo  {journal} {Supercond. Sci. Tech.}\ }\textbf {\bibinfo
  {volume} {29}},\ \bibinfo {pages} {125006} (\bibinfo {year}
  {2016})}\BibitemShut {NoStop}%
\bibitem [{\citenamefont {Li}\ \emph {et~al.}(2016)\citenamefont {Li},
  \citenamefont {Chao},\ and\ \citenamefont {Lee}}]{WJLi16PRB}%
  \BibitemOpen
  \bibfield  {author} {\bibinfo {author} {\bibfnamefont {W.-J.}\ \bibnamefont
  {Li}}, \bibinfo {author} {\bibfnamefont {S.-P.}\ \bibnamefont {Chao}}, \ and\
  \bibinfo {author} {\bibfnamefont {T.-K.}\ \bibnamefont {Lee}},\ }\bibfield
  {title} {\enquote {\bibinfo {title} {Theoretical study of large
  proximity-induced $s$-wave-like pairing from a $d$-wave superconductor},}\
  }\href {\doibase 10.1103/PhysRevB.93.035140} {\bibfield  {journal} {\bibinfo
  {journal} {Phys. Rev. B}\ }\textbf {\bibinfo {volume} {93}},\ \bibinfo
  {pages} {035140} (\bibinfo {year} {2016})}\BibitemShut {NoStop}%
\bibitem [{\citenamefont {Wu}\ \emph {et~al.}(2016)\citenamefont {Wu},
  \citenamefont {Qin}, \citenamefont {Liang}, \citenamefont {Fan},\ and\
  \citenamefont {Hu}}]{XXWu16PRB}%
  \BibitemOpen
  \bibfield  {author} {\bibinfo {author} {\bibfnamefont {X.}~\bibnamefont
  {Wu}}, \bibinfo {author} {\bibfnamefont {S.}~\bibnamefont {Qin}}, \bibinfo
  {author} {\bibfnamefont {Y.}~\bibnamefont {Liang}}, \bibinfo {author}
  {\bibfnamefont {H.}~\bibnamefont {Fan}}, \ and\ \bibinfo {author}
  {\bibfnamefont {J.}~\bibnamefont {Hu}},\ }\bibfield  {title} {\enquote
  {\bibinfo {title} {{Topological characters in
  $\mathrm{Fe}({\mathrm{Te}}_{1\ensuremath{-}x}{\mathrm{Se}}_{x})$ thin
  films}},}\ }\href {\doibase 10.1103/PhysRevB.93.115129} {\bibfield  {journal}
  {\bibinfo  {journal} {Phys. Rev. B}\ }\textbf {\bibinfo {volume} {93}},\
  \bibinfo {pages} {115129} (\bibinfo {year} {2016})}\BibitemShut {NoStop}%
\bibitem [{\citenamefont {Wang}\ \emph {et~al.}(2015)\citenamefont {Wang},
  \citenamefont {Zhang}, \citenamefont {Xu}, \citenamefont {Zeng},
  \citenamefont {Miao}, \citenamefont {Xu}, \citenamefont {Qian}, \citenamefont
  {Weng}, \citenamefont {Richard}, \citenamefont {Fedorov}, \citenamefont
  {Ding}, \citenamefont {Dai},\ and\ \citenamefont {Fang}}]{ZJWang15PRB}%
  \BibitemOpen
  \bibfield  {author} {\bibinfo {author} {\bibfnamefont {Z.}~\bibnamefont
  {Wang}}, \bibinfo {author} {\bibfnamefont {P.}~\bibnamefont {Zhang}},
  \bibinfo {author} {\bibfnamefont {G.}~\bibnamefont {Xu}}, \bibinfo {author}
  {\bibfnamefont {L.~K.}\ \bibnamefont {Zeng}}, \bibinfo {author}
  {\bibfnamefont {H.}~\bibnamefont {Miao}}, \bibinfo {author} {\bibfnamefont
  {X.}~\bibnamefont {Xu}}, \bibinfo {author} {\bibfnamefont {T.}~\bibnamefont
  {Qian}}, \bibinfo {author} {\bibfnamefont {H.}~\bibnamefont {Weng}}, \bibinfo
  {author} {\bibfnamefont {P.}~\bibnamefont {Richard}}, \bibinfo {author}
  {\bibfnamefont {A.~V.}\ \bibnamefont {Fedorov}}, \bibinfo {author}
  {\bibfnamefont {H.}~\bibnamefont {Ding}}, \bibinfo {author} {\bibfnamefont
  {X.}~\bibnamefont {Dai}}, \ and\ \bibinfo {author} {\bibfnamefont
  {Z.}~\bibnamefont {Fang}},\ }\bibfield  {title} {\enquote {\bibinfo {title}
  {{Topological nature of the ${\mathrm{FeSe}}_{0.5}{\mathrm{Te}}_{0.5}$
  superconductor}},}\ }\href {\doibase 10.1103/PhysRevB.92.115119} {\bibfield
  {journal} {\bibinfo  {journal} {Phys. Rev. B}\ }\textbf {\bibinfo {volume}
  {92}},\ \bibinfo {pages} {115119} (\bibinfo {year} {2015})}\BibitemShut
  {NoStop}%
\bibitem [{\citenamefont {Zhou}\ \emph {et~al.}(2019)\citenamefont {Zhou},
  \citenamefont {Gao},\ and\ \citenamefont {Wang}}]{TZhou19PRB}%
  \BibitemOpen
  \bibfield  {author} {\bibinfo {author} {\bibfnamefont {T.}~\bibnamefont
  {Zhou}}, \bibinfo {author} {\bibfnamefont {Y.}~\bibnamefont {Gao}}, \ and\
  \bibinfo {author} {\bibfnamefont {Z.~D.}\ \bibnamefont {Wang}},\ }\bibfield
  {title} {\enquote {\bibinfo {title} {Detecting competing orders through the
  edge states in the heterostructures with high-${T}_{c}$ superconductors},}\
  }\href {\doibase 10.1103/PhysRevB.99.104517} {\bibfield  {journal} {\bibinfo
  {journal} {Phys. Rev. B}\ }\textbf {\bibinfo {volume} {99}},\ \bibinfo
  {pages} {104517} (\bibinfo {year} {2019})}\BibitemShut {NoStop}%
\bibitem [{\citenamefont {Zareapour}\ \emph {et~al.}(2012)\citenamefont
  {Zareapour}, \citenamefont {Hayat}, \citenamefont {Zhao}, \citenamefont
  {Kreshchuk}, \citenamefont {Jain}, \citenamefont {Kwok}, \citenamefont {Lee},
  \citenamefont {Cheong}, \citenamefont {Xu}, \citenamefont {Yang} \emph
  {et~al.}}]{Zareapour12nacom}%
  \BibitemOpen
  \bibfield  {author} {\bibinfo {author} {\bibfnamefont {P.}~\bibnamefont
  {Zareapour}}, \bibinfo {author} {\bibfnamefont {A.}~\bibnamefont {Hayat}},
  \bibinfo {author} {\bibfnamefont {S.~Y.~F.}\ \bibnamefont {Zhao}}, \bibinfo
  {author} {\bibfnamefont {M.}~\bibnamefont {Kreshchuk}}, \bibinfo {author}
  {\bibfnamefont {A.}~\bibnamefont {Jain}}, \bibinfo {author} {\bibfnamefont
  {D.~C.}\ \bibnamefont {Kwok}}, \bibinfo {author} {\bibfnamefont
  {N.}~\bibnamefont {Lee}}, \bibinfo {author} {\bibfnamefont {S.-W.}\
  \bibnamefont {Cheong}}, \bibinfo {author} {\bibfnamefont {Z.}~\bibnamefont
  {Xu}}, \bibinfo {author} {\bibfnamefont {A.}~\bibnamefont {Yang}},  \emph
  {et~al.},\ }\bibfield  {title} {\enquote {\bibinfo {title}
  {{Proximity-induced high-temperature superconductivity in the topological
  insulators Bi$_2$Se$_3$ and Bi$_2$Te$_3$}},}\ }\href {\doibase
  10.1038/ncomms2042} {\bibfield  {journal} {\bibinfo  {journal} {Nat.
  Commun.}\ }\textbf {\bibinfo {volume} {3}},\ \bibinfo {pages} {1056}
  (\bibinfo {year} {2012})}\BibitemShut {NoStop}%
\bibitem [{\citenamefont {Wang}\ \emph {et~al.}(2013)\citenamefont {Wang},
  \citenamefont {Ding}, \citenamefont {Fedorov}, \citenamefont {Yao},
  \citenamefont {Li}, \citenamefont {Lv}, \citenamefont {Zhao}, \citenamefont
  {Zhang}, \citenamefont {Xu}, \citenamefont {Schneeloch} \emph
  {et~al.}}]{EWang13npyhs}%
  \BibitemOpen
  \bibfield  {author} {\bibinfo {author} {\bibfnamefont {E.}~\bibnamefont
  {Wang}}, \bibinfo {author} {\bibfnamefont {H.}~\bibnamefont {Ding}}, \bibinfo
  {author} {\bibfnamefont {A.~V.}\ \bibnamefont {Fedorov}}, \bibinfo {author}
  {\bibfnamefont {W.}~\bibnamefont {Yao}}, \bibinfo {author} {\bibfnamefont
  {Z.}~\bibnamefont {Li}}, \bibinfo {author} {\bibfnamefont {Y.-F.}\
  \bibnamefont {Lv}}, \bibinfo {author} {\bibfnamefont {K.}~\bibnamefont
  {Zhao}}, \bibinfo {author} {\bibfnamefont {L.-G.}\ \bibnamefont {Zhang}},
  \bibinfo {author} {\bibfnamefont {Z.}~\bibnamefont {Xu}}, \bibinfo {author}
  {\bibfnamefont {J.}~\bibnamefont {Schneeloch}},  \emph {et~al.},\ }\bibfield
  {title} {\enquote {\bibinfo {title} {{Fully gapped topological surface states
  in Bi$_2$Se$_3$ films induced by a d-wave high-temperature
  superconductor}},}\ }\href {\doibase 10.1038/nphys2744} {\bibfield  {journal}
  {\bibinfo  {journal} {Nat. Phys.}\ }\textbf {\bibinfo {volume} {9}},\
  \bibinfo {pages} {621} (\bibinfo {year} {2013})}\BibitemShut {NoStop}%
\bibitem [{\citenamefont {Zhao}\ \emph {et~al.}(2018)\citenamefont {Zhao},
  \citenamefont {Rachmilowitz}, \citenamefont {Ren}, \citenamefont {Han},
  \citenamefont {Schneeloch}, \citenamefont {Zhong}, \citenamefont {Gu},
  \citenamefont {Wang},\ and\ \citenamefont {Zeljkovic}}]{HZhao18PRB}%
  \BibitemOpen
  \bibfield  {author} {\bibinfo {author} {\bibfnamefont {H.}~\bibnamefont
  {Zhao}}, \bibinfo {author} {\bibfnamefont {B.}~\bibnamefont {Rachmilowitz}},
  \bibinfo {author} {\bibfnamefont {Z.}~\bibnamefont {Ren}}, \bibinfo {author}
  {\bibfnamefont {R.}~\bibnamefont {Han}}, \bibinfo {author} {\bibfnamefont
  {J.}~\bibnamefont {Schneeloch}}, \bibinfo {author} {\bibfnamefont
  {R.}~\bibnamefont {Zhong}}, \bibinfo {author} {\bibfnamefont
  {G.}~\bibnamefont {Gu}}, \bibinfo {author} {\bibfnamefont {Z.}~\bibnamefont
  {Wang}}, \ and\ \bibinfo {author} {\bibfnamefont {I.}~\bibnamefont
  {Zeljkovic}},\ }\bibfield  {title} {\enquote {\bibinfo {title}
  {{Superconducting proximity effect in a topological insulator using Fe(Te,
  Se)}},}\ }\href {\doibase 10.1103/PhysRevB.97.224504} {\bibfield  {journal}
  {\bibinfo  {journal} {Phys. Rev. B}\ }\textbf {\bibinfo {volume} {97}},\
  \bibinfo {pages} {224504} (\bibinfo {year} {2018})}\BibitemShut {NoStop}%
\bibitem [{\citenamefont {Perconte}\ \emph {et~al.}(2018)\citenamefont
  {Perconte}, \citenamefont {Cuellar}, \citenamefont {Moreau-Luchaire},
  \citenamefont {Piquemal-Banci}, \citenamefont {Galceran}, \citenamefont
  {Kidambi}, \citenamefont {Martin}, \citenamefont {Hofmann}, \citenamefont
  {Bernard}, \citenamefont {Dlubak} \emph {et~al.}}]{Perconte18Nphys}%
  \BibitemOpen
  \bibfield  {author} {\bibinfo {author} {\bibfnamefont {D.}~\bibnamefont
  {Perconte}}, \bibinfo {author} {\bibfnamefont {F.~A.}\ \bibnamefont
  {Cuellar}}, \bibinfo {author} {\bibfnamefont {C.}~\bibnamefont
  {Moreau-Luchaire}}, \bibinfo {author} {\bibfnamefont {M.}~\bibnamefont
  {Piquemal-Banci}}, \bibinfo {author} {\bibfnamefont {R.}~\bibnamefont
  {Galceran}}, \bibinfo {author} {\bibfnamefont {P.~R}\ \bibnamefont
  {Kidambi}}, \bibinfo {author} {\bibfnamefont {M.-B.}\ \bibnamefont {Martin}},
  \bibinfo {author} {\bibfnamefont {S.}~\bibnamefont {Hofmann}}, \bibinfo
  {author} {\bibfnamefont {R.}~\bibnamefont {Bernard}}, \bibinfo {author}
  {\bibfnamefont {B.}~\bibnamefont {Dlubak}},  \emph {et~al.},\ }\bibfield
  {title} {\enquote {\bibinfo {title} {{Tunable Klein-like tunnelling of
  high-temperature superconducting pairs into graphene}},}\ }\href {\doibase
  10.1038/nphys4278} {\bibfield  {journal} {\bibinfo  {journal} {Nat. Phys.}\
  }\textbf {\bibinfo {volume} {14}},\ \bibinfo {pages} {25} (\bibinfo {year}
  {2018})}\BibitemShut {NoStop}%
\bibitem [{\citenamefont {Xu}\ \emph {et~al.}(2014)\citenamefont {Xu},
  \citenamefont {Liu}, \citenamefont {Richardella}, \citenamefont {Belopolski},
  \citenamefont {Alidoust}, \citenamefont {Neupane}, \citenamefont {Bian},
  \citenamefont {Samarth},\ and\ \citenamefont {Hasan}}]{SYXu14PRB}%
  \BibitemOpen
  \bibfield  {author} {\bibinfo {author} {\bibfnamefont {S.-Y.}\ \bibnamefont
  {Xu}}, \bibinfo {author} {\bibfnamefont {C.}~\bibnamefont {Liu}}, \bibinfo
  {author} {\bibfnamefont {A.}~\bibnamefont {Richardella}}, \bibinfo {author}
  {\bibfnamefont {I.}~\bibnamefont {Belopolski}}, \bibinfo {author}
  {\bibfnamefont {N.}~\bibnamefont {Alidoust}}, \bibinfo {author}
  {\bibfnamefont {M.}~\bibnamefont {Neupane}}, \bibinfo {author} {\bibfnamefont
  {G.}~\bibnamefont {Bian}}, \bibinfo {author} {\bibfnamefont {N.}~\bibnamefont
  {Samarth}}, \ and\ \bibinfo {author} {\bibfnamefont {M.~Z.}\ \bibnamefont
  {Hasan}},\ }\bibfield  {title} {\enquote {\bibinfo {title} {Fermi-level
  electronic structure of a topological-insulator/cuprate-superconductor based
  heterostructure in the superconducting proximity effect regime},}\ }\href
  {\doibase 10.1103/PhysRevB.90.085128} {\bibfield  {journal} {\bibinfo
  {journal} {Phys. Rev. B}\ }\textbf {\bibinfo {volume} {90}},\ \bibinfo
  {pages} {085128} (\bibinfo {year} {2014})}\BibitemShut {NoStop}%
\bibitem [{\citenamefont {Yilmaz}\ \emph {et~al.}(2014)\citenamefont {Yilmaz},
  \citenamefont {Pletikosi\ifmmode~\acute{c}\else \'{c}\fi{}}, \citenamefont
  {Weber}, \citenamefont {Sadowski}, \citenamefont {Gu}, \citenamefont
  {Caruso}, \citenamefont {Sinkovic},\ and\ \citenamefont
  {Valla}}]{Yilmaz14PRL}%
  \BibitemOpen
  \bibfield  {author} {\bibinfo {author} {\bibfnamefont {T.}~\bibnamefont
  {Yilmaz}}, \bibinfo {author} {\bibfnamefont {I.}~\bibnamefont
  {Pletikosi\ifmmode~\acute{c}\else \'{c}\fi{}}}, \bibinfo {author}
  {\bibfnamefont {A.~P.}\ \bibnamefont {Weber}}, \bibinfo {author}
  {\bibfnamefont {J.~T.}\ \bibnamefont {Sadowski}}, \bibinfo {author}
  {\bibfnamefont {G.~D.}\ \bibnamefont {Gu}}, \bibinfo {author} {\bibfnamefont
  {A.~N.}\ \bibnamefont {Caruso}}, \bibinfo {author} {\bibfnamefont
  {B.}~\bibnamefont {Sinkovic}}, \ and\ \bibinfo {author} {\bibfnamefont
  {T.}~\bibnamefont {Valla}},\ }\bibfield  {title} {\enquote {\bibinfo {title}
  {{Absence of a Proximity Effect for a Thin-Films of a
  ${\mathrm{Bi}}_{2}{\mathrm{Se}}_{3}$ Topological Insulator Grown on Top of a
  ${\mathrm{Bi}}_{2}{\mathrm{Sr}}_{2}{\mathrm{CaCu}}_{2}{\mathrm{O}}_{8+\ensuremath{\delta}}$
  Cuprate Superconductor}},}\ }\href {\doibase 10.1103/PhysRevLett.113.067003}
  {\bibfield  {journal} {\bibinfo  {journal} {Phys. Rev. Lett.}\ }\textbf
  {\bibinfo {volume} {113}},\ \bibinfo {pages} {067003} (\bibinfo {year}
  {2014})}\BibitemShut {NoStop}%
\bibitem [{\citenamefont {Qian}\ \emph {et~al.}(2014)\citenamefont {Qian},
  \citenamefont {Liu}, \citenamefont {Fu},\ and\ \citenamefont
  {Li}}]{XFQian14science}%
  \BibitemOpen
  \bibfield  {author} {\bibinfo {author} {\bibfnamefont {X.}~\bibnamefont
  {Qian}}, \bibinfo {author} {\bibfnamefont {J.}~\bibnamefont {Liu}}, \bibinfo
  {author} {\bibfnamefont {L.}~\bibnamefont {Fu}}, \ and\ \bibinfo {author}
  {\bibfnamefont {J.}~\bibnamefont {Li}},\ }\bibfield  {title} {\enquote
  {\bibinfo {title} {Quantum spin hall effect in two-dimensional transition
  metal dichalcogenides},}\ }\href {\doibase 10.1126/science.1256815}
  {\bibfield  {journal} {\bibinfo  {journal} {Science}\ }\textbf {\bibinfo
  {volume} {346}},\ \bibinfo {pages} {1344} (\bibinfo {year}
  {2014})}\BibitemShut {NoStop}%
\bibitem [{\citenamefont {Tang}\ \emph {et~al.}(2017)\citenamefont {Tang},
  \citenamefont {Zhang}, \citenamefont {Wong}, \citenamefont {Pedramrazi},
  \citenamefont {Tsai}, \citenamefont {Jia}, \citenamefont {Moritz},
  \citenamefont {Claassen}, \citenamefont {Ryu}, \citenamefont {Kahn} \emph
  {et~al.}}]{SJTang17nphys}%
  \BibitemOpen
  \bibfield  {author} {\bibinfo {author} {\bibfnamefont {S.}~\bibnamefont
  {Tang}}, \bibinfo {author} {\bibfnamefont {C.}~\bibnamefont {Zhang}},
  \bibinfo {author} {\bibfnamefont {D.}~\bibnamefont {Wong}}, \bibinfo {author}
  {\bibfnamefont {Z.}~\bibnamefont {Pedramrazi}}, \bibinfo {author}
  {\bibfnamefont {H.-Z.}\ \bibnamefont {Tsai}}, \bibinfo {author}
  {\bibfnamefont {C.}~\bibnamefont {Jia}}, \bibinfo {author} {\bibfnamefont
  {B.}~\bibnamefont {Moritz}}, \bibinfo {author} {\bibfnamefont
  {M.}~\bibnamefont {Claassen}}, \bibinfo {author} {\bibfnamefont
  {H.}~\bibnamefont {Ryu}}, \bibinfo {author} {\bibfnamefont {S.}~\bibnamefont
  {Kahn}},  \emph {et~al.},\ }\bibfield  {title} {\enquote {\bibinfo {title}
  {{Quantum spin Hall state in monolayer 1T'-WTe$_2$}},}\ }\href {\doibase
  10.1038/nphys4174} {\bibfield  {journal} {\bibinfo  {journal} {Nat. Phys.}\
  }\textbf {\bibinfo {volume} {13}},\ \bibinfo {pages} {683} (\bibinfo {year}
  {2017})}\BibitemShut {NoStop}%
\bibitem [{\citenamefont {Fei}\ \emph {et~al.}(2017)\citenamefont {Fei},
  \citenamefont {Palomaki}, \citenamefont {Wu}, \citenamefont {Zhao},
  \citenamefont {Cai}, \citenamefont {Sun}, \citenamefont {Nguyen},
  \citenamefont {Finney}, \citenamefont {Xu},\ and\ \citenamefont
  {Cobden}}]{ZYFei17nphys}%
  \BibitemOpen
  \bibfield  {author} {\bibinfo {author} {\bibfnamefont {Z.}~\bibnamefont
  {Fei}}, \bibinfo {author} {\bibfnamefont {T.}~\bibnamefont {Palomaki}},
  \bibinfo {author} {\bibfnamefont {S.}~\bibnamefont {Wu}}, \bibinfo {author}
  {\bibfnamefont {W.}~\bibnamefont {Zhao}}, \bibinfo {author} {\bibfnamefont
  {X.}~\bibnamefont {Cai}}, \bibinfo {author} {\bibfnamefont {B.}~\bibnamefont
  {Sun}}, \bibinfo {author} {\bibfnamefont {P.}~\bibnamefont {Nguyen}},
  \bibinfo {author} {\bibfnamefont {J.}~\bibnamefont {Finney}}, \bibinfo
  {author} {\bibfnamefont {X.}~\bibnamefont {Xu}}, \ and\ \bibinfo {author}
  {\bibfnamefont {D.~H.}\ \bibnamefont {Cobden}},\ }\bibfield  {title}
  {\enquote {\bibinfo {title} {{Edge conduction in monolayer WTe$_2$}},}\
  }\href {\doibase 10.1038/nphys4091} {\bibfield  {journal} {\bibinfo
  {journal} {Nat. Phys.}\ }\textbf {\bibinfo {volume} {13}},\ \bibinfo {pages}
  {677} (\bibinfo {year} {2017})}\BibitemShut {NoStop}%
\bibitem [{\citenamefont {Wu}\ \emph {et~al.}(2018)\citenamefont {Wu},
  \citenamefont {Fatemi}, \citenamefont {Gibson}, \citenamefont {Watanabe},
  \citenamefont {Taniguchi}, \citenamefont {Cava},\ and\ \citenamefont
  {Jarillo-Herrero}}]{SFWu18science}%
  \BibitemOpen
  \bibfield  {author} {\bibinfo {author} {\bibfnamefont {S.}~\bibnamefont
  {Wu}}, \bibinfo {author} {\bibfnamefont {V.}~\bibnamefont {Fatemi}}, \bibinfo
  {author} {\bibfnamefont {Q.~D.}\ \bibnamefont {Gibson}}, \bibinfo {author}
  {\bibfnamefont {K.}~\bibnamefont {Watanabe}}, \bibinfo {author}
  {\bibfnamefont {T.}~\bibnamefont {Taniguchi}}, \bibinfo {author}
  {\bibfnamefont {R.~J.}\ \bibnamefont {Cava}}, \ and\ \bibinfo {author}
  {\bibfnamefont {P.}~\bibnamefont {Jarillo-Herrero}},\ }\bibfield  {title}
  {\enquote {\bibinfo {title} {{Observation of the quantum spin Hall effect up
  to 100 kelvin in a monolayer crystal}},}\ }\href
  {https://science.sciencemag.org/content/359/6371/76} {\bibfield  {journal}
  {\bibinfo  {journal} {Science}\ }\textbf {\bibinfo {volume} {359}},\ \bibinfo
  {pages} {76} (\bibinfo {year} {2018})}\BibitemShut {NoStop}%
\bibitem [{\citenamefont {Chen}\ \emph {et~al.}(2018)\citenamefont {Chen},
  \citenamefont {Pai}, \citenamefont {Chan}, \citenamefont {Sun}, \citenamefont
  {Xu}, \citenamefont {Lin}, \citenamefont {Chou}, \citenamefont {Fedorov},\
  and\ \citenamefont {Chiang}}]{PChen18ncomm}%
  \BibitemOpen
  \bibfield  {author} {\bibinfo {author} {\bibfnamefont {P}~\bibnamefont
  {Chen}}, \bibinfo {author} {\bibfnamefont {W.~W.}\ \bibnamefont {Pai}},
  \bibinfo {author} {\bibfnamefont {Y.-H.}\ \bibnamefont {Chan}}, \bibinfo
  {author} {\bibfnamefont {W.-L.}\ \bibnamefont {Sun}}, \bibinfo {author}
  {\bibfnamefont {C.-Z.}\ \bibnamefont {Xu}}, \bibinfo {author} {\bibfnamefont
  {D.-S.}\ \bibnamefont {Lin}}, \bibinfo {author} {\bibfnamefont {M.~Y.}\
  \bibnamefont {Chou}}, \bibinfo {author} {\bibfnamefont {A.-V.}\ \bibnamefont
  {Fedorov}}, \ and\ \bibinfo {author} {\bibfnamefont {T.-C.}\ \bibnamefont
  {Chiang}},\ }\bibfield  {title} {\enquote {\bibinfo {title} {{Large
  quantum-spin-Hall gap in single-layer 1T'-WSe$_2$}},}\ }\href {\doibase
  10.1038/s41467-018-04395-2} {\bibfield  {journal} {\bibinfo  {journal} {Nat.
  Commun.}\ }\textbf {\bibinfo {volume} {9}},\ \bibinfo {pages} {2003}
  (\bibinfo {year} {2018})}\BibitemShut {NoStop}%
\bibitem [{\citenamefont {Weng}\ \emph {et~al.}(2015)\citenamefont {Weng},
  \citenamefont {Ranjbar}, \citenamefont {Liang}, \citenamefont {Song},
  \citenamefont {Khazaei}, \citenamefont {Yunoki}, \citenamefont {Arai},
  \citenamefont {Kawazoe}, \citenamefont {Fang},\ and\ \citenamefont
  {Dai}}]{HMWeng15PRB}%
  \BibitemOpen
  \bibfield  {author} {\bibinfo {author} {\bibfnamefont {H.}~\bibnamefont
  {Weng}}, \bibinfo {author} {\bibfnamefont {A.}~\bibnamefont {Ranjbar}},
  \bibinfo {author} {\bibfnamefont {Y.}~\bibnamefont {Liang}}, \bibinfo
  {author} {\bibfnamefont {Z.}~\bibnamefont {Song}}, \bibinfo {author}
  {\bibfnamefont {M.}~\bibnamefont {Khazaei}}, \bibinfo {author} {\bibfnamefont
  {S.}~\bibnamefont {Yunoki}}, \bibinfo {author} {\bibfnamefont
  {M.}~\bibnamefont {Arai}}, \bibinfo {author} {\bibfnamefont {Y.}~\bibnamefont
  {Kawazoe}}, \bibinfo {author} {\bibfnamefont {Z.}~\bibnamefont {Fang}}, \
  and\ \bibinfo {author} {\bibfnamefont {X.}~\bibnamefont {Dai}},\ }\bibfield
  {title} {\enquote {\bibinfo {title} {{Large-gap two-dimensional topological
  insulator in oxygen functionalized MXene}},}\ }\href {\doibase
  10.1103/PhysRevB.92.075436} {\bibfield  {journal} {\bibinfo  {journal} {Phys.
  Rev. B}\ }\textbf {\bibinfo {volume} {92}},\ \bibinfo {pages} {075436}
  (\bibinfo {year} {2015})}\BibitemShut {NoStop}%
\bibitem [{\citenamefont {Si}\ \emph {et~al.}(2016)\citenamefont {Si},
  \citenamefont {Jin}, \citenamefont {Zhou}, \citenamefont {Sun},\ and\
  \citenamefont {Liu}}]{SChen16NL}%
  \BibitemOpen
  \bibfield  {author} {\bibinfo {author} {\bibfnamefont {C.}~\bibnamefont
  {Si}}, \bibinfo {author} {\bibfnamefont {K.-H.}\ \bibnamefont {Jin}},
  \bibinfo {author} {\bibfnamefont {J.}~\bibnamefont {Zhou}}, \bibinfo {author}
  {\bibfnamefont {Z.}~\bibnamefont {Sun}}, \ and\ \bibinfo {author}
  {\bibfnamefont {F.}~\bibnamefont {Liu}},\ }\bibfield  {title} {\enquote
  {\bibinfo {title} {{Large-Gap Quantum Spin Hall State in MXenes: d-Band
  Topological Order in a Triangular Lattice}},}\ }\href {\doibase
  10.1021/acs.nanolett.6b03118} {\bibfield  {journal} {\bibinfo  {journal}
  {Nano Lett.}\ }\textbf {\bibinfo {volume} {16}},\ \bibinfo {pages} {6584}
  (\bibinfo {year} {2016})}\BibitemShut {NoStop}%
\bibitem [{\citenamefont {Reis}\ \emph {et~al.}(2017)\citenamefont {Reis},
  \citenamefont {Li}, \citenamefont {Dudy}, \citenamefont {Bauernfeind},
  \citenamefont {Glass}, \citenamefont {Hanke}, \citenamefont {Thomale},
  \citenamefont {Sch{\"a}fer},\ and\ \citenamefont {Claessen}}]{Reis17science}%
  \BibitemOpen
  \bibfield  {author} {\bibinfo {author} {\bibfnamefont {F.}~\bibnamefont
  {Reis}}, \bibinfo {author} {\bibfnamefont {G.}~\bibnamefont {Li}}, \bibinfo
  {author} {\bibfnamefont {L.}~\bibnamefont {Dudy}}, \bibinfo {author}
  {\bibfnamefont {M.}~\bibnamefont {Bauernfeind}}, \bibinfo {author}
  {\bibfnamefont {S.}~\bibnamefont {Glass}}, \bibinfo {author} {\bibfnamefont
  {W.}~\bibnamefont {Hanke}}, \bibinfo {author} {\bibfnamefont
  {R.}~\bibnamefont {Thomale}}, \bibinfo {author} {\bibfnamefont
  {J.}~\bibnamefont {Sch{\"a}fer}}, \ and\ \bibinfo {author} {\bibfnamefont
  {R.}~\bibnamefont {Claessen}},\ }\bibfield  {title} {\enquote {\bibinfo
  {title} {{Bismuthene on a SiC substrate: A candidate for a high-temperature
  quantum spin Hall material}},}\ }\href {\doibase 10.1126/science.aai8142}
  {\bibfield  {journal} {\bibinfo  {journal} {Science}\ }\textbf {\bibinfo
  {volume} {357}},\ \bibinfo {pages} {287} (\bibinfo {year}
  {2017})}\BibitemShut {NoStop}%
\bibitem [{\citenamefont {Hsu}\ \emph {et~al.}(2015)\citenamefont {Hsu},
  \citenamefont {Huang}, \citenamefont {Chuang}, \citenamefont {Kuo},
  \citenamefont {Liu}, \citenamefont {Lin},\ and\ \citenamefont
  {Bansil}}]{Hsu15NJP}%
  \BibitemOpen
  \bibfield  {author} {\bibinfo {author} {\bibfnamefont {C.-H.}\ \bibnamefont
  {Hsu}}, \bibinfo {author} {\bibfnamefont {Z.-Q.}\ \bibnamefont {Huang}},
  \bibinfo {author} {\bibfnamefont {F.-C.}\ \bibnamefont {Chuang}}, \bibinfo
  {author} {\bibfnamefont {C.-C.}\ \bibnamefont {Kuo}}, \bibinfo {author}
  {\bibfnamefont {Y.-T.}\ \bibnamefont {Liu}}, \bibinfo {author} {\bibfnamefont
  {H.}~\bibnamefont {Lin}}, \ and\ \bibinfo {author} {\bibfnamefont
  {A.}~\bibnamefont {Bansil}},\ }\bibfield  {title} {\enquote {\bibinfo {title}
  {{The nontrivial electronic structure of Bi/Sb honeycombs on {SiC}(0001)}},}\
  }\href {\doibase 10.1088/1367-2630/17/2/025005} {\bibfield  {journal}
  {\bibinfo  {journal} {New J. Phys.}\ }\textbf {\bibinfo {volume} {17}},\
  \bibinfo {pages} {025005} (\bibinfo {year} {2015})}\BibitemShut {NoStop}%
\bibitem [{\citenamefont {Wrasse}\ and\ \citenamefont
  {Schmidt}(2014)}]{Wrasse14NL}%
  \BibitemOpen
  \bibfield  {author} {\bibinfo {author} {\bibfnamefont {E.~O.}\ \bibnamefont
  {Wrasse}}\ and\ \bibinfo {author} {\bibfnamefont {T.~M.}\ \bibnamefont
  {Schmidt}},\ }\bibfield  {title} {\enquote {\bibinfo {title} {{Prediction of
  Two-Dimensional Topological Crystalline Insulator in PbSe Monolayer}},}\
  }\href {\doibase 10.1021/nl502481f} {\bibfield  {journal} {\bibinfo
  {journal} {Nano Lett.}\ }\textbf {\bibinfo {volume} {14}},\ \bibinfo {pages}
  {5717} (\bibinfo {year} {2014})}\BibitemShut {NoStop}%
\bibitem [{\citenamefont {Liu}\ \emph {et~al.}(2015)\citenamefont {Liu},
  \citenamefont {Qian},\ and\ \citenamefont {Fu}}]{JWLiu15NL}%
  \BibitemOpen
  \bibfield  {author} {\bibinfo {author} {\bibfnamefont {J.}~\bibnamefont
  {Liu}}, \bibinfo {author} {\bibfnamefont {X.}~\bibnamefont {Qian}}, \ and\
  \bibinfo {author} {\bibfnamefont {L.}~\bibnamefont {Fu}},\ }\bibfield
  {title} {\enquote {\bibinfo {title} {{Crystal Field Effect Induced
  Topological Crystalline Insulators In Monolayer IV¨CVI Semiconductors}},}\
  }\href {\doibase 10.1021/acs.nanolett.5b00308} {\bibfield  {journal}
  {\bibinfo  {journal} {Nano Lett.}\ }\textbf {\bibinfo {volume} {15}},\
  \bibinfo {pages} {2657} (\bibinfo {year} {2015})}\BibitemShut {NoStop}%
\bibitem [{\citenamefont {Wan}\ \emph {et~al.}(2017)\citenamefont {Wan},
  \citenamefont {Yao}, \citenamefont {Sun}, \citenamefont {Liu},\ and\
  \citenamefont {Zhang}}]{WHWan17AM}%
  \BibitemOpen
  \bibfield  {author} {\bibinfo {author} {\bibfnamefont {W.}~\bibnamefont
  {Wan}}, \bibinfo {author} {\bibfnamefont {Y.}~\bibnamefont {Yao}}, \bibinfo
  {author} {\bibfnamefont {L.}~\bibnamefont {Sun}}, \bibinfo {author}
  {\bibfnamefont {C.-C.}\ \bibnamefont {Liu}}, \ and\ \bibinfo {author}
  {\bibfnamefont {F.}~\bibnamefont {Zhang}},\ }\bibfield  {title} {\enquote
  {\bibinfo {title} {{Topological, valleytronic, and optical properties of
  monolayer PbS}},}\ }\href {\doibase 10.1002/adma.201604788} {\bibfield
  {journal} {\bibinfo  {journal} {Adv. Mater.}\ }\textbf {\bibinfo {volume}
  {29}},\ \bibinfo {pages} {1604788} (\bibinfo {year} {2017})}\BibitemShut
  {NoStop}%
\bibitem [{\citenamefont {Bernevig}\ \emph {et~al.}(2006)\citenamefont
  {Bernevig}, \citenamefont {Hughes},\ and\ \citenamefont
  {Zhang}}]{Bernevig06science}%
  \BibitemOpen
  \bibfield  {author} {\bibinfo {author} {\bibfnamefont {B.~A.}\ \bibnamefont
  {Bernevig}}, \bibinfo {author} {\bibfnamefont {T.~L.}\ \bibnamefont
  {Hughes}}, \ and\ \bibinfo {author} {\bibfnamefont {S.-C.}\ \bibnamefont
  {Zhang}},\ }\bibfield  {title} {\enquote {\bibinfo {title} {{Quantum Spin
  Hall Effect and Topological Phase Transition in HgTe Quantum Wells}},}\
  }\href {\doibase 10.1126/science.1133734} {\bibfield  {journal} {\bibinfo
  {journal} {Science}\ }\textbf {\bibinfo {volume} {314}},\ \bibinfo {pages}
  {1757} (\bibinfo {year} {2006})}\BibitemShut {NoStop}%
\bibitem [{\citenamefont {Stewart}(2011)}]{Stewart11RMP}%
  \BibitemOpen
  \bibfield  {author} {\bibinfo {author} {\bibfnamefont {G.~R.}\ \bibnamefont
  {Stewart}},\ }\bibfield  {title} {\enquote {\bibinfo {title}
  {Superconductivity in iron compounds},}\ }\href {\doibase
  10.1103/RevModPhys.83.1589} {\bibfield  {journal} {\bibinfo  {journal} {Rev.
  Mod. Phys.}\ }\textbf {\bibinfo {volume} {83}},\ \bibinfo {pages} {1589}
  (\bibinfo {year} {2011})}\BibitemShut {NoStop}%
\bibitem [{\citenamefont {Hirschfeld}\ \emph {et~al.}(2011)\citenamefont
  {Hirschfeld}, \citenamefont {Korshunov},\ and\ \citenamefont
  {Mazin}}]{Hirschfeld11RPP}%
  \BibitemOpen
  \bibfield  {author} {\bibinfo {author} {\bibfnamefont {P.~J.}\ \bibnamefont
  {Hirschfeld}}, \bibinfo {author} {\bibfnamefont {M.~M.}\ \bibnamefont
  {Korshunov}}, \ and\ \bibinfo {author} {\bibfnamefont {I.~I.}\ \bibnamefont
  {Mazin}},\ }\bibfield  {title} {\enquote {\bibinfo {title} {{Gap symmetry and
  structure of Fe-based superconductors}},}\ }\href {\doibase
  10.1088/0034-4885/74/12/124508} {\bibfield  {journal} {\bibinfo  {journal}
  {Rep. Prog. Phys.}\ }\textbf {\bibinfo {volume} {74}},\ \bibinfo {pages}
  {124508} (\bibinfo {year} {2011})}\BibitemShut {NoStop}%
\bibitem [{\citenamefont {Zhang}\ \emph {et~al.}(2018)\citenamefont {Zhang},
  \citenamefont {Yaji}, \citenamefont {Hashimoto}, \citenamefont {Ota},
  \citenamefont {Kondo}, \citenamefont {Okazaki}, \citenamefont {Wang},
  \citenamefont {Wen}, \citenamefont {Gu}, \citenamefont {Ding},\ and\
  \citenamefont {Shin}}]{PZhang18science}%
  \BibitemOpen
  \bibfield  {author} {\bibinfo {author} {\bibfnamefont {P.}~\bibnamefont
  {Zhang}}, \bibinfo {author} {\bibfnamefont {K.}~\bibnamefont {Yaji}},
  \bibinfo {author} {\bibfnamefont {T.}~\bibnamefont {Hashimoto}}, \bibinfo
  {author} {\bibfnamefont {Y.}~\bibnamefont {Ota}}, \bibinfo {author}
  {\bibfnamefont {T.}~\bibnamefont {Kondo}}, \bibinfo {author} {\bibfnamefont
  {K.}~\bibnamefont {Okazaki}}, \bibinfo {author} {\bibfnamefont
  {Z.}~\bibnamefont {Wang}}, \bibinfo {author} {\bibfnamefont {J.}~\bibnamefont
  {Wen}}, \bibinfo {author} {\bibfnamefont {G.~D.}\ \bibnamefont {Gu}},
  \bibinfo {author} {\bibfnamefont {H.}~\bibnamefont {Ding}}, \ and\ \bibinfo
  {author} {\bibfnamefont {S.}~\bibnamefont {Shin}},\ }\bibfield  {title}
  {\enquote {\bibinfo {title} {Observation of topological superconductivity on
  the surface of an iron-based superconductor},}\ }\href {\doibase
  10.1126/science.aan4596} {\bibfield  {journal} {\bibinfo  {journal}
  {Science}\ }\textbf {\bibinfo {volume} {360}},\ \bibinfo {pages} {182}
  (\bibinfo {year} {2018})}\BibitemShut {NoStop}%
\bibitem [{\citenamefont {Wang}\ \emph
  {et~al.}(2018{\natexlab{b}})\citenamefont {Wang}, \citenamefont {Kong},
  \citenamefont {Fan}, \citenamefont {Chen}, \citenamefont {Zhu}, \citenamefont
  {Liu}, \citenamefont {Cao}, \citenamefont {Sun}, \citenamefont {Du},
  \citenamefont {Schneeloch}, \citenamefont {Zhong}, \citenamefont {Gu},
  \citenamefont {Fu}, \citenamefont {Ding},\ and\ \citenamefont
  {Gao}}]{DFWang18science}%
  \BibitemOpen
  \bibfield  {author} {\bibinfo {author} {\bibfnamefont {D.}~\bibnamefont
  {Wang}}, \bibinfo {author} {\bibfnamefont {L.}~\bibnamefont {Kong}}, \bibinfo
  {author} {\bibfnamefont {P.}~\bibnamefont {Fan}}, \bibinfo {author}
  {\bibfnamefont {H.}~\bibnamefont {Chen}}, \bibinfo {author} {\bibfnamefont
  {S.}~\bibnamefont {Zhu}}, \bibinfo {author} {\bibfnamefont {W.}~\bibnamefont
  {Liu}}, \bibinfo {author} {\bibfnamefont {L.}~\bibnamefont {Cao}}, \bibinfo
  {author} {\bibfnamefont {Y.}~\bibnamefont {Sun}}, \bibinfo {author}
  {\bibfnamefont {S.}~\bibnamefont {Du}}, \bibinfo {author} {\bibfnamefont
  {J.}~\bibnamefont {Schneeloch}}, \bibinfo {author} {\bibfnamefont
  {R.}~\bibnamefont {Zhong}}, \bibinfo {author} {\bibfnamefont
  {G.}~\bibnamefont {Gu}}, \bibinfo {author} {\bibfnamefont {L.}~\bibnamefont
  {Fu}}, \bibinfo {author} {\bibfnamefont {H.}~\bibnamefont {Ding}}, \ and\
  \bibinfo {author} {\bibfnamefont {H.-J.}\ \bibnamefont {Gao}},\ }\bibfield
  {title} {\enquote {\bibinfo {title} {{Evidence for Majorana bound states in
  an iron-based superconductor}},}\ }\href {\doibase 10.1126/science.aao1797}
  {\bibfield  {journal} {\bibinfo  {journal} {Science}\ }\textbf {\bibinfo
  {volume} {362}},\ \bibinfo {pages} {333} (\bibinfo {year}
  {2018}{\natexlab{b}})}\BibitemShut {NoStop}%
\bibitem [{\citenamefont {Zhang}\ \emph
  {et~al.}(2019{\natexlab{b}})\citenamefont {Zhang}, \citenamefont {Wang},
  \citenamefont {Wu}, \citenamefont {Yaji}, \citenamefont {Ishida},
  \citenamefont {Kohama}, \citenamefont {Dai}, \citenamefont {Sun},
  \citenamefont {Bareille}, \citenamefont {Kuroda} \emph
  {et~al.}}]{PZhang19nphys}%
  \BibitemOpen
  \bibfield  {author} {\bibinfo {author} {\bibfnamefont {P.}~\bibnamefont
  {Zhang}}, \bibinfo {author} {\bibfnamefont {Z.}~\bibnamefont {Wang}},
  \bibinfo {author} {\bibfnamefont {X.}~\bibnamefont {Wu}}, \bibinfo {author}
  {\bibfnamefont {K.}~\bibnamefont {Yaji}}, \bibinfo {author} {\bibfnamefont
  {Y.}~\bibnamefont {Ishida}}, \bibinfo {author} {\bibfnamefont
  {Y.}~\bibnamefont {Kohama}}, \bibinfo {author} {\bibfnamefont
  {G.}~\bibnamefont {Dai}}, \bibinfo {author} {\bibfnamefont {Y.}~\bibnamefont
  {Sun}}, \bibinfo {author} {\bibfnamefont {C.}~\bibnamefont {Bareille}},
  \bibinfo {author} {\bibfnamefont {K.}~\bibnamefont {Kuroda}},  \emph
  {et~al.},\ }\bibfield  {title} {\enquote {\bibinfo {title} {{Multiple
  topological states in iron-based superconductors}},}\ }\href {\doibase
  10.1038/s41567-018-0280-z} {\bibfield  {journal} {\bibinfo  {journal} {Nat.
  Phys.}\ }\textbf {\bibinfo {volume} {15}},\ \bibinfo {pages} {41} (\bibinfo
  {year} {2019}{\natexlab{b}})}\BibitemShut {NoStop}%
\bibitem [{Note5()}]{Note5}%
  \BibitemOpen
  \bibinfo {note} {Hard-wall boundary conditions are applicable because of the quadratic terms in our low-energy model.}\BibitemShut {Stop}%
\bibitem [{\citenamefont {Zhang}\ \emph {et~al.}(2016)\citenamefont {Zhang},
  \citenamefont {Lu},\ and\ \citenamefont {Shen}}]{ZhangSB16NJP}%
  \BibitemOpen
  \bibfield  {author} {\bibinfo {author} {\bibfnamefont {S.-B.}\ \bibnamefont
  {Zhang}}, \bibinfo {author} {\bibfnamefont {H.-Z.}\ \bibnamefont {Lu}}, \
  and\ \bibinfo {author} {\bibfnamefont {S.-Q.}\ \bibnamefont {Shen}},\
  }\bibfield  {title} {\enquote {\bibinfo {title} {Linear magnetoconductivity
  in an intrinsic topological {Weyl} semimetal},}\ }\href
  {http://stacks.iop.org/1367-2630/18/i=5/a=053039} {\bibfield  {journal}
  {\bibinfo  {journal} {New J. Phys.}\ }\textbf {\bibinfo {volume} {18}},\
  \bibinfo {pages} {053039} (\bibinfo {year} {2016})}\BibitemShut {NoStop}%
\bibitem [{Note1()}]{Note1}%
  \BibitemOpen
  \bibinfo {note} {For convenience, we assume the pairing potential $\Delta
  ({\protect \bf k})$ independent of chemical potential $\mu $. This should be
  justified since $\Delta ({\protect \bf k})$ is induced via the proximity
  effect.}\BibitemShut {Stop}%
\bibitem [{Sup()}]{SupplementalMaterial}%
  \BibitemOpen
  \href@noop {} {\bibinfo  {journal} {See the Supplemental Material for the
  details}\ }\BibitemShut {NoStop}%
\bibitem [{\citenamefont {Jackiw}\ and\ \citenamefont
  {Rebbi}(1976)}]{Jackiw76PRD}%
  \BibitemOpen
\bibfield  {journal} {  }\bibfield  {author} {\bibinfo {author} {\bibfnamefont
  {R.}~\bibnamefont {Jackiw}}\ and\ \bibinfo {author} {\bibfnamefont
  {C.}~\bibnamefont {Rebbi}},\ }\bibfield  {title} {\enquote {\bibinfo {title}
  {Solitons with fermion number 1/2},}\ }\href {\doibase
  10.1103/PhysRevD.13.3398} {\bibfield  {journal} {\bibinfo  {journal} {Phys.
  Rev. D}\ }\textbf {\bibinfo {volume} {13}},\ \bibinfo {pages} {3398}
  (\bibinfo {year} {1976})}\BibitemShut {NoStop}%
  \bibitem [{\citenamefont {Benalcazar}\ \emph
  {et~al.}(2017{\natexlab{a}})\citenamefont {Benalcazar}, \citenamefont
  {Bernevig},\ and\ \citenamefont {Hughes}}]{Benalcazar17science}%
  \BibitemOpen
  \bibfield  {author} {\bibinfo {author} {\bibfnamefont {W.~A.}\ \bibnamefont
  {Benalcazar}}, \bibinfo {author} {\bibfnamefont {B.~A.}\ \bibnamefont
  {Bernevig}}, \ and\ \bibinfo {author} {\bibfnamefont {T.~L.}\ \bibnamefont
  {Hughes}},\ }\bibfield  {title} {\enquote {\bibinfo {title} {Quantized
  electric multipole insulators},}\ }\href {\doibase 10.1126/science.aah6442}
  {\bibfield  {journal} {\bibinfo  {journal} {Science}\ }\textbf {\bibinfo
  {volume} {357}},\ \bibinfo {pages} {61} (\bibinfo {year}
  {2017}{\natexlab{a}})}\BibitemShut {NoStop}%
\bibitem [{\citenamefont {Benalcazar}\ \emph
  {et~al.}(2017{\natexlab{b}})\citenamefont {Benalcazar}, \citenamefont
  {Bernevig},\ and\ \citenamefont {Hughes}}]{Benalcazar17PRB}%
  \BibitemOpen
  \bibfield  {author} {\bibinfo {author} {\bibfnamefont {W.~A.}\ \bibnamefont
  {Benalcazar}}, \bibinfo {author} {\bibfnamefont {B.~A.}\ \bibnamefont
  {Bernevig}}, \ and\ \bibinfo {author} {\bibfnamefont {T.~L.}\ \bibnamefont
  {Hughes}},\ }\bibfield  {title} {\enquote {\bibinfo {title} {{Electric
  multipole moments, topological multipole moment pumping, and chiral hinge
  states in crystalline insulators}},}\ }\href {\doibase
  10.1103/PhysRevB.96.245115} {\bibfield  {journal} {\bibinfo  {journal} {Phys.
  Rev. B}\ }\textbf {\bibinfo {volume} {96}},\ \bibinfo {pages} {245115}
  (\bibinfo {year} {2017}{\natexlab{b}})}\BibitemShut {NoStop}%
\bibitem [{\citenamefont {Song}\ \emph {et~al.}(2017)\citenamefont {Song},
  \citenamefont {Fang},\ and\ \citenamefont {Fang}}]{ZDSong17PRL}%
  \BibitemOpen
  \bibfield  {author} {\bibinfo {author} {\bibfnamefont {Z.}~\bibnamefont
  {Song}}, \bibinfo {author} {\bibfnamefont {Z.}~\bibnamefont {Fang}}, \ and\
  \bibinfo {author} {\bibfnamefont {C.}~\bibnamefont {Fang}},\ }\bibfield
  {title} {\enquote {\bibinfo {title} {{$(d\ensuremath{-}2)$-Dimensional Edge
  States of Rotation Symmetry Protected Topological States}},}\ }\href
  {\doibase 10.1103/PhysRevLett.119.246402} {\bibfield  {journal} {\bibinfo
  {journal} {Phys. Rev. Lett.}\ }\textbf {\bibinfo {volume} {119}},\ \bibinfo
  {pages} {246402} (\bibinfo {year} {2017})}\BibitemShut {NoStop}%
\bibitem [{Note2()}]{Note2}%
  \BibitemOpen
  \bibinfo {note} {The bulk gap is given by $m_{\protect \text {g}}=m_{0}$ if
  $v_{x(y)}^{2}>2m_{0}m_{x(y)}$ and $m_{\protect \text {g}}=|v_{x(y)}|\protect
  [m_{0}/m_{x(y)}-v_{x(y)}^{2}/4m_{x(y)}^{2}]^{1/2}$ otherwise.}\BibitemShut
  {Stop}%
\bibitem [{Note3()}]{Note3}%
  \BibitemOpen
  \bibinfo {note} {$\Delta _{\protect \text {eff}}^{x(y)}$ can also switch sign
  at a $\mu $ larger than the bulk gap, which happens in the case of QSHI with
  small inverted gaps $2m_{0}m<v^{2}$. In this case, the edge states coexist
  with bulk states, and Majorana corner states persist even the bulk is not
  insulating.}\BibitemShut {Stop}%
\bibitem [{\citenamefont {Liu}\ \emph {et~al.}(2008)\citenamefont {Liu},
  \citenamefont {Hughes}, \citenamefont {Qi}, \citenamefont {Wang},\ and\
  \citenamefont {Zhang}}]{CXLiu08PRLb}%
  \BibitemOpen
  \bibfield  {author} {\bibinfo {author} {\bibfnamefont {C.}~\bibnamefont
  {Liu}}, \bibinfo {author} {\bibfnamefont {T.~L.}\ \bibnamefont {Hughes}},
  \bibinfo {author} {\bibfnamefont {X.-L.}\ \bibnamefont {Qi}}, \bibinfo
  {author} {\bibfnamefont {K.}~\bibnamefont {Wang}}, \ and\ \bibinfo {author}
  {\bibfnamefont {S.-C.}\ \bibnamefont {Zhang}},\ }\bibfield  {title} {\enquote
  {\bibinfo {title} {{Quantum Spin Hall Effect in Inverted Type-II
  Semiconductors}},}\ }\href {\doibase 10.1103/PhysRevLett.100.236601}
  {\bibfield  {journal} {\bibinfo  {journal} {Phys. Rev. Lett.}\ }\textbf
  {\bibinfo {volume} {100}},\ \bibinfo {pages} {236601} (\bibinfo {year}
  {2008})}\BibitemShut {NoStop}%
\bibitem [{\citenamefont {Knez}\ \emph {et~al.}(2011)\citenamefont {Knez},
  \citenamefont {Du},\ and\ \citenamefont {Sullivan}}]{Knez11prl}%
  \BibitemOpen
  \bibfield  {author} {\bibinfo {author} {\bibfnamefont {I.}~\bibnamefont
  {Knez}}, \bibinfo {author} {\bibfnamefont {R.~R.}\ \bibnamefont {Du}}, \ and\
  \bibinfo {author} {\bibfnamefont {G.}~\bibnamefont {Sullivan}},\ }\bibfield
  {title} {\enquote {\bibinfo {title} {{Evidence for Helical Edge Modes in
  Inverted $\mathrm{InAs}/\mathrm{GaSb}$ Quantum Wells}},}\ }\href {\doibase
  10.1103/PhysRevLett.107.136603} {\bibfield  {journal} {\bibinfo  {journal}
  {Phys. Rev. Lett.}\ }\textbf {\bibinfo {volume} {107}},\ \bibinfo {pages}
  {136603} (\bibinfo {year} {2011})}\BibitemShut {NoStop}%
\bibitem [{\citenamefont {Krishtopenko}\ and\ \citenamefont
  {Teppe}(2018)}]{Krishtopenkoeaap18science}%
  \BibitemOpen
  \bibfield  {author} {\bibinfo {author} {\bibfnamefont {S.~S.}\ \bibnamefont
  {Krishtopenko}}\ and\ \bibinfo {author} {\bibfnamefont {F.}~\bibnamefont
  {Teppe}},\ }\bibfield  {title} {\enquote {\bibinfo {title} {{Quantum spin
  Hall insulator with a large bandgap, Dirac fermions, and bilayer graphene
  analog}},}\ }\href {\doibase 10.1126/sciadv.aap7529} {\bibfield  {journal}
  {\bibinfo  {journal} {Sci. Adv.}\ }\textbf {\bibinfo {volume} {4}},\ \bibinfo
  {pages} {eaap7529} (\bibinfo {year} {2018})}\BibitemShut {NoStop}%
  \bibitem [{Note6()}]{Note6}%
  \BibitemOpen
  \bibinfo {note} {In the customary regularization, we replace $k_{x(y)}\rightarrow\sin k_{x(y)} $ and $k^2_{x(y)}\rightarrow 2[1-\cos k_{x(y)}]$ in the model \eqref{eq:General-model} and obtain a tight-binding lattice model. We also perform the calculation for a different lattice model and obtain the same features, see the Supplemental Material.}\BibitemShut {Stop}%
\bibitem [{\citenamefont {Asano}(2001)}]{Asano01PRB}%
  \BibitemOpen
  \bibfield  {author} {\bibinfo {author} {\bibfnamefont {Y.}~\bibnamefont
  {Asano}},\ }\bibfield  {title} {\enquote {\bibinfo {title} {{Numerical method
  for dc Josephson current between d-wave superconductors}},}\ }\href {\doibase
  10.1103/PhysRevB.63.052512} {\bibfield  {journal} {\bibinfo  {journal} {Phys.
  Rev. B}\ }\textbf {\bibinfo {volume} {63}},\ \bibinfo {pages} {052512}
  (\bibinfo {year} {2001})}\BibitemShut {NoStop}%
\bibitem [{\citenamefont {Mart\'{\i}n-Rodero}\ \emph
  {et~al.}(1994)\citenamefont {Mart\'{\i}n-Rodero}, \citenamefont
  {Garc\'{\i}a-Vidal},\ and\ \citenamefont {Levy~Yeyati}}]{Rodero94PRL}%
  \BibitemOpen
  \bibfield  {author} {\bibinfo {author} {\bibfnamefont {A.}~\bibnamefont
  {Mart\'{\i}n-Rodero}}, \bibinfo {author} {\bibfnamefont {F.~J.}\ \bibnamefont
  {Garc\'{\i}a-Vidal}}, \ and\ \bibinfo {author} {\bibfnamefont
  {A.}~\bibnamefont {Levy~Yeyati}},\ }\bibfield  {title} {\enquote {\bibinfo
  {title} {{Microscopic theory of Josephson mesoscopic constrictions}},}\
  }\href {\doibase 10.1103/PhysRevLett.72.554} {\bibfield  {journal} {\bibinfo
  {journal} {Phys. Rev. Lett.}\ }\textbf {\bibinfo {volume} {72}},\ \bibinfo
  {pages} {554} (\bibinfo {year} {1994})}\BibitemShut {NoStop}%
\bibitem [{\citenamefont {Furusaki}(1994)}]{Furusaki94PB}%
  \BibitemOpen
  \bibfield  {author} {\bibinfo {author} {\bibfnamefont {A.}~\bibnamefont
  {Furusaki}},\ }\bibfield  {title} {\enquote {\bibinfo {title} {{DC Josephson
  effect in dirty SNS junctions: numerical study}},}\ }\href {\doibase
  10.1016/0921-4526(94)90061-2} {\bibfield  {journal} {\bibinfo  {journal}
  {Phys. B}\ }\textbf {\bibinfo {volume} {203}},\ \bibinfo {pages} {214}
  (\bibinfo {year} {1994})}\BibitemShut {NoStop}%
\bibitem [{\citenamefont {Fu}\ and\ \citenamefont {Kane}(2009)}]{LFu09PRB}%
  \BibitemOpen
  \bibfield  {author} {\bibinfo {author} {\bibfnamefont {L.}~\bibnamefont
  {Fu}}\ and\ \bibinfo {author} {\bibfnamefont {C.~L.}\ \bibnamefont {Kane}},\
  }\bibfield  {title} {\enquote {\bibinfo {title} {{Josephson current and noise
  at a superconductor/quantum-spin-Hall-insulator/ superconductor junction}},}\
  }\href {\doibase 10.1103/PhysRevB.79.161408} {\bibfield  {journal} {\bibinfo
  {journal} {Phys. Rev. B}\ }\textbf {\bibinfo {volume} {79}},\ \bibinfo
  {pages} {161408(R)} (\bibinfo {year} {2009})}\BibitemShut {NoStop}%
\bibitem [{\citenamefont {Cr\'epin}\ and\ \citenamefont
  {Trauzettel}(2014)}]{Crepin14PRL}%
  \BibitemOpen
  \bibfield  {author} {\bibinfo {author} {\bibfnamefont {F.}~\bibnamefont
  {Cr\'epin}}\ and\ \bibinfo {author} {\bibfnamefont {B.}~\bibnamefont
  {Trauzettel}},\ }\bibfield  {title} {\enquote {\bibinfo {title} {{Parity
  Measurement in Topological Josephson Junctions}},}\ }\href {\doibase
  10.1103/PhysRevLett.112.077002} {\bibfield  {journal} {\bibinfo  {journal}
  {Phys. Rev. Lett.}\ }\textbf {\bibinfo {volume} {112}},\ \bibinfo {pages}
  {077002} (\bibinfo {year} {2014})}\BibitemShut {NoStop}%
\bibitem [{\citenamefont {Li}\ \emph {et~al.}(2018)\citenamefont {Li},
  \citenamefont {Zhang},\ and\ \citenamefont {Shen}}]{CALi18PRB}%
  \BibitemOpen
  \bibfield  {author} {\bibinfo {author} {\bibfnamefont {C.-A.}\ \bibnamefont
  {Li}}, \bibinfo {author} {\bibfnamefont {S.-B.}\ \bibnamefont {Zhang}}, \
  and\ \bibinfo {author} {\bibfnamefont {S.-Q.}\ \bibnamefont {Shen}},\
  }\bibfield  {title} {\enquote {\bibinfo {title} {{Hidden edge Dirac point and
  robust quantum edge transport in InAs/GaSb quantum wells}},}\ }\href
  {\doibase 10.1103/PhysRevB.97.045420} {\bibfield  {journal} {\bibinfo
  {journal} {Phys. Rev. B}\ }\textbf {\bibinfo {volume} {97}},\ \bibinfo
  {pages} {045420} (\bibinfo {year} {2018})}\BibitemShut {NoStop}%
\bibitem [{\citenamefont {Beenakker}(1991)}]{Beenakker91PRL}%
  \BibitemOpen
  \bibfield  {author} {\bibinfo {author} {\bibfnamefont {C.~W.~J.}\
  \bibnamefont {Beenakker}},\ }\bibfield  {title} {\enquote {\bibinfo {title}
  {{Universal Limit of Critical-Current Fluctuations in Mesoscopic Josephson
  Junctions}},}\ }\href {\doibase 10.1103/PhysRevLett.67.3836} {\bibfield
  {journal} {\bibinfo  {journal} {Phys. Rev. Lett.}\ }\textbf {\bibinfo
  {volume} {67}},\ \bibinfo {pages} {3836} (\bibinfo {year}
  {1991})}\BibitemShut {NoStop}%
\bibitem [{\citenamefont {Fu}\ and\ \citenamefont {Kane}(2008)}]{Fu08PRL}%
  \BibitemOpen
  \bibfield  {author} {\bibinfo {author} {\bibfnamefont {L.}~\bibnamefont
  {Fu}}\ and\ \bibinfo {author} {\bibfnamefont {C.~L.}\ \bibnamefont {Kane}},\
  }\bibfield  {title} {\enquote {\bibinfo {title} {{Superconducting Proximity
  Effect and Majorana Fermions at the Surface of a Topological Insulator}},}\
  }\href {\doibase 10.1103/PhysRevLett.100.096407} {\bibfield  {journal}
  {\bibinfo  {journal} {Phys. Rev. Lett.}\ }\textbf {\bibinfo {volume} {100}},\
  \bibinfo {pages} {096407} (\bibinfo {year} {2008})}\BibitemShut {NoStop}%
\bibitem [{\citenamefont {Lutchyn}\ \emph {et~al.}(2010)\citenamefont
  {Lutchyn}, \citenamefont {Sau},\ and\ \citenamefont
  {Das~Sarma}}]{Lutchyn10PRL}%
  \BibitemOpen
  \bibfield  {author} {\bibinfo {author} {\bibfnamefont {R.~M.}\ \bibnamefont
  {Lutchyn}}, \bibinfo {author} {\bibfnamefont {J.~D.}\ \bibnamefont {Sau}}, \
  and\ \bibinfo {author} {\bibfnamefont {S.}~\bibnamefont {Das~Sarma}},\
  }\bibfield  {title} {\enquote {\bibinfo {title} {{Majorana Fermions and a
  Topological Phase Transition in Semiconductor-Superconductor
  Heterostructures}},}\ }\href {\doibase 10.1103/PhysRevLett.105.077001}
  {\bibfield  {journal} {\bibinfo  {journal} {Phys. Rev. Lett.}\ }\textbf
  {\bibinfo {volume} {105}},\ \bibinfo {pages} {077001} (\bibinfo {year}
  {2010})}\BibitemShut {NoStop}%
\bibitem [{\citenamefont {Sau}\ \emph {et~al.}(2010)\citenamefont {Sau},
  \citenamefont {Lutchyn}, \citenamefont {Tewari},\ and\ \citenamefont
  {Das~Sarma}}]{Sau10PRL}%
  \BibitemOpen
  \bibfield  {author} {\bibinfo {author} {\bibfnamefont {Jay~D.}\ \bibnamefont
  {Sau}}, \bibinfo {author} {\bibfnamefont {Roman~M.}\ \bibnamefont {Lutchyn}},
  \bibinfo {author} {\bibfnamefont {Sumanta}\ \bibnamefont {Tewari}}, \ and\
  \bibinfo {author} {\bibfnamefont {S.}~\bibnamefont {Das~Sarma}},\ }\bibfield
  {title} {\enquote {\bibinfo {title} {{Generic New Platform for Topological
  Quantum Computation Using Semiconductor Heterostructures}},}\ }\href
  {\doibase 10.1103/PhysRevLett.104.040502} {\bibfield  {journal} {\bibinfo
  {journal} {Phys. Rev. Lett.}\ }\textbf {\bibinfo {volume} {104}},\ \bibinfo
  {pages} {040502} (\bibinfo {year} {2010})}\BibitemShut {NoStop}%
\bibitem [{\citenamefont {Oreg}\ \emph {et~al.}(2010)\citenamefont {Oreg},
  \citenamefont {Refael},\ and\ \citenamefont {von Oppen}}]{Oreg10PRL}%
  \BibitemOpen
  \bibfield  {author} {\bibinfo {author} {\bibfnamefont {Y.}~\bibnamefont
  {Oreg}}, \bibinfo {author} {\bibfnamefont {G.}~\bibnamefont {Refael}}, \ and\
  \bibinfo {author} {\bibfnamefont {F.}~\bibnamefont {von Oppen}},\ }\bibfield
  {title} {\enquote {\bibinfo {title} {{Helical Liquids and Majorana Bound
  States in Quantum Wires}},}\ }\href {\doibase 10.1103/PhysRevLett.105.177002}
  {\bibfield  {journal} {\bibinfo  {journal} {Phys. Rev. Lett.}\ }\textbf
  {\bibinfo {volume} {105}},\ \bibinfo {pages} {177002} (\bibinfo {year}
  {2010})}\BibitemShut {NoStop}%
\bibitem [{\citenamefont {Liu}\ and\ \citenamefont
  {Zhang}(2013)}]{CXLiu13models}%
  \BibitemOpen
  \bibfield  {author} {\bibinfo {author} {\bibfnamefont {C.}~\bibnamefont
  {Liu}}\ and\ \bibinfo {author} {\bibfnamefont {S.-C.}\ \bibnamefont
  {Zhang}},\ }\bibfield  {title} {\enquote {\bibinfo {title} {Models and
  materials for topological insulators},}\ }in\ \href {\doibase
  10.1016/B978-0-444-63314-9.00003-2} {\emph {\bibinfo {booktitle}
  {Contemporary Concepts of Condensed Matter Science}}},\ Vol.~\bibinfo
  {volume} {6}\ (\bibinfo  {publisher} {Elsevier},\ \bibinfo {year} {2013})\
  pp.\ \bibinfo {pages} {59--89}\BibitemShut {NoStop}%
\bibitem [{Note7()}]{Note7}%
  \BibitemOpen
  \bibinfo {note} {This corresponds to a specific axis of the crystal as discussed in Ref.\ \citep{XFQian14science}}\BibitemShut {Stop}%
\bibitem [{\citenamefont {Lee}\ \emph {et~al.}(1981)\citenamefont {Lee},\ and\ \citenamefont {Fisher}}]{PALee81PRL}%
  \BibitemOpen
  \bibfield  {author} {\bibinfo {author} {\bibfnamefont {P. A.}~\bibnamefont
  {Lee}},\ and\
  \bibinfo {author} {\bibfnamefont {D. S.}~\bibnamefont {Fisher}},\ }\bibfield
  {title} {\enquote {\bibinfo {title} {{Anderson Localization in Two Dimensions}},}\ }\href {\doibase 10.1103/PhysRevLett.47.882}
  {\bibfield  {journal} {\bibinfo  {journal} {Phys. Rev. Lett.}\ }\textbf
  {\bibinfo {volume} {47}},\ \bibinfo {pages} {882} (\bibinfo {year}
  {1981})}\BibitemShut {NoStop}%
\bibitem [{\citenamefont {Yu}\ \emph {et~al.}(2010)\citenamefont {Yu},
  \citenamefont {Benervig},\ and\ \citenamefont {Dai}}]{RYu11PRB}%
  \BibitemOpen
  \bibfield  {author} {\bibinfo {author} {\bibfnamefont {R.}~\bibnamefont
  {Yu}}, \bibinfo {author} {\bibfnamefont {X. L.}~\bibnamefont {Qi}}, \bibinfo {author} {\bibfnamefont {A.}~\bibnamefont {Benervig}}, \bibinfo {author} {\bibfnamefont {Z.}~\bibnamefont {Fang}}, \ and\
  \bibinfo {author} {\bibfnamefont {X.}~\bibnamefont {Dai}},\ }\bibfield
  {title} {\enquote {\bibinfo {title} {{Equivalent expression of ${\mathbb{Z}}_{2}$ topological invariant for band insulators using the non-Abelian Berry connection}},}\ }\href {\doibase 10.1103/PhysRevB.84.075119}
  {\bibfield  {journal} {\bibinfo  {journal} {Phys. Rev. B}\ }\textbf
  {\bibinfo {volume} {84}},\ \bibinfo {pages} {075119} (\bibinfo {year}
  {2011})}\BibitemShut {NoStop}%
\bibitem [{\citenamefont {Fidkowski}\ \emph {et~al.}(2010)\citenamefont {Jackson},\ and\ \citenamefont {Klich}}]{Fidkowski11PRL}%
  \BibitemOpen
  \bibfield  {author} {\bibinfo {author} {\bibfnamefont {L.}~\bibnamefont
  {Fidkowski}}, \bibinfo {author} {\bibfnamefont {T. S.}~\bibnamefont {Jackson}}, \ and\
  \bibinfo {author} {\bibfnamefont {I.}~\bibnamefont {Klich}},\ }\bibfield
  {title} {\enquote {\bibinfo {title} {{Model Characterization of Gapless Edge Modes of Topological Insulators Using Intermediate Brillouin-Zone Functions}},}\ }\href {\doibase 10.1103/PhysRevLett.107.036601}
  {\bibfield  {journal} {\bibinfo  {journal} {Phys. Rev. Lett.}\ }\textbf
  {\bibinfo {volume} {107}},\ \bibinfo {pages} {036601} (\bibinfo {year}
  {2011})}\BibitemShut {NoStop}%
\end{thebibliography}
\end{document}